# Temperature-dependent EXAFS measurements of the Pb L3-edge allow quantification of the anharmonicity of the lead-halide bond of chlorine-substituted methylammonium (MA) lead triiodide

Götz Schuck[1]*, Daniel M. Többens[1], Dirk Wallacher[1], Nico Grimm[1], Tong Sy Tien[2] and Susan Schorr[1,3]

[1] Helmholtz-Zentrum Berlin für Materialien und Energie, Hahn-Meitner-Platz 1, 14109 Berlin, Germany

[2] Department of Basic Sciences, University of Fire Prevention and Fighting, 243 Khuat Duy Tien, Hanoi 120602, Vietnam

[3] Institut für Geologische Wissenschaften, Freie Universität Berlin, Malteserstr. 74, 12249 Berlin, Germany

## ABSTRACT

This article reports on studies of chlorine-substituted $MAPbI_3$ using combined temperature-dependent XRD synchrotron and Pb-L3 edge EXAFS to analyze the anharmonicity of the lead halide bond. The EXAFS parameters were described in the orthorhombic phase by an Einstein or $T^2$ type behavior, which was then compared with the experimental EXAFS parameters of the tetragonal/cubic phase. In the orthorhombic phase, it was observed that the asymmetry of the pair distribution function (cumulant $C_3$) in $MAPbCl_3$ is much lower than in $MAPbI_3$. Compared with the behavior in the orthorhombic phase, the anharmonicity changed after the phase transition to the room temperature phase, with $MAPbCl_3$ showing an increased anharmonicity and $MAPbI_3$ a decrease. The differences between $MAPbI_3$ and 2% chlorine substitution were small, both in the orthorhombic and tetragonal phases. By determining the structural parameters required to convert the effective force constants $k_0$ and $k_3$ resulting from the EXAFS analysis into the Morse potential parameters α and D, we could establish that our results agree with other experimental findings. Moreover, by using XRD we found that the $[PbX_6]$ octahedra shrink slightly in the tetragonal phase of $MAPbI_3$ and $MAPbI_{2.94}Cl_{0.06}$, towards increasing temperatures. This behavior in the tetragonal phase is related to the dominant negative tension effects observed by EXAFS.

## 1. INTRODUCTION

The evolution of halide perovskite photovoltaics [1] from their early beginnings [2] to advanced tandem solar cells with 30% efficiency [3] is very impressive. However, despite this extraordinary development, many fundamental questions still need to be answered in this research area.[4,5] We have made it our task to understand the fundamental properties of chlorine substituted $MAPbI_3$ perovskites as much as possible since chlorine substitution in methylammonium ($[CH_3NH_3]^+$, MA) lead trihalide ($X_3$; X = I, Cl) hybrid perovskites leads to an improvement of material stability and also to an improvement of electronic properties.[6,7] In doing so, we rely on basic crystallographic knowledge that was recently gained in our research group on hybrid perovskites.[8-11] This includes the realization that the solubility of chlorine in $MAPbI_3$ is limited, so that a maximum of 3 mol% chlorine can be incorporated in $MAPbI_3$ and a maximum of 1 mol% iodine in $MAPbCl_3$.[11] A basic understanding of the crystallographic properties of $MAPbI_{3-x}Cl_x$ requires knowledge of the temperature-dependent phase transformations that are taking place and that are caused by ordering phenomena of the MA molecule. The phase transformations in $MAPbX_3$ are largely determined by the interaction of the MA molecule with the $[PbX_6]$ octahedra framework structure. Thus, the cubic aristotype crystal structure (space group $Pm\bar{3}m$) of $MAPbI_3$,[12,13] where the MA molecule rotates freely, changes into a tetragonal structure (most commonly used *I*4/*mcm*) at 330 K,[8,10] in which rotation of the MA molecule is much more restricted (there is still a disorder in the crystal structure), and then changes into the ordered orthorhombic structure (*Pnma*) at 161 K,[14-16] in which the crystal structure is fully ordered. A similar phase transformation evolution is observed for $MAPbI_{2.94}Cl_{0.06}$, where slightly lower phase transformation temperatures were found compared to $MAPbI_3$ (326 K, cubic → tetragonal, and 155 K, tetragonal → orthorhombic).[17] For the chlorine halide, the cubic $Pm\bar{3}m$ crystal structure was observed down to a temperature of 177 K.[18] For lower temperatures, however, there are ambiguous observations in the literature. The intermediate structure of $MAPbCl_3$ (temperature range from 177 K to 172 K), has been described as tetragonal and eventually modulated [19], but could not be observed by others.[20] The low temperature modification is described as orthorhombic, space group *Pnma*.[21] Although this is the same space group symmetry as in the low temperature phase of $MAPbI_3$, the crystal structure differs because in $MAPbCl_3$, compared to the cubic structure, all three axes are doubled. FTIR investigations showed that, in the orthorhombic phase (T < 161 K for $MAPbI_3$ and T < 172 K for $MAPbCl_3$), the iodide is more strongly influenced by hydrogen bonding than the chloride.[22] Our QENS results on $MAPbI_3$, $MAPbCl_3$, and $MAPbI_{2.94}Cl_{0.06}$ showed that chlorine substitution in the orthorhombic phase leads to a weakening of the hydrogen bridge bonds since the characteristic relaxation times of the threefold MA molecule jump rotation around the C–N axis at 70 K in $MAPbCl_3$ (135 ps) and $MAPbI_{2.94}Cl_{0.06}$ (485 ps) are much shorter than in $MAPbI_3$ (1635 ps).[17] The structural counterpart of the changes in the dynamics of MA molecules caused by chlorine substitution is the influence of chlorine substitution on the tilting and distortion of the $[PbX_6]$ octahedra, especially in the temperature range of the orthorhombic-tetragonal phase transformation. In fact, phase transitions in lead-halide perovskites are rather directly linked to the flexibility of the overall inorganic framework, i.e. the anharmonicity of the octahedral tilting, which may be for example quantified with the Pb-X-Pb or X-Pb-X in plane angle.[23-26] Further important aspects of the anharmonicity of the inorganic framework in lead halide perovskites are the asymmetries in the Pb-X pair distribution function observed in the analysis of the local structure.[27-29] The different entropy contributions (stochastic structural fluctuations, anharmonicity and lattice softness) in hybrid perovskites are now considered to have a bigger influence on the optoelectronic properties than the dynamics of the MA molecule.[5] This finding is crucial since the optoelectronic properties are of greatest importance for the application as photovoltaic material. Many of the optoelectronic properties are, in turn, strongly influenced by the ordering processes of the orthorhombic/tetragonal phase transformation. The prominent importance of the anharmonicity of the halide motion becomes clear when one considers that, for example, an iodine shift of 0.4 Å leads to a considerable decrease in the maximum allowed charge carrier mobility (about 160 $cm^2\ V^{-1}\ s^{-1}$),[30] which can be interpreted as dipolar fluctuations of the charges.[5] Since the short-range order of the central lead atom is modified by the change of the octahedron geometry, one way to examine the anharmonicity of the lead-halide bond is X-ray absorption fine structure (XAFS). XAFS investigations of $MAPbX_3$ perovskites have so far been mostly applied to thin film samples, mainly to investigate the formation of precursors [31-33] or to study degrading processes.[34] To this end, the X-ray absorption near edge structure (XANES) region of the Pb L3-edge was mainly investigated, but more and more publications have been investigating the extended X-ray absorption fine structure (EXAFS) region to better understand the lead-halide bonds.[35,36] Because the interatomic distance, in our case the lead-halide distance, is the fundamental physical quantity, EXAFS is particularly sensitive (for details on the EXAFS methodology see section 2.3.2). It should be mentioned that Pb L3-edge EXAFS studies were available before interest in hydride perovskites increased in recent years.[37] Although there are some publications dealing with the EXAFS edges of the anion, with the work being confined to bromine,[38] we want to limit ourselves to the Pb L-3 edge for better comparability. Besides the consideration of the temperature dependence of the EXAFS Debye-Waller factor,[39] the framework of the temperature-dependent EXAFS cumulant analysis developed in the last decades [40-45] allows further statements about the properties of the investigated atomic pairs since statements about the anharmonicity can also be made.[46]

Through a combination of temperature-dependent EXAFS Pb L3-edge-investigations and temperature-dependent synchrotron XRD studies, the influence of the chlorine substitution on the EXAFS Debye-Waller factor, the distortions of the $[PbX_6]$ octahedra and thus the anharmonicity of the Pb-X bonds was investigated in three different samples: $MAPbI_3$, $MAPbCl_3$, and $MAPbI_{2.94}Cl_{0.06}$. The experimental determination of the effective force constants that govern the anharmonicity of the lead-halide bond enables the corresponding Morse parameters to be obtained, which allows comparison with other experimental results and theoretical calculations, such as DFT calculations.

## 2. EXPERIMENTAL DETAILS

### 2.1 Synthesis and sample handling

Small single crystals of $MAPbI_3$, $MAPbCl_3$, and $MAPbI_{2.94}Cl_{0.06}$, which were further ground in an agate mortar and used in powder form, were synthesized as described previously.[9,11,22] After synthesis, the powder was tested for phase purity using X-ray diffraction (XRD) (Bruker D8 powder diffractometer with sample spinner with Bragg–Brentano geometry, and Cu Ka1 radiation) and no impurity phases were detected. The samples were handled in the glove box or under an inert gas atmosphere (argon).

### 2.2 Synchrotron XRD and EXAFS measurements

The temperature-dependent Pb L3-edge (13.035 keV) EXAFS and the synchrotron XRD measurements were both performed on the KMC-2 beam line [47] at BESSY II, Helmholtz-Zentrum, Berlin, Germany. This beam line operates a graded SiGe monochromator constructed of two independent crystals (energy resolution of $E/\Delta E = 4000$). The beam intensity at the KMC-2 is stabilized to an accuracy of 0.3%. The two experiments, EXAFS and XRD, could be conducted because the KMC-2 beam line comprises two end stations: "XANES" and "Diffraction". X-ray powder diffraction data in the range of $4° < 2\Theta < 94.4°$ were collected at the "Diffraction" end station of KMC-2 beam line using a radiation energy of 8048 eV ($\lambda = 1.5406(1)$ Å). A modified Gifford-McMahon (GM) closed-cycle cryocooler, in-house label TMP-CCR-HXR, configured with a double Kapton cupola and helium exchange gas, was used. The samples were placed in a flat sample holder, and mounted in symmetric reflection geometry. Diffraction and EXAFS measurements were carried out in BESSY II TopUp mode at 298 mA.

For the EXAFS experiments, the cryostat was mounted at the Cryo-EXAFS stage (Fig. S1) inside a three-axis translation motor stage for fine tuning of the sample position. The EXAFS experiments were conducted in transmission geometry. The sample powder was weighed for 1.6 absorption lengths (for $MAPbI_3$: 16 mg/cm$^2$), so that an edge step of approx. 1.1 was achieved. The powder was then brushed onto Kapton tape and then folded and subsequently fixed in a plastic frame (Fig. S1). The EXAFS measurements in the orthorhombic phase were performed for $MAPbI_3$ at temperatures of 20, 45 K, and between 70 and 160 K in temperature steps of 5 K. For $MAPbI_{2.94}Cl_{0.06}$ EXAFS data was measured at temperatures of 20, 35, 50, 65, 80, 90, 95 K, and in the range from 105 to 165 K in 5 K steps. For $MAPbCl_3$, EXAFS measurements at 20, 50 and 80 K were taken, and between 110 and 160 K measurements were taken every 10 K. In addition, measurements were also carried out in the tetragonal/cubic phases (for $MAPbI_3$ at 195, 230 and 265 K; for $MAPbI_{2.94}Cl_{0.06}$ at 170, 175, 180, 190, 200, 220, and 260 K; and for $MAPbCl_3$ at 195, 230, 265 K). Both in EXAFS measurements (all samples) and in the synchrotron XRD measurements ($MAPbI_3$ and $MAPbI_{2.94}Cl_{0.06}$), the temperature was increased from the lowest temperatures (usually 20K) to higher temperatures (room temperature). A different temperature protocol of the synchrotron XRD measurements was used only for the low temperature structure of $MAPbCl_3$ here, the sample was cooled down to 180 K and then further cooled. For $MAPbCl_3$ low temperature phase the data were collected in temperature steps of -2 K until 140 K and in steps of -5 K until 35 K. The temperature-dependent synchrotron XRD measurements for $MAPbI_3$, $MAPbI_{2.94}Cl_{0.06}$ were performed at low temperatures at 35, 50 and 80 K, in the range from 90 K to 180 K in 5 K steps and in the tetragonal/cubic range (also for the cubic phase of $MAPbCl_3$) at temperatures of 190, 200, 220, 260 and 300 K (for $MAPbI_{2.94}Cl_{0.06}$ additionally at 320 and 340 K).

Each temperature step of the synchrotron XRD measurements took 2-5 minutes to heat up (or to cool down for the low temperature structure of $MAPbCl_3$), followed by 15 minutes equilibration and 37 minutes data collection, for a total of 54 minutes.

### 2.3 DATA ANALYSIS

#### 2.3.1 Synchrotron XRD

Structural analysis of the data was done using F.O.X. for simulated annealing,[48] and Fullprof for Rietveld refinement (Fullprof.2k. version 5.30 & 7.20).[49] The MA cation was treated throughout the analysis as a rigid body, with geometry taken from previous DFT optimization.[22] The refinement of anisotropic temperature factors for the halide atoms could not be done consistently for all data sets (for details, see supplement (section MSRD and MSD), so that in most cases isotropic temperature factors were refined with the Rietveld method. Details of the analysis of the orthorhombic $MAPbCl_3$ phase and individual results are given in supporting information.

#### 2.3.2 EXAFS

EXAFS covers an energy range from 30 to 1000 eV above the absorption edge. This allows the local structural environment of the absorber atom to be investigated. In this context, the coordination numbers N, the interatomic distances and the variation or distribution of these distances are important. The interatomic distance to the nearest neighbor atoms $R_{EXAFS}$ is the basic physical quantity to which the EXAFS is sensitive. The mean square relative displacement (MSRD) - i.e. the EXAFS Debye-Waller factor $\sigma^2$ - is important. The EXAFS Debye-Waller factor corresponds to the variation of the atomic distances. The relationship (also called "EXAFS formula") [50] of N, $R_{EXAFS}$ and $\sigma^2$ will be discussed later in this section. First, we want to show the way from the measured absorbance spectra to the EXAFS analysis (Fig. 1). The goal here is to Fourier transform the absorption "fine structure" $\chi(k)$ data after data normalization and background subtraction to obtain the radial distance distribution. The Athena software from the Demeter 0.9.26 software suite [51] was used for standardization and energy calibration of the EXAFS data.[50] Starting point of the EXAFS data treatment was the determination of the respective $E_0$ values of the temperature-dependent EXAFS spectra for the three analyzed samples. The first derivatives of the normalized sample spectra, and the corresponding normalized reference spectra (Pb foil) were exported from Athena, and the $E_0$ positions (first maximum) were determined by fitting a Gaussian function with fityk [52].

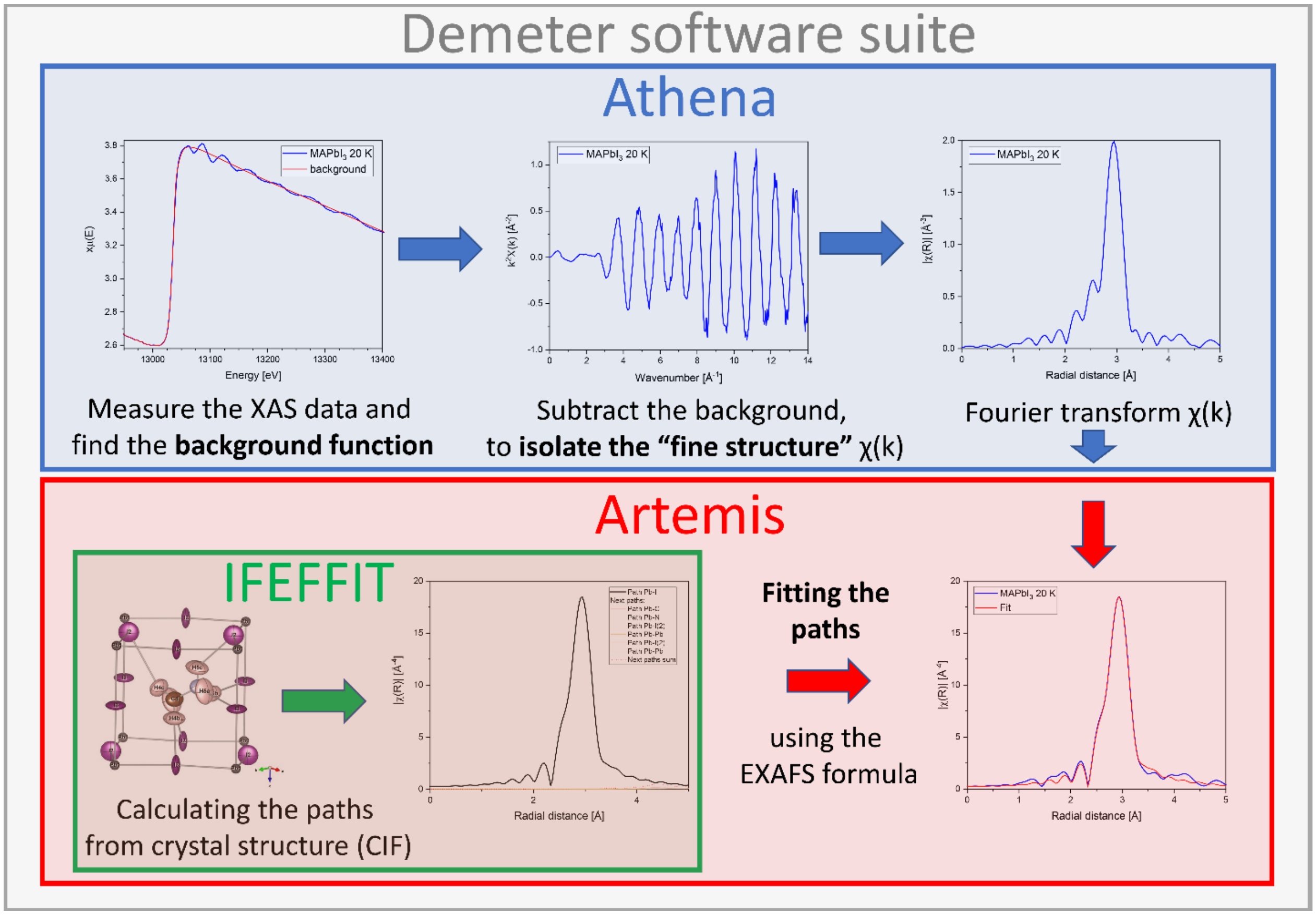

**Fig. 1** Simplified EXAFS "Workflow" overview, EXAFS data processing with Demeter 0.9.26 software suite, fitting the paths calculated from IFEFFIT with the EXAFS formula (as described in this section) in Artemis. Crystallographic Information File (CIF).

We chose this way because in Athena no error values for the energy calibration are available, with the way we chose beam damage or effects due to phase transitions could be excluded because $E_0$ remained constant within the $E_0$ variance of about 0.1 eV over the whole temperature range for all three samples. For the energy calibration, 13035 eV was used for the Pb reference. The following averaged $E_0$ values were determined for $MAPbI_3$: $\overline{E_0}$ = 13035.39(9) eV, for $MAPbI_{2.94}Cl_{0.06}$: $\overline{E_0}$ = 13036.17(6) eV and for $MAPbCl_3$: $\overline{E_0}$ = 13036.53(8) eV (Fig. S2). After the $E_0$ calibration (Fig. S3), the data for the respective temperature could be merged and a background correction ($R_{bkg}$ = 1.5 for all three samples)[53] was applied in Athena (Fig. 1). The actual EXAFS fits in Artemis were then performed in radial space (R range: 2 - 4 Å for $MAPbI_3$ and $MAPbI_{2.94}Cl_{0.06}$; and for $MAPbCl_3$ with an of R range: 1.5 - 3.0 Å) with multiple k weights (k = 1, 2, and 3) on the Fourier transformed data (k range: 3 - 14 $Å^{-1}$ for $MAPbI_3$ and $MAPbI_{2.94}Cl_{0.06}$; for $MAPbCl_3$ the k range is 3 - 12 $Å^{-1}$). In Artemis, the parameter $S_0^2$ (amplitude reduction factor), $R_{EXAFS}$ (distance from the absorbing atom to the scattering atom), N (degeneracy), $\sigma^2$(mean square radial displacement), $C_3$ (asymmetry parameter) and $C_4$ (symmetrical flattening or sharpening) of the EXAFS formula are used as fit parameters and thus the paths i (absorbing atom / scattering atom), calculated with the software framework IFEFFIT [51] (within Artemis, Fig. 1), are fitted individually. The EXAFS formula is:[41]

$$\chi(k) = S_0^2 \sum_i N_i \frac{f_i(k)}{kR_{EXAFS,i}^2} e^{-\frac{2R_{EXAFS,i}}{\lambda(k)}} e^{-2k^2\sigma_i^2+\frac{2}{3}k^4 C_{4,i}} \times \sin\left(2kR_{EXAFS,i} - \frac{4\sigma_i^2 k}{R_{EXAFS,i}}\left(1+\frac{R_{EXAFS,i}}{\lambda}\right) - \frac{4}{3}k^3 C_{3,i} + \delta_i(k)\right), \quad (1)$$

(k: photoelectron wavenumber, f: scattering amplitude, λ: photoelectron inelastic mean free path, δ: corresponding (back-)scattering phase). Furthermore, the EXAFS formula also includes the cumulant expansion, whereby the third cumulant $C_3$ corresponds to the asymmetry (skewness) of the pair distribution function and the fourth cumulant $C_4$ to a symmetrical flattening or sharpening of the pair distribution function. For the derivation of the EXAFS formula and the extension of the EXAFS formula by the cumulant analysis, we refer to the literature.[40,41,50,54,55]

## 3. RESULTS AND DISCUSSION

### 3.1 Synchrotron XRD

Based on the crystal structures for $MAPbI_3$ (T > 330 K cubic,[56] T > 161 K tetragonal,[8] and T < 161 K orthorhombic [14]) and $MAPbCl_3$ (T > 177 K cubic,[13] T > 171 K tetragonal, and T < 171 K orthorhombic with doubled unit cell [21]) known from literature, the temperature-dependent synchrotron XRD of the three samples was refined using the Rietveld method.

The Rietveld results of $MAPbI_3$ show very good agreement with the studies of Whitfield et al. (Fig. 2, S10a, and S11).[15]

$MAPbCl_3$ at temperatures above 170 K shows a cubic crystal structure, space group $Pm\overline{3}m$, which was refined using the structure model of Baikie.[13] Upon further cooling, at 170 K the appearance of a non-cubic phase is evident from beginning peak splitting and additional weak reflections and at 172 K the cubic phase had vanished completely. In order to describe this phase an orthorhombic lattice of a ≈ 8.0 Å, b ≈ 11.3 Å, c ≈ 7.9 Å (orthorhombic phase O2) was necessary (Fig. S4, S7, S10c, S11 and Table S1). This is in contrast to older descriptions, but in agreement with recent observations.[20,57] The intermediate, tetragonal phase is absent, and the observed orthorhombic phase is different. A satisfactory description of the peak intensities was only possible in space group *Pnma*. It should be noted that this fully-ordered structure is the same as observed for the low-temperature phases of $MAPbI_3$ and $MAPbBr_3$. At 150 K and below (Fig. S5 and S6) a second phase appears as a minor fraction, which can be described with good agreement by the known low-temperature structure of $MAPbCl_3$, which also has space group *Pnma*, but a larger unit cell of a ≈ 11.2 Å, b ≈ 11.3 Å, c ≈ 11.3 Å (orthorhombic phase O1).[21] However, this fraction only reached about 20% even at the lowest temperature of 30 K; the major part of the sample remained in the O2 structure with the smaller unit cell. We did not test it extensively, but from the comparison of a number of experiments with different samples it is probable that the formation of the large-cell structure is hindered in finely ground samples (Fig. S8). Measurements with coarser samples seem to form larger fractions of this phase, up to close to 100% with a very coarse sample. However, due to extremely bad particle statistics, no structure refinement was possible from these data. We wish to point out in particular that samples prepared for EXAFS experiments are very finely ground and thus have to be assumed to form nearly exclusively the small-cell form O2 at low temperatures. All attempts to describe all peaks with one phase failed. A few extra peaks appearing at 150 K and below remained unexplained. This could indicate a third fraction or unidentified structural variations in one of the two phases.

Based on the crystal structures of $MAPbI_3$, a temperature-dependent crystal structure investigation was carried out for $MAPbI_{2.94}Cl_{0.06}$ (Fig. 2, S9, S10b, S11 and Table S2). For the Rietveld refinement of $MAPbI_{2.94}Cl_{0.06}$, the chlorine and iodine occupation on the two anion positions was not refined to reduce the number of free parameters since, according to Franz et al.,[11] the composition can be determined from the lattice constants.

The results that are relevant to the analysis of EXAFS data are shown in Fig. 2. For all three samples, the unit cell volume increases with increasing temperature. The behavior at the orthorhombic/tetragonal phase transition is very similar for $MAPbI_3$ and $MAPbI_{2.94}Cl_{0.06}$. In both cases, a discontinuous jump of about 2% occurs (Fig. 2a). For $MAPbCl_3$ (Fig. 2b), a similar behavior in the temperature range of the orthorhombic/cubic phase transformation cannot be observed from the literature data and, from our Rietveld results, one would formerly see a continuous course here. The temperature behavior of the averaged lead-halide distance $R_{XRD}$, which can be derived from the Rietveld refinement, is interesting and of greater importance for the EXAFS analysis. The XRD lead-halide distances are averaged because the EXAFS data presented here cannot resolve the XRD observed distances (see Fig. S14, S15 and section 3.2 EXAFS). In contrast to the unit cell volume, which increases with increasing temperature, the lead-halide distances (Fig. 2c and 2d) at the transition from the orthorhombic to the tetragonal phase decrease with increasing temperature for all three investigated samples. For $MAPbI_3$ and $MAPbI_{2.94}Cl_{0.06}$ we observe a discontinuous decrease of the averaged lead-halide distance by about -0.01 Å in the region of the phase transition from the orthorhombic to the tetragonal phase (150 -165 K), based on our Rietveld results, a similar behavior can also be observed for $MAPbCl_3$, also here we can observe a discontinuous decrease of the averaged lead-chlorine distance by -0.012 Å near the orthorhombic to tetragonal phase transition (168 K - 170 K).

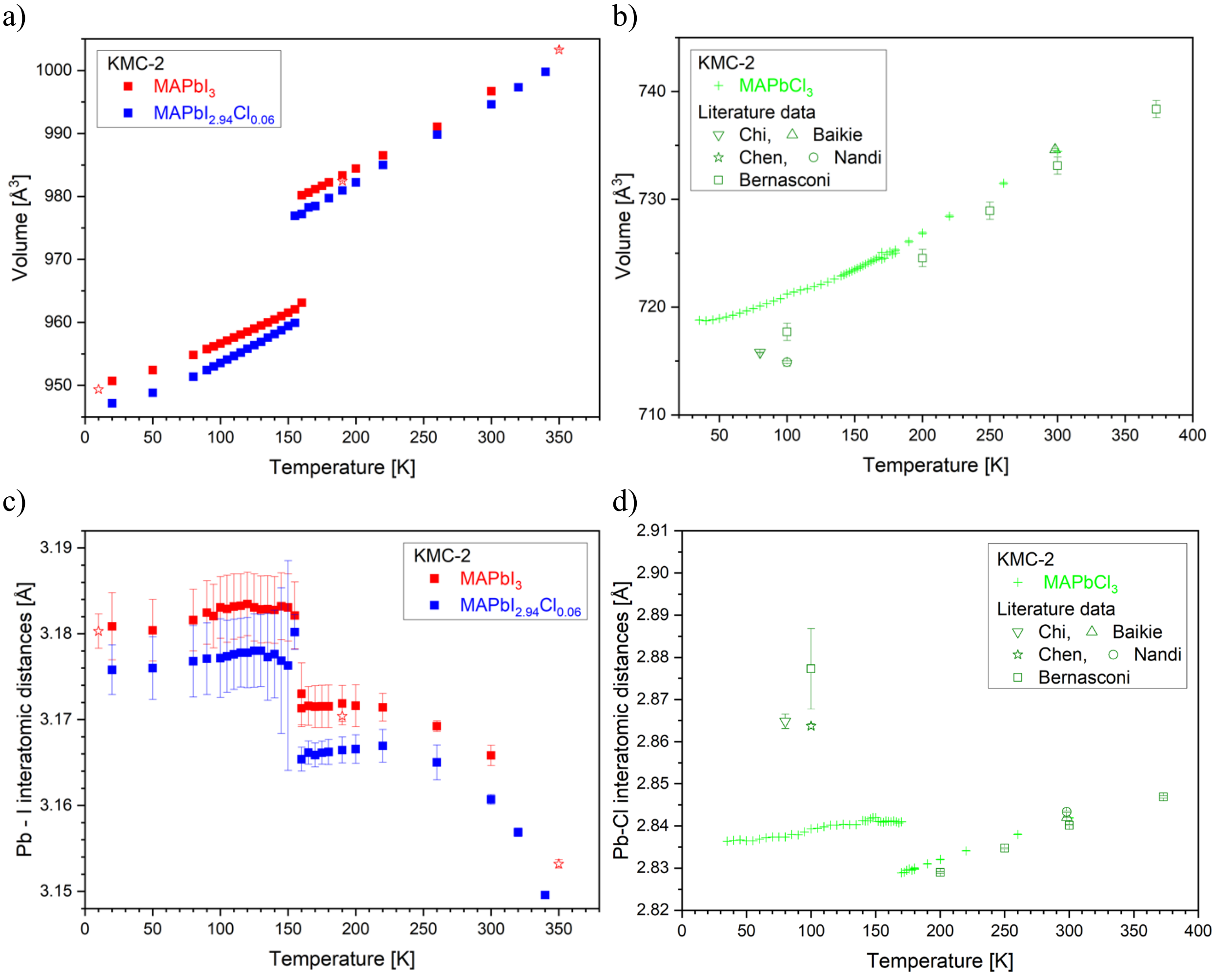

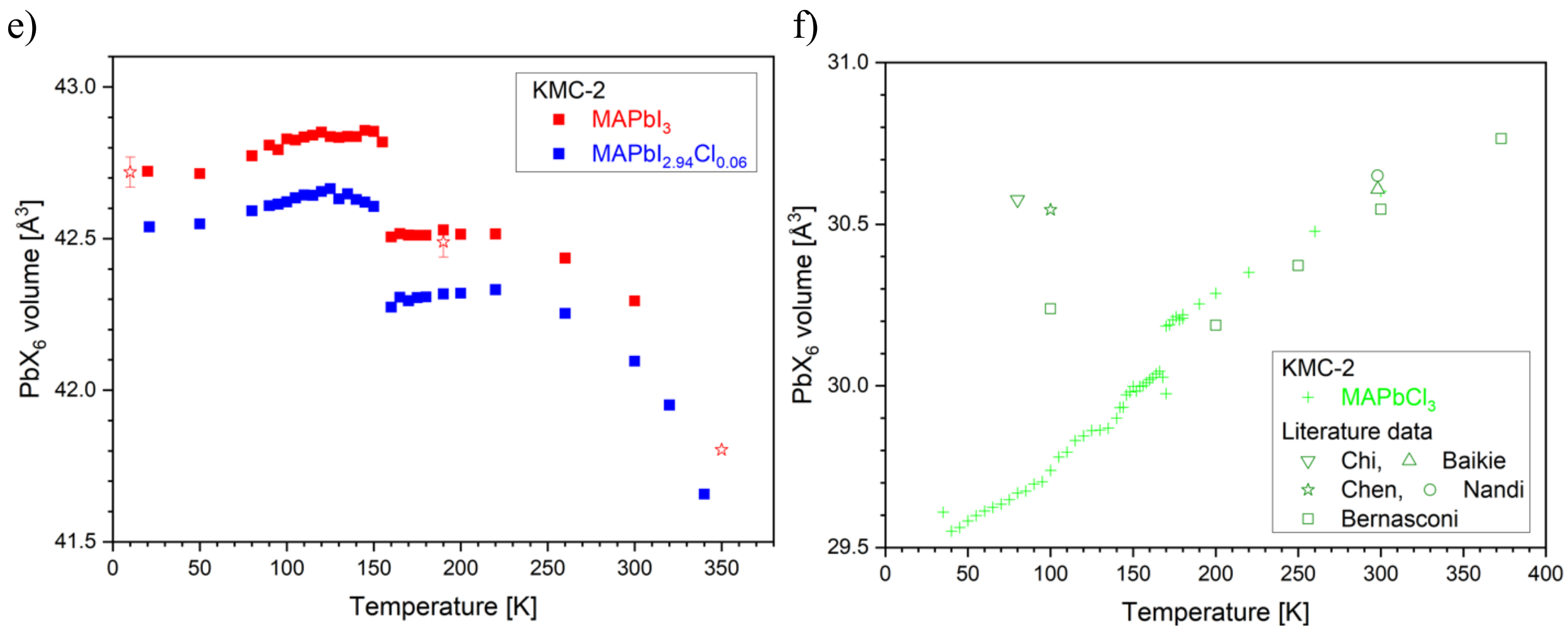

**Fig. 2** Results of the Rietveld analysis of synchrotron XRD data of $MAPbI_{3-x}Cl_x$. Temperature dependence of: the unit cell volume (a and b), averaged Pb-X interatomic distance $R_{XRD}$ (c and d), and $PbX_6$ octahedral volume (values from VESTA,[58] e and f) for $MAPbI_3$ (red color), $MAPbI_{2.94}Cl_{0.06}$ (blue color), and $MAPbCl_3$ (green color). $MAPbCl_3$ measured at KMC-2 (solid symbols) in comparison with literature values (open symbols; triangle pointing down: [21], triangle pointing up: [13], star: [59], square: [18], circle: [60]. For details on the deviation of Pb-Cl distances in the orthorhombic phase, see Fig. S15. Cubic volumes were converted into the pseudo tetragonal. For $MAPbI_3$ in comparison with literature values from Whitfield et al. [15] (red open stars).

In contrast to $MAPbI_3$ and $MAPbI_{2.94}Cl_{0.06}$, where the Pb-X distances significantly decrease towards the cubic phase with increasing temperature (by about -0.02 Å in the temperature range from 165 K to 260 K), a different behavior was observed in the tetragonal phase of $MAPbCl_3$, where the Pb-X distances increase with increasing temperature (by about +0.013 Å in the temperature range from 170 K to 300 K). Also remarkable are the changes of the lead-halide octahedra volumes in $MAPbI_3$ and $MAPbI_{2.94}Cl_{0.06}$. Unlike to the crystal lattice volume, which increases with increasing temperature, the volume of octahedra decreases with increasing temperature. This shrinking of the lead-halide octahedra with increasing temperatures cannot be observed in $MAPbCl_3$. Here, the volumes of the $[PbCl_6]$ octahedra at room temperature become even larger than in the case of low temperature regions. Through our diffraction investigations we can clarify that with increasing temperature the $[PbX_6]$ octahedra shrinks in the tetragonal room temperature phase of $MAPbI_3$ and $MAPbI_{2.94}Cl_{0.06}$ before the phase transition into the cubic phase, similar to tetragonal $SrZrO_3$ perovskite [61]. In the case of $SrZrO_3$, it was concluded that the $[ZrO_6]$ octahedra is not rigid.[61]

### 3.2 EXAFS

In the $k^3$-weighted Pb L3-edge EXAFS data in k-space shown in Fig. 3 and in the Fourier transforms of the EXAFS signal in Fig. 4 and S17, the temperature influence becomes already obvious. Also, large differences between $MAPbI_3$ and $MAPbI_{2.94}Cl_{0.06}$ (these two EXAFS findings look very similar) on the one hand, and $MAPbCl_3$ on the other hand are visible. In the orthorhombic phase, the intensities of $MAPbI_3$ and $MAPbI_{2.94}Cl_{0.06}$ decrease massively from 20 K to 150 K, especially in the range of larger k-values (between 8 and 15 $Å^{-1}$) as temperatures rise. For $MAPbCl_3$, this effect cannot be observed in this manner. However, we can also observe clearly that at higher temperatures above the tetragonal-orthorhombic phase transformation, only minor changes occur (Fig. 3 and 4). Based on our results from the Rietveld analysis of the synchrotron XRD data we conclude that only one lead-halide distance has to be considered for the development of a theoretical model to describe our temperature-dependent Pb L3-edge EXAFS data in the orthorhombic low temperature phase for $MAPbI_3$, $MAPbI_{2.94}Cl_{0.06}$ and $MAPbCl_3$. This is because that the distance resolution $\Delta r$ is limited by the observable k-space $\Delta k$ of our EXAFS measurements since the following applies: [62,63]

$$\Delta r = \pi/2\Delta k. \quad (2)$$

In our case this means that with a usable k-space of $\Delta k = 11$ $Å^{-1}$ (interval $k = 3 - 14$ $Å^{-1}$), $\Delta r$ of less than 0.143 Å can no longer be resolved unambiguously. The standard deviation $\sigma r$ from averaging the three different lead-iodine distances in $MAPbI_3$ (Fig. S14) and $MAPbI_{2.94}Cl_{0.06}$ determined in the Rietveld analysis is a maximum of 0.022 Å in the orthorhombic phase and is even smaller by an order of magnitude in the tetragonal phase.

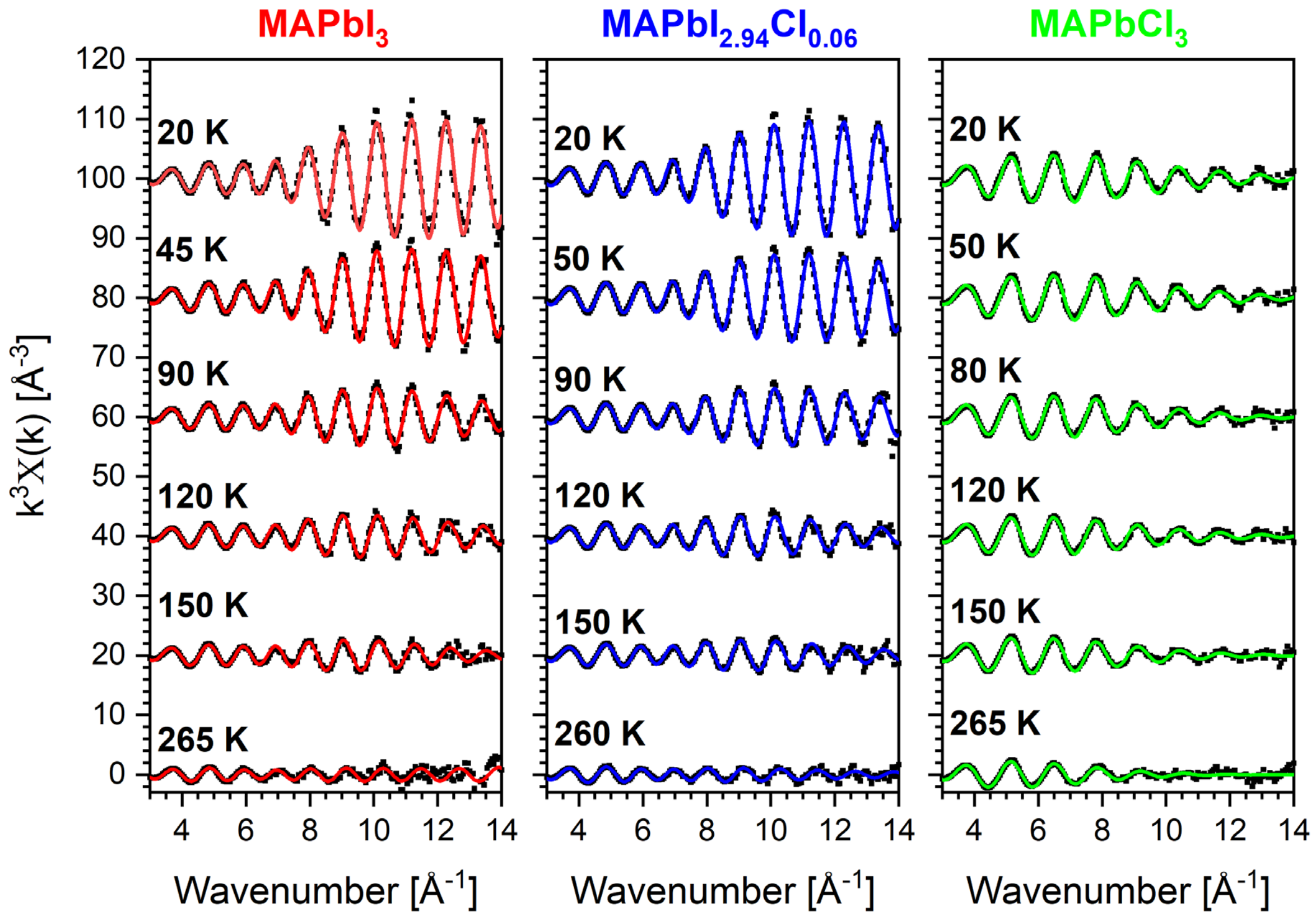

**Fig. 3** The temperature-dependent measured $k^3$-weighted Pb L3-edge EXAFS data in k-space (black squares) and the respective best fits for $MAPbI_3$ (red lines), for $MAPbI_{2.94}Cl_{0.06}$ (blue lines) and for $MAPbCl_3$ (green lines). For the six temperatures presented, an offset was chosen for ease of presentation, but this is the same for all measured samples. Note that the values of the ordinate are the same in all three figures.

Also, from crystal structure analyzes of $MAPbCl_3$ in the orthorhombic phase O1 known from literature,[21,59] we can conclude that the six different lead-chlorine distances ($\sigma r$ = 0.095 Å) (Fig. S15) cannot be resolved with our EXAFS data, since it is even more difficult here that only a k range from 3 to 12 $Å^{-1}$ can be used due to the weaker oscillations ($\Delta r$ < 0.174 Å). The orthorhombic phase observed here (phase O2) has an even smaller averaged lead-chlorine bond length splitting of $\sigma r$ = 0.05 Å, so that it is justified to assume only one averaged distance.

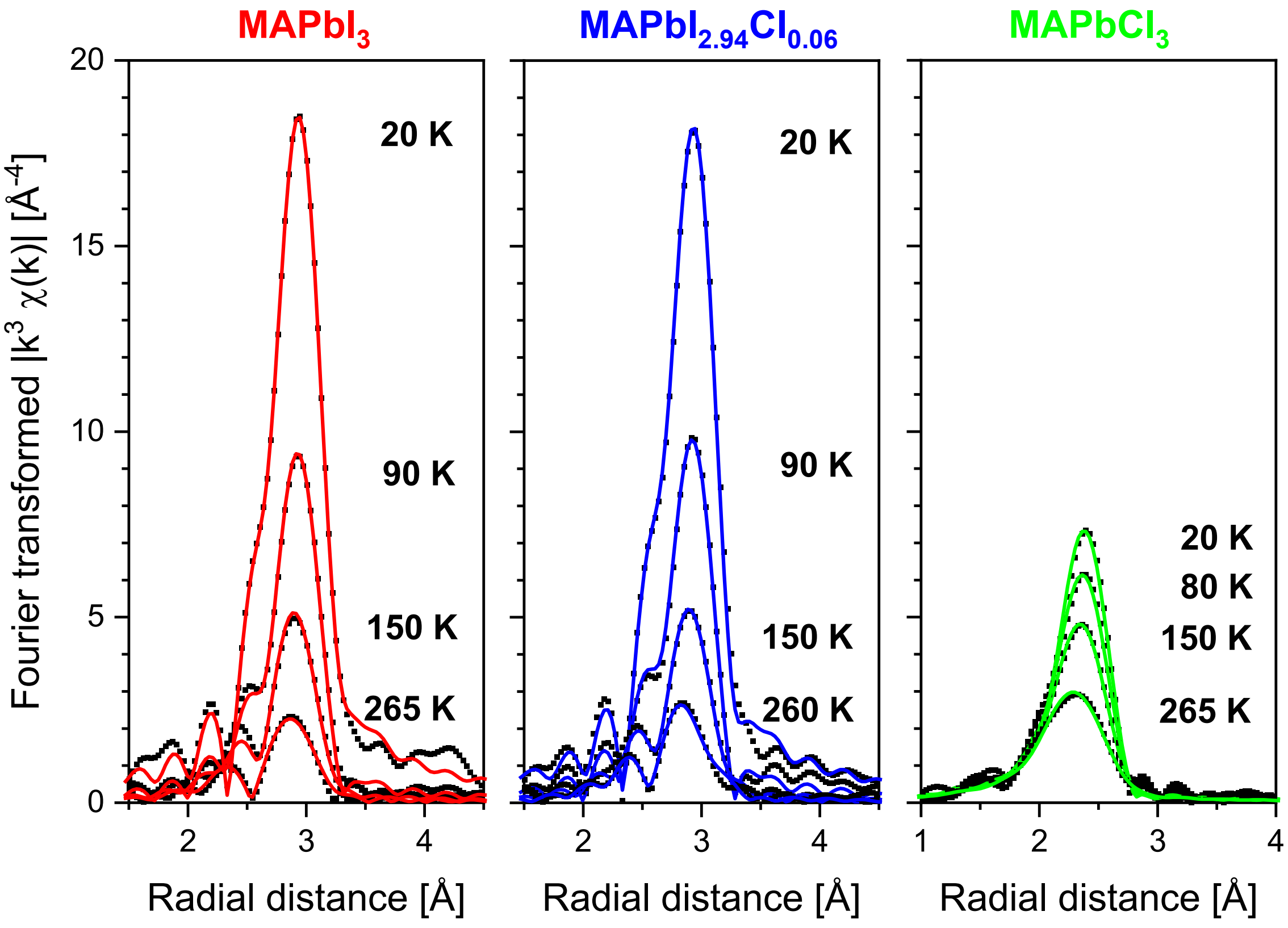

**Fig. 4** The temperature-dependent Fourier transforms of the EXAFS signal (black squares are the moduli) and the respective best fits for $MAPbI_3$ (red lines), for $MAPbI_{2.94}Cl_{0.06}$ (blue lines) and for $MAPbCl_3$ (green lines). The Fourier transforms have been made in the interval k = 3–14 $Å^{-1}$, with a $k^3$ weighting and a Hanning window for all spectra. Note that the values of the ordinate are the same in all three figures. The real and imaginary parts of the Fourier transform are shown in Fig. S17.

These considerations contrast with the investigations of Ishida et al. [37] where two lead-iodine distances were assumed for the orthorhombic phase of $MAPbI_3$ and four lead-chlorine distances for the orthorhombic phase of $MAPbCl_3$. However, when we look at the results of this investigation we quickly see that this analysis leads to a lead-halide distance that is way too small (for example, a value of 3.140(2) Å is given for 11 K results of $MAPbI_3$)[37]. It is significantly smaller than the smallest Pb-I distance of $MAPbI_3$ of 3.177(2) Å (20 K; Pb-I3 orthogonal symmetry equivalent in Fig. S14) obtained from the synchrotron XRD analysis. Also, for $MAPbCl_3$, Ishida et al. obtained distances that were too small, even when using 4 paths (2 with N = 2 and 2 with N = 1, with N = degeneracy of the path). These much too small (non-physical) distances observed by Ishida et al.,[37] (especially for $MAPbCl_3$) were only obtained because it was assumed that a separation of the lead-halide distances should be possible with the k-range used at the time, which is obviously not the case. Because of this observation we examined the temperature-dependent EXAFS data of our three samples with only one lead-halide distance $R_{EXAFS}$ (one path of the first path Pb-X with N = 6) and accordingly with only one EXAFS Debye Waller value ($\sigma^2$). Thereby, we investigated Pb L3-edge EXAFS data in the orthorhombic phase (between 20 K and 155 K) but also in tetragonal/cubic phases for comparison (up to 300 K). The range in R space (R: 2 - 4 Å for $MAPbI_3$ and $MAPbI_{2.94}Cl_{0.06}$, and R: 1.5 - 3.0 Å for $MAPbCl_3$) of the Fourier transformed EXAFS spectra that was used for the analysis ensured that only the lead-halide distance of the first path Pb-X(1) in the fitted region was considered (in the case of $MAPbI_3$ the second nearest neighbor of the central lead atom is too far away to contribute significantly, around 4.87 Å for the C atom and around 4.92 Å for the N atom of the MA molecule, but also for the chlorine compound (O2 phase at 35 K) the next neighbors were sufficiently distant, 4.38 Å for the N atom and 4.39 Å for the C atom). Visualization of the scattering amplitudes (calculated by Artemis via IFEFFIT) and fits of the 20 K EXAFS data using the next paths (Fig. S12 and S13) showed the following: The intensity ratios for $MAPbI_3$ in the R range 2 - 4 Å are such that 99.0% of the scattering come from the first path Pb-I(1) (98.2% for $MAPbCl_3$ in the R range of 1.55 - 3 Å) and only 1% from all other paths (1.8% for $MAPbCl_3$). For Pb-I(1) (resp. Pb-Cl(1)) the fit results were identical and showed that the first FT peak (Pb-X(1)) is not influenced by the next paths. Alternatively, an EXAFS analysis with the software EDA instead of Artemis could have been considered as it was carried out for $ReO_3$,[64] but we limited our work to the chosen procedure since an additional EXAFS analysis would have gone beyond the scope of this work. All fits were performed in R space. The third and fourth cumulants (C3 and C4) also had to be used to describe the EXAFS data, as they improved both the Chi2 values and the R-factors of the fits (Fig. S16).

$S_0^2$ and $\Delta E_0$ were determined at the lowest temperature (20 K) and then fixed for all other temperatures (as done by others in temperature-dependent studies of EXAFS Debye Waller factors).[65] $\Delta E_0$ and $S_0^2$ can influence the absolute values of $R_{EXAFS}$, $C_2$, and $C_3$. The $\Delta E_0$ value strongly affects the absolute values of $R_{EXAFS}$ and $C_3$, and $S_0^2$ can influence the absolute values of $C_2$. To avoid these influences in the temperature-dependence of the EXAFS cumulants, the relative values $\Delta R_{EXAFS}$, $\Delta\sigma^2$ ($\Delta C_2$), $\Delta C_3$ in relation to the lowest temperature were used as the analysis strategy [66-70] since the temperature development of the relative values is generally only very slightly influenced by $\Delta E_0$ or $S_0^2$. A relative value in this context is the difference between the absolute value and the value at lowest temperature. Fig. 5 shows the relative interatomic distances $\Delta R_{EXAFS}$ resulting from the fits of the temperature-dependent EXAFS data. These EXAFS results are visualized together with the averaged relative interatomic distances resulting from synchrotron XRD. Note that EXAFS-derived relative distances $\Delta R_{EXAFS}$ differ from $\Delta R_{XRD}$. They are generally larger and the discontinuous change at the transition into disordered phases is smaller. Both effects result from the apparent shortening of $\Delta R_{XRD}$, which is derived from averaged positions of atoms with strong anisotropic displacement. Rietveld analysis did not allow the reliable refinement of anisotropic displacement parameters. EXAFS, however, allows to determine true values for interatomic distances and anisotropic displacement.

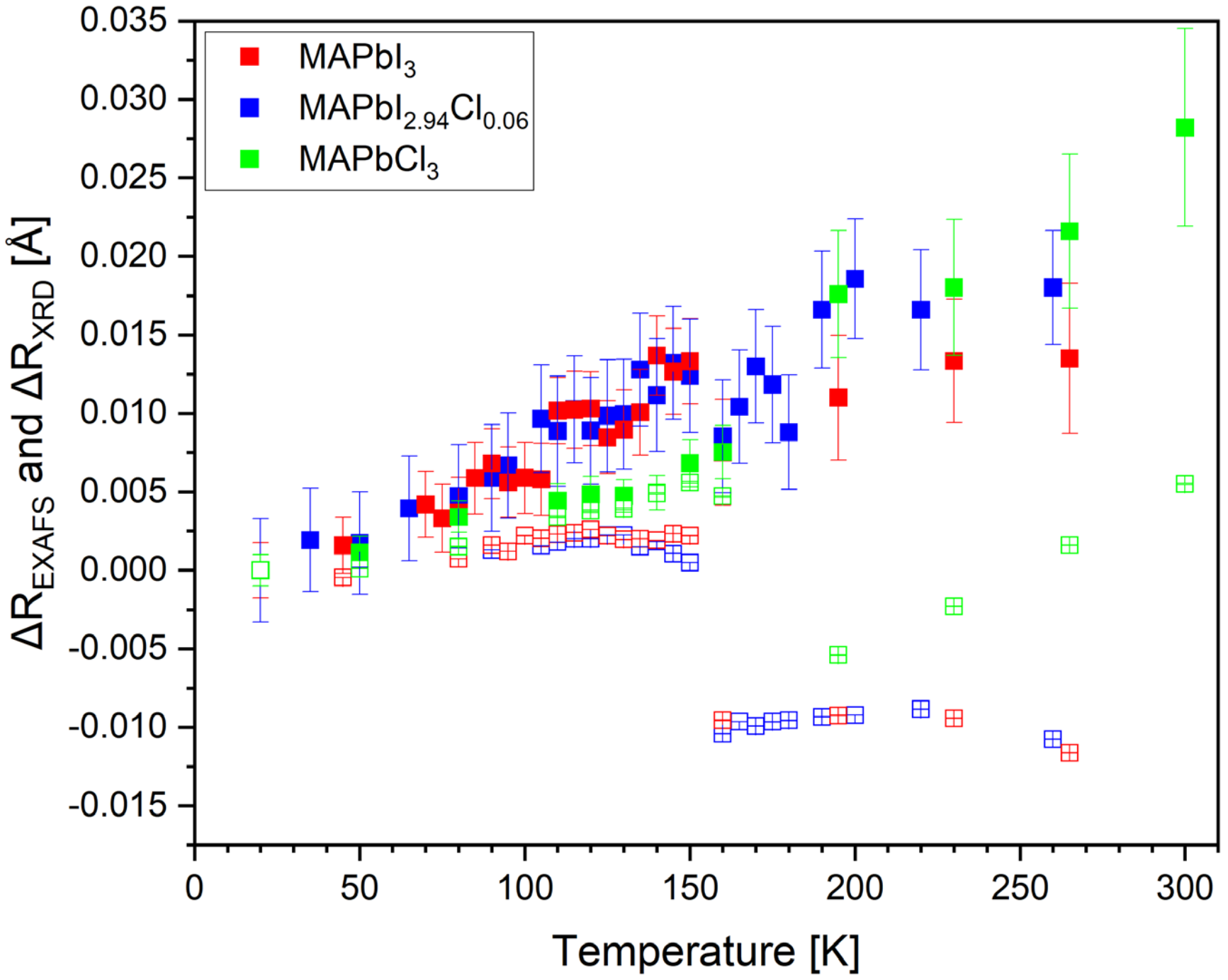

**Fig. 5** Temperature-dependent relative interatomic distances $\Delta R_{EXAFS}$ of $MAPbI_3$ (red), $MAPbI_{2.94}Cl_{0.06}$ (blue), and $MAPbCl_3$ (green). The $\Delta R_{EXAFS}$ values determined with EXAFS (solid symbols) are compared with the relative averaged $\Delta R_{XRD}$ values determined with XRD (open symbols).

To determine the temperature-dependent behavior of $C_2$, the temperature-dependence of the relative $\Delta C_2$ is fitted with an Einstein behavior as described by Dalba and Fornasini.[69,70] According to Bunker,[63] the Einstein model is given by the following relation:

$$C_2(T) = \frac{\hbar^2}{\mu k_B}\frac{1}{2\theta_E}\coth\left(\frac{\theta_E}{2T}\right) + \sigma_{static}^2 \quad (3)$$

where $\mu = m_1m_2/(m_1+m_2)$ [g/mol] is the reduced mass of the absorber/scatterer pair, $\Theta_E$ is the Einstein temperature and $\sigma_{static}^2$ is the static component which, however, does not have to be considered here, as we have proceeded from the relative MSRD. The resulting temperature-dependent behavior of $C_2$ is shown in Fig. 6a. The Einstein behavior was applied to all compositions in the temperature range of the orthorhombic structure. For $MAPbI_3$, an Einstein temperature $\Theta_E$ of 97.0(3) K was determined for the behavior of $C_2$ in the orthorhombic phase, for $MAPbI_{2.94}Cl_{0.06}$ an $\Theta_E$ of 97.8(4) K and for $MAPbCl_3$ an $\Theta_E$ of 180.5(8) K. From the Einstein temperature, a direct derivation of the effective force constant $k_0$ could be made since the following applies:

$$k_0 = \Theta_E^2 \mu \tag{4}$$

$$k_0 = (\mu * 1.660539*10^{-27}\ \text{kg} * (2\pi * \nu_E)^2)/16.022\ [\text{eV/Å}^2] \tag{5}$$

where [1 N/m = 1/16.022 eV/Å$^2$], [1 THz = 47.9924 K], and $\nu_E$: Einstein frequency [THz] (Table 1). The effective force constant $k_0$ is not to be confused with the force constant of lattice dynamic models.[39] For the three components we obtained the $k_0$ values given in Table 1. The difference of the temperature-dependent behavior of $C_2$ (and thus also $\nu_E$ and $k_0$) between $MAPbI_3$ and $MAPbCl_3$ was significant (Fig. 6a). $MAPbI_{2.94}Cl_{0.06}$ caused only small differences compared to $MAPbI_3$, thus $\nu_E$ and $k_0$ hardly differed (within the observed errors) for these two components. In general, it could be seen that the Einstein model determined for the orthorhombic phase also describes the temperature-dependent behavior of $C_2$ for the tetragonal/cubic phase well (Fig. 6a). In principle, the temperature-dependent behavior of $C_2$ (parallel mean square relative displacement: "parallel MSRD") could be compared directly with the parallel MSD (parallel mean square displacement) determined from diffraction experiments.[64,71] The parallel MSD corresponded to the averaged principal direction of the anisotropic temperature ellipsoids that were aligned parallel to the lead-halide bond.[71] However, since anisotropic temperature factors (ADP) [72] must be considered for the investigated halide atoms in the hybrid perovskites, but these could not be determined for all temperatures and components from the carried out Rietveld refinements (and are also not usable and consistent in the literature), we reference here to a summary of the present results in the supplement (section MSRD and MSD) and also refer to later investigations with a parameterized Rietveld refinement [73] approach. From the temperature-dependent comparison of the relative interatomic distances $\Delta R_{EXAFS}$ and the averaged relative interatomic distances $\Delta R_{XRD}$ resulting from synchrotron XRD, however, it was also possible to infer the relative "perpendicular MSRD" ($\Delta\langle\Delta u^2_{perp}\rangle$), which corresponded to the vibrations perpendicular to the lead-halide bond direction (in-plane vibrations). The following relationship applied:[67]

$$\Delta\langle\Delta u^2_{perp}\rangle = 2 * R_{XRD} (\Delta R_{EXAFS} - \Delta R_{XRD}), \quad (6)$$

which can be calculated from the experimental data shown in Fig. 2 and 5. The absolute values of the perpendicular MSRD were then obtained, analogously to the parallel MSRD, through fitting of $\Delta\langle\Delta u^2_{perp}\rangle$ with a correlated Einstein model [69,70] in the temperature range of the orthorhombic phase, although the differing dimensionality ($\Delta u^2_{perp} = \Delta u^2_x + \Delta u^2_y$) makes it necessary to use the following term:[39,71]

$$< \Delta u^2_{perp} > (T) = \frac{\hbar^2}{\mu k_B} \frac{1}{\theta_{E(perp)}} \coth\left(\frac{\theta_{E(perp)}}{2T}\right) + \sigma^2_{static(perp)} \quad (7)$$

However, the static component $\sigma^2_{static(perp)}$ does not need to be considered here, as we have proceeded from the relative perpendicular MSRD. Analogous to $k_0$, the corresponding perpendicular effective force constant $k_\perp$ from the correlated Einstein model could be calculated. The determined Einstein temperatures $\Theta_{E(perp)}$ (Table 1) described the behavior of the perpendicular MSRD in the orthorhombic phase well (Fig. 6b), whereby the differences between $MAPbI_3$ and $MAPbI_{2.94}Cl_{0.06}$ are somewhat larger than for the parallel MSRD (Fig. 6a). Again, the temperature-dependent behavior of $MAPbCl_3$ was clearly different from that of $MAPbI_3$. For all three components, after the phase transition to the tetragonal/cubic phase, significantly larger values were observed for the perpendicular MSRD than predicted by the determined Einstein behavior. This effect was particularly large in the case of $MAPbCl_3$ since a jump-like change in the perpendicular MSRD was observed here.

a)

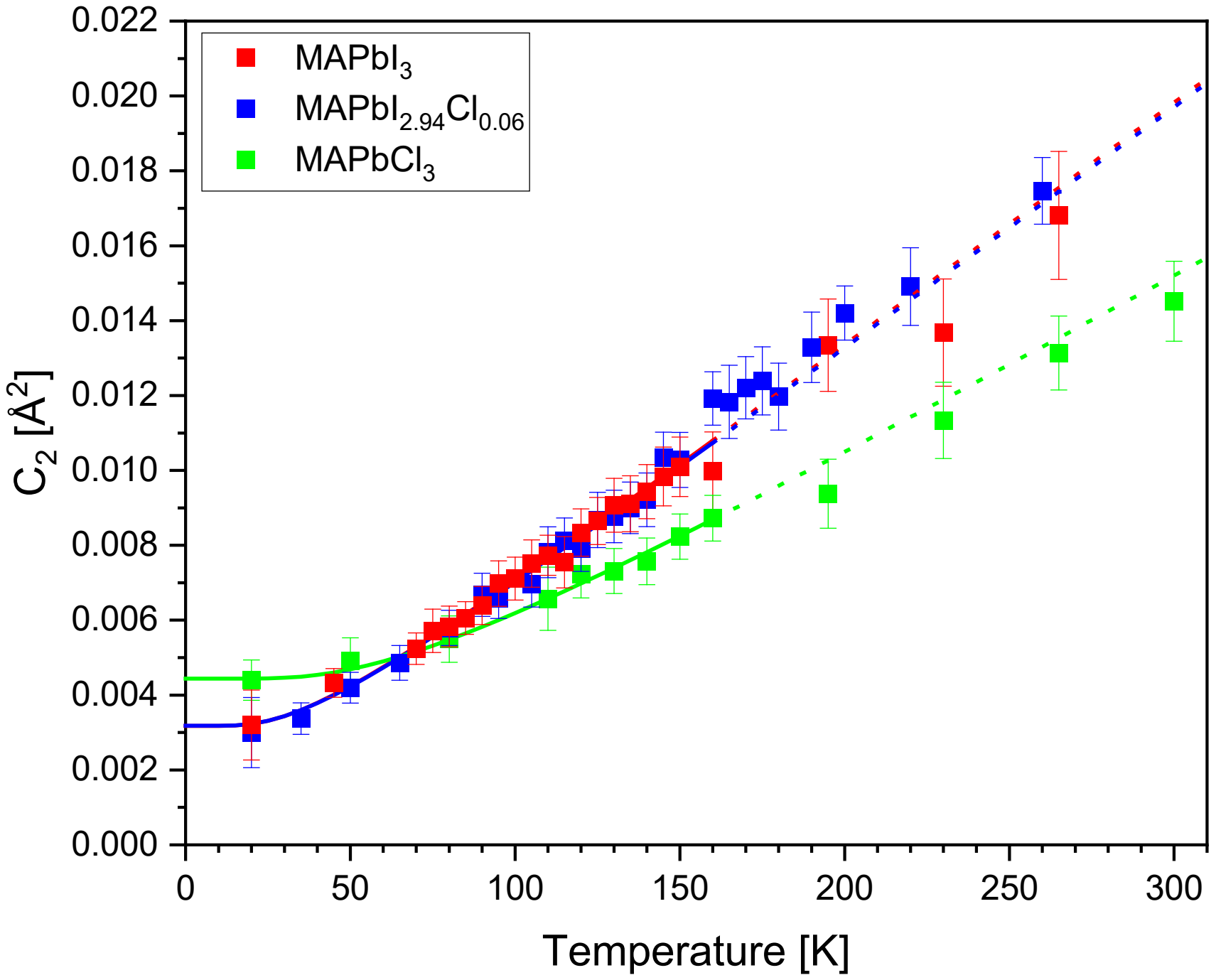

b)

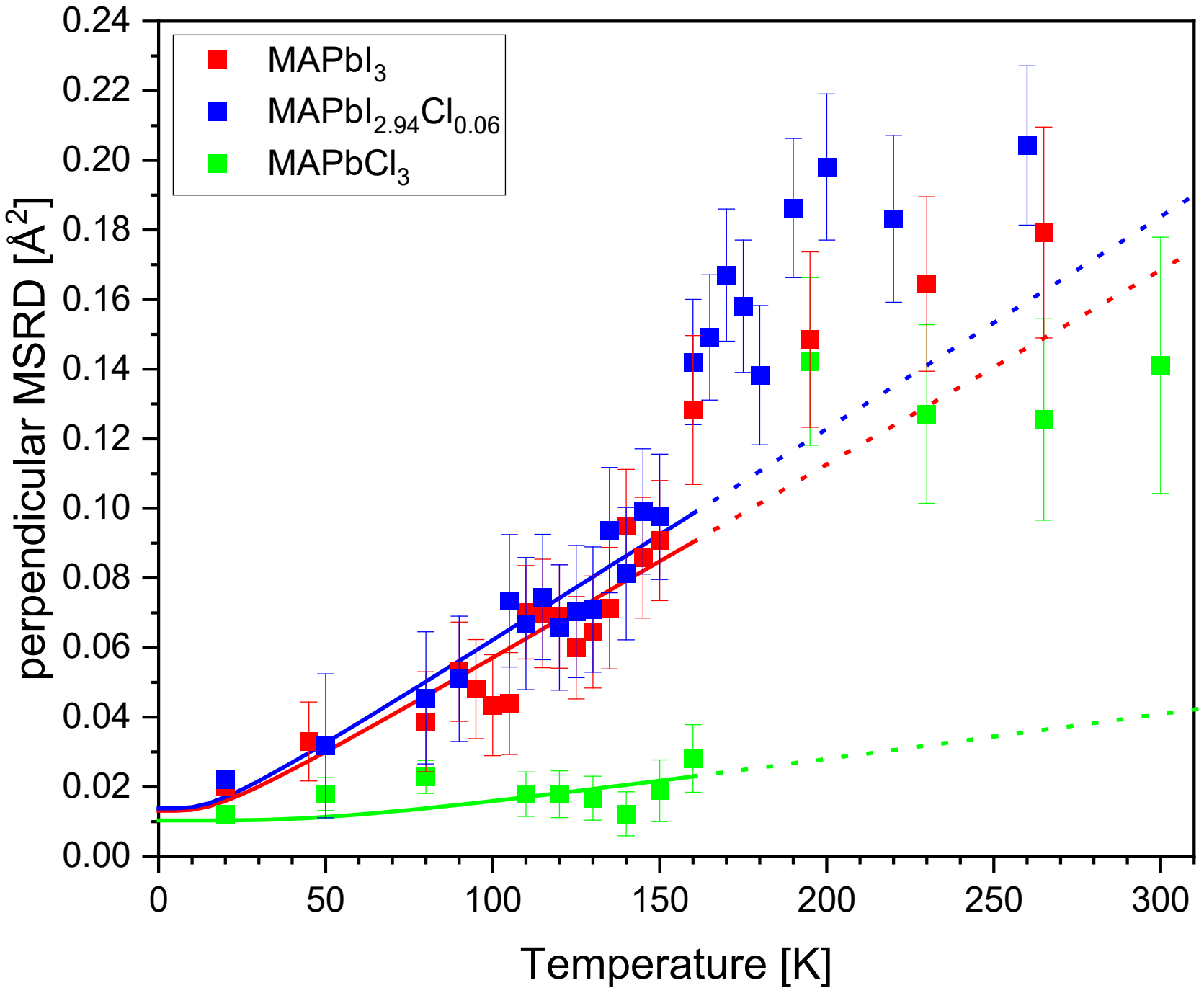

**Fig. 6** Temperature-dependent MSRD of $MAPbI_3$ (red), $MAPbI_{2.94}Cl_{0.06}$ (blue) and $MAPbCl_3$ (green). In addition to $C_2$ (parallel MSRD) in a), the perpendicular MSRD is shown in b), which was calculated based on $\Delta R_{XRD}$ and $\Delta R_{EXAFS}$. In the orthorhombic temperature range, the MSRD values were fitted with Einstein behavior (solid line). Extrapolations of the low-temperature, Einstein model fits to higher temperatures, are shown as dotted lines.

For the relationship between perpendicular MSRD and the perpendicular MSD obtained from diffraction measurements, please refer to the comments in the supplement (section MSRD and MSD).

The ratio γ = perpendicular MSRD / parallel MSRD describes the degree of anisotropy of the relative vibrations.[67] The ratio γ was usually not constant as a function of temperature since different correlated Einstein relations were determined for the parallel MSRD and the perpendicular MSRD. For purely isotropic vibrations in an ideal Debye crystal, the ratio γ = 2, [66,67,74] but since our samples showed very clear anisotropy, larger γ ratios were obtained (Fig. 7).

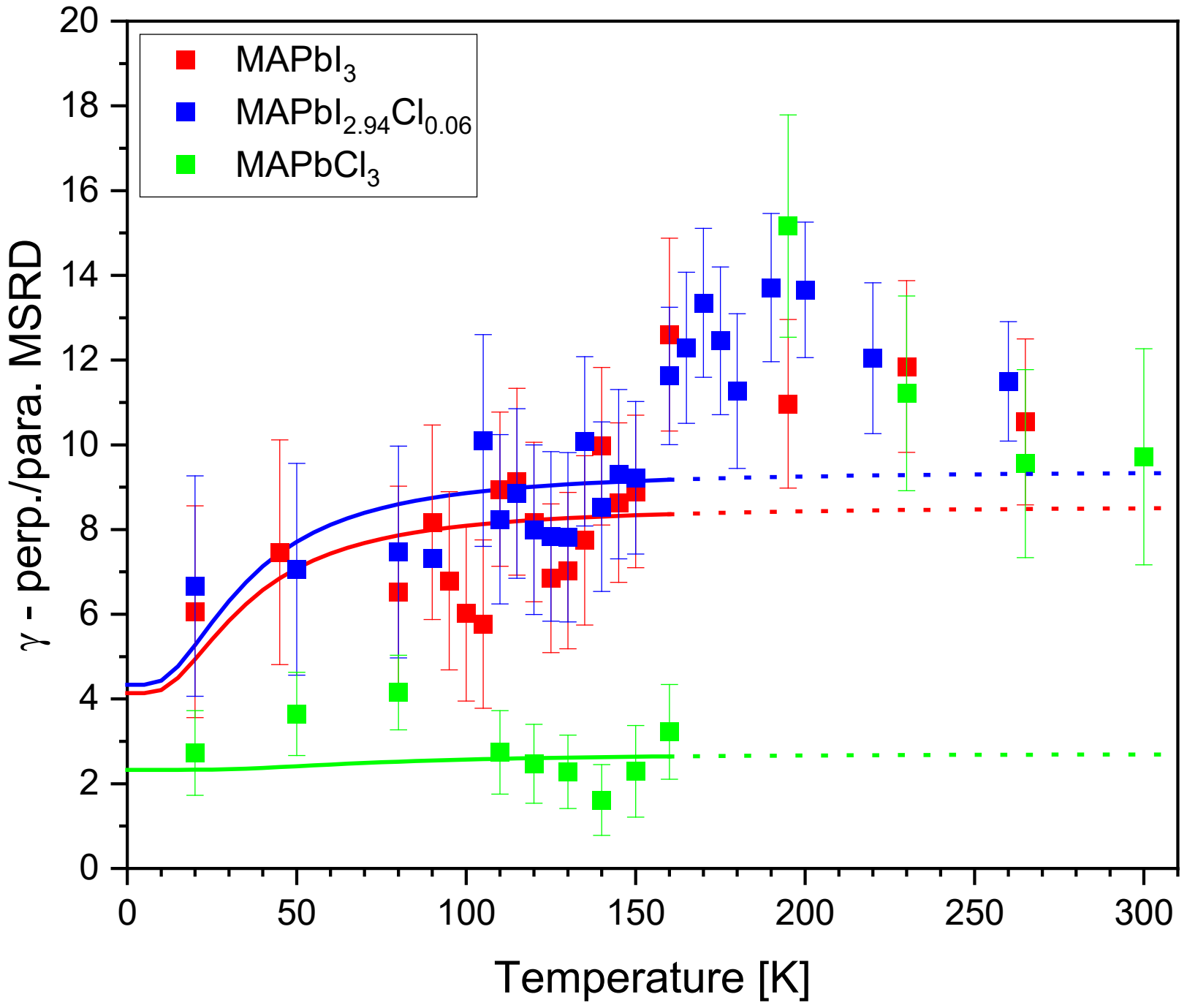

**Fig. 7** Ratio γ = perpendicular MSRD / parallel MSRD of $MAPbI_3$ (red), $MAPbI_{2.94}Cl_{0.06}$ (blue) and $MAPbCl_3$ (green). γ describes the degree of anisotropy of the relative vibrations. In the orthorhombic temperature range, the MSRD values were fitted with Einstein behavior, the solid lines show the resulting γ dependence (dotted lines: extrapolations of the low-temperature Einstein model). The uncertainty for γ was determined using the Gaussian error propagation method.[75]

Here, $MAPbI_3$ and $MAPbI_{2.94}Cl_{0.06}$ already showed very clear anisotropy in the orthorhombic phases, whereas an almost isotropic behavior (γ only slightly greater than 2) was observed for $MAPbCl_3$. While the ratio γ is temperature-dependent, the ratio $\xi = k_0/k_\perp$ provides a temperature-independent description of the degree of anisotropy of the relative vibrations,[67] which is valid for the whole temperature range of the orthorhombic phase (Table 1). Here, ratios of ξ for $MAPbI_3$ 4.26(8) and for $MAPbI_{2.94}Cl_{0.06}$ 4.60(5) were significantly larger than for $MAPbCl_3$ 1.38(9), showing larger anisotropy of vibrations. After transition to the tetragonal and cubic phase, $MAPbI_3$ and $MAPbI_{2.94}Cl_{0.06}$ showed a deviation from the Einstein behavior, with higher γ.

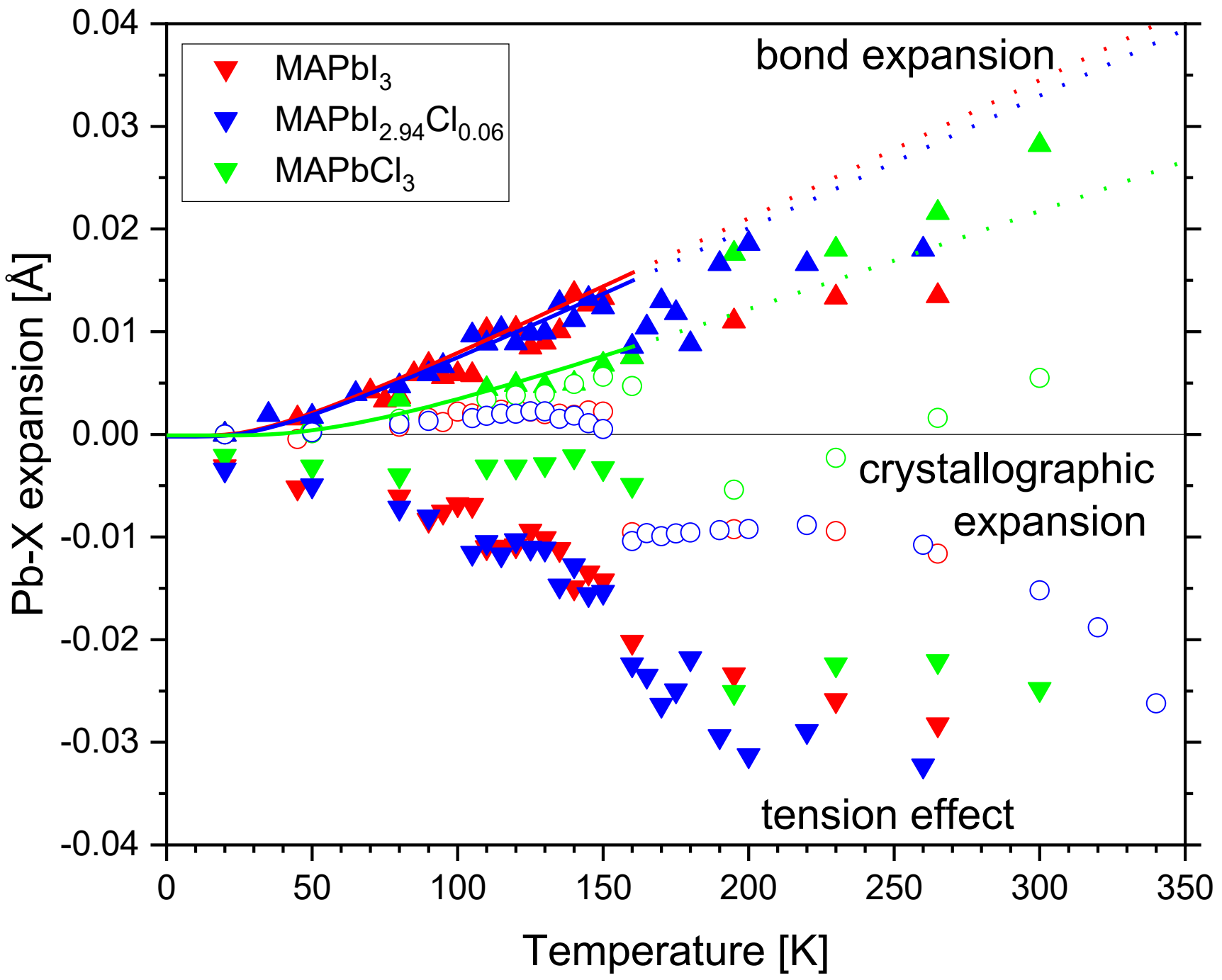

**Fig. 8** Pb-X expansion of $MAPbI_3$ (red), $MAPbI_{2.94}Cl_{0.06}$ (blue) and $MAPbCl_3$ (green). Positive bond expansion $\Delta R_{EXAFS}$ (solid upward triangles) and negative tension effect (solid downward triangles) result in the overall crystallographic expansion $\Delta R_{XRD}$ (open circles). For error bars compare Fig. 5 and 6b.

The behavior of $MAPbCl_3$ changed abruptly and became similar to the clear anisotropy of the other two components. The temperature-dependent behavior of the perpendicular MSRD $<\Delta u_{perp}^2>$ can illustrate the negative tension effect in the lead-halide bond (Fig. 8), whereby the different lead-halide bond characteristics in the orthorhombic phase on one hand and in the tetragonal/cubic phase on the other hand became apparent. Tension is defined as - $<\Delta u_{perp}^2>/2R_{XRD}$.[76] The positive bond expansion and the negative tension effect of the lead-halide bond are roughly equal in all three samples in the orthorhombic phase, resulting in the crystallographic bond length observed by XRD being nearly temperature-independent. However, in the disordered tetragonal/cubic phases, strong negative tension effects result in a shortening of $R_{XRD}$. In contrast to $MAPbI_3$ and $MAPbI_{2.94}Cl_{0.06}$, however, in $MAPbCl_3$ the positive bond expansion upon heating becomes predominate.

Now that we discussed the behavior of parallel and perpendicular MSRD, we turn to the cumulant $C_3$. The EXAFS cumulant $C_3$ measures the asymmetry of the distance distribution, i.e. the occurrence of shorter or longer intervals. This distance distribution asymmetry is determined by the anharmonicity of the crystal potential.[67] By analyzing $C_3$, we gained direct experimental access to the anharmonicity of the lead-halide bond. According to Fornasini et al.,[77] the third cumulant is important because their temperature dependence gives interesting clues about the local dynamical behavior (asymmetry of the pair distribution function)[78]. Here, a $C_3$ of 0 would correspond to a symmetrical Gaussian shape of the pair distribution function. A negative $C_3$ corresponds to an asymmetric Gaussian distribution with more weight on the shorter bonds and a positive $C_3$ corresponds to an asymmetric Gaussian behavior with more weight on the longer bonds of the pair distribution function.

**Table 1** Results of Einstein temperatures and frequencies for parallel ($\Theta_{E\parallel}$, $\nu_{E\parallel}$) and perpendicular MSRD ($\Theta_{E\perp}$, $\nu_{E\perp}$), results of the determination of effective force constants $k_\parallel$ (corresponds to $k_0$), $k_3$ and $k_\perp$ of the temperature-dependent behavior of the $C_2$ and $C_3$ cumulant (based on a classical approximation as discussed by Yokoyama et al. [65] and Fornasini et al. [77]) and perpendicular MSRD. The two ratios $|k_3|/k_0$ (potential anharmonicity)[77] and $\xi = k_\parallel/k_\perp$ (degree of anisotropy of the relative vibrations)[76] are also given. All values are determined in each case for the orthorhombic phase.

| | $MAPbI_3$ | $MAPbI_{2.94}Cl_{0.06}$ | $MAPbCl_3$ |
|---|---|---|---|
| μ | 78.7017 | 77.9944 | 30.2709 |
| $\Theta_{E\parallel}$ [K] | 97.0(3) | 97.8(4) | 180.5(8) |
| $\nu_{E\parallel}$ [THz] | 2.021(5) | 2.037(9) | 3.76(2) |
| $k_\parallel$ [eV/Å$^2$] | 1.315(3) | 1.324(6) | 1.751(9) |
| $k_3$ [eV/Å$^3$] | -0.88(2) | -0.90(2) | -1.14(5) |
| $|k_3|/k_0$ | 0.67(2) | 0.68(2) | 0.65(3) |
| $\Theta_{E\perp}$ [K] | 46.9(8) | 45.1(6) | 155(9) |
| $\nu_{E\perp}$ [THz] | 0.98(2) | 0.95(1) | 3.2(2) |
| $k_\perp$ [eV/Å$^2$] | 0.309(6) | 0.288(3) | 1.27(8) |
| $k_\parallel/k_\perp$ | 4.26(8) | 4.60(5) | 1.38(9) |

For the analysis of the Artemis Fit results, we used the relative values $\Delta C_3$ with respect to the lowest temperature for the temperature-dependent $C_3$ cumulants, as we did for $R_{EXAFS}$ and $C_2$. $\Delta C_3$ was fitted [67] by a simplified version of the classical approximation for cumulants valid for higher temperatures: [65,77]

$$C_3(T) = -\frac{6k_3}{k_0^3}(k_B T)^2, \tag{8}$$

here, $k_B$ [eV·K$^{-1}$] is the Boltzmann constant, $k_0$ is the effective (harmonic) force constant determined with $C_2$(T), and the higher order (anharmonic) force constant $k_3$ [eV/Å$^3$]. According to Frenkel et al.,[79] our used approximation for higher temperatures for $C_3$(T) is already valid for RbBr at 125 K, so that we can assume that an adjustment of $\Delta C_3$ in the orthorhombic phases (up to 160 K) is also meaningful. As a result of our analysis, we obtained the temperature dependence of $C_3$ shown in Fig. 9 and the force constants $k_3$ in Table 1. We observe a positive $C_3$ for all three components, which corresponds to an asymmetric Gaussian behavior with more weight to the longer Pb-X bonds of the pair distribution function. Since $C_3$ in the orthorhombic phase is larger for $MAPbI_3$ and $MAPbI_{2.94}Cl_{0.06}$ than for $MAPbCl_3$, we also observe that $MAPbI_3$ and $MAPbI_{2.94}Cl_{0.06}$ exhibit a more asymmetric form of the pair distribution function, i.e. more weight to the longer Pb-X bonds, than $MAPbCl_3$. In general, the assumed $T^2$-like behavior described the experimental values of all three components in the orthorhombic phase quite well. In relation to this, we note that the determined $k_3$ also described the relative thermal expansion $\Delta a$ very well, because of the asymmetry of the effective potential ($a = -3k_3C_2/k_0$) [67] as shown in Fig. 8 (as a solid line for the orthorhombic phase). But back to the discussion of $C_3$: $MAPbI_3$ and $MAPbI_{2.94}Cl_{0.06}$ hardly differed in their $C_3$ behavior in the orthorhombic phase. Towards higher temperatures, in the tetragonal phase, $MAPbI_{2.94}Cl_{0.06}$ showed slightly higher $C_3$ values than $MAPbI_3$. At the same time, both components had slightly lower $C_3$ values than the theoretical $T^2$-like curve. $MAPbCl_3$ showed lower $C_3$ values than $MAPbI_3$ and $MAPbI_{2.94}Cl_{0.06}$ in the orthorhombic phase, but in the cubic phase higher $C_3$ values than the theoretical $T^2$-like curve could be observed for $MAPbCl_3$. The cumulant $C_4$ was utilized in the Fit (Fig. S23) but not analyzed further, because according to Yokoyama et al. this cumulant contributes little to the temperature dependence of the effective force constants and can therefore be neglected in the discussion of the anharmonicity.[65] Now that we had obtained the two effective force constants $k_0$ and $k_3$, we could deal with the effective pair potential $V(x)$ which is given by the following relationship:[77]

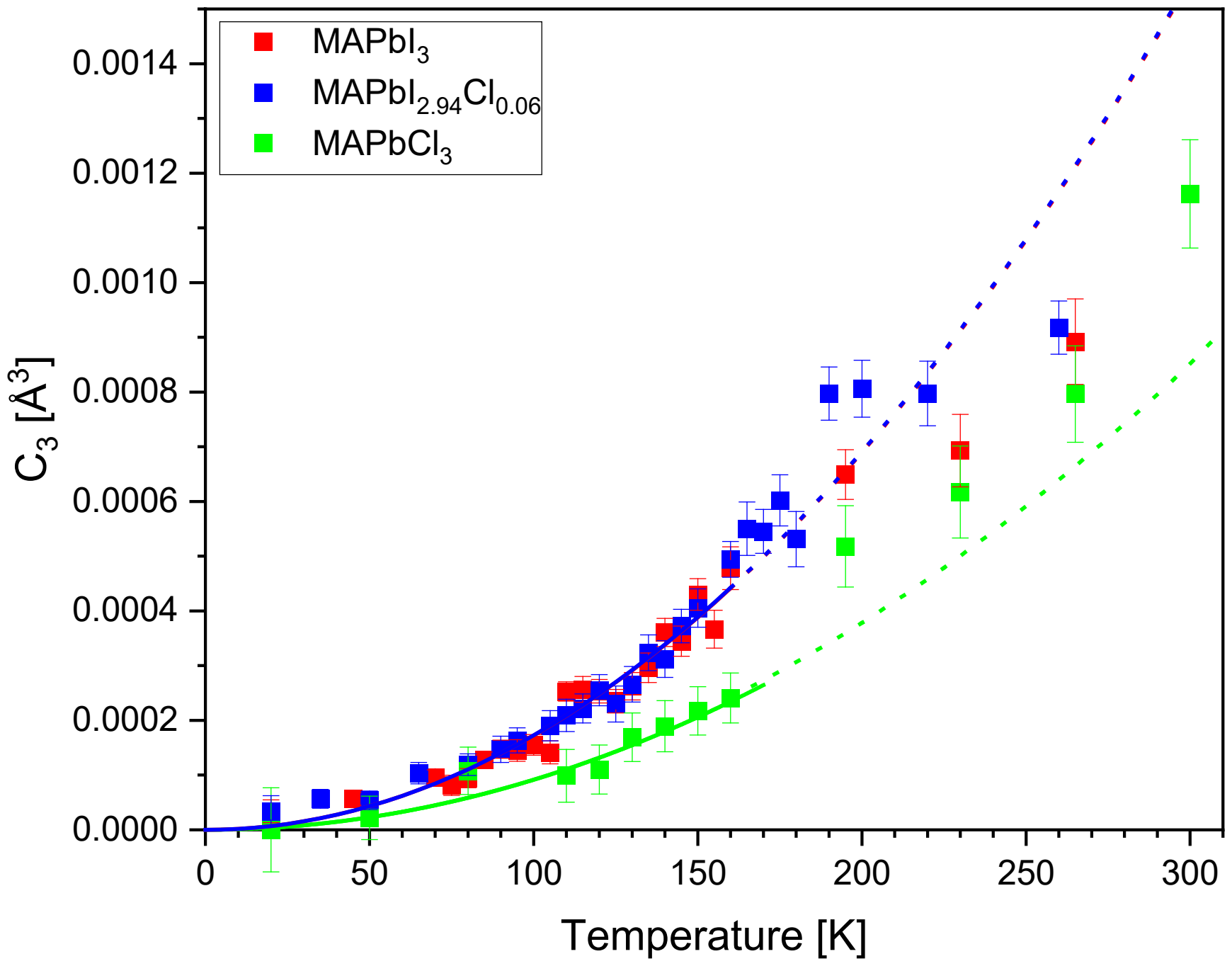

**Fig. 9** $C_3$(T) of $MAPbI_3$ (red), $MAPbI_{2.94}Cl_{0.06}$ (blue) and $MAPbCl_3$ (green). The solid lines show the corresponding $T^2$-like classical approximation for $C_3$(T) for the orthorhombic phase. The extrapolation of the low-temperature $T^2$-like behavior of $C_3$(T) to higher temperatures is shown as a dotted line.

$$V(x) = \frac{1}{2}k_0x^2 + k_3x^3 + \cdots, \tag{9}$$

where $x = r - r_0$ is the deviation of the interatomic distance from the position of the potential minimum. The effective potentials for the orthorhombic phase shown in Fig. 10a could each be compared with the corresponding harmonic potentials, in which case only $k_0$ was considered:

$$V(x) = \frac{1}{2}k_0x^2. \tag{10}$$

The potential anharmonicity,[77] measured by the ratio $|k_3|/k_0$, was indeed quite similar for all three components ($MAPbI_3$: 0.67(2), $MAPbI_{2.94}Cl_{0.06}$: 0.68(2), and for $MAPbCl_3$: 0.65(3)). Yet, the three components differed in the various effects of anharmonicity (thermal expansion of the bond $C_1$ ($R_{EXAFS}$) and the asymmetry of the distribution of the distances $C_3$, Fig. 8 and Fig. 9) as already seen. It is worth mentioning that the deviation of $\Delta C_1$ (bond expansion) and $\Delta a$ (relative thermal expansion because of to the asymmetry of the effective potential) in our study was very small (Fig.

8), which to our knowledge has not been observed in the literature before. On the contrary, it is usually observed that Δa is always smaller than $\Delta C_1$.[77] The difference $r_v$ between $\Delta C_1$ and Δa is then regarded as a shift of the effective potential minimum. We now want to go one step further and make the connections between our results and an experimental Morse potential determined for $MAPbI_3$. The experimental results we will refer to here were obtained from temperature-dependent neutron total scattering experiments to elucidate the local structure of $MAPbI_3$.[28]As in the same research group's study of $CsPbI_3$,[29] reverse Monte Carlo modeling was performed for $MAPbI_3$ to construct configurations of atoms consistent with both local and long-range structures. For $MAPbI_3$, Liu could determine the following Morse potential parameters D (dissociation energy) and α (width of the potential) for the lead-iodine bond: D = 0.201(17) eV (Note: the value of D = 0.176 eV given by Liu is not correct, Dove, M. T. personal communication, March 9, 2021) and α = 1.72(6) Å$^{-1}$. The Morse potential parameters could be directly compared with the effective potential determined here, since according to Hung et al. it is possible to calculate the Morse potential parameters D and α from the effective force constants $k_0$ and $k_3$ by solving the following two equation system:[80]

$$k_0 \cong 2DS_2\alpha^2 \tag{11}$$

$$k_3 = S_3 D\alpha^3 \tag{12}$$

$S_2$ and $S_3$ are structure parameters that depend on the environment of the central atom of the local structure under investigation. In our case, we needed these structure parameters for an octahedral environment. To do this, we used the method described by Hung et al. to derive $S_2$ and $S_3$:[81]

$$V_{eff}(x) = V(x) + \sum_{i=1,2}\sum_{j\neq i} V\left(\frac{\mu}{M_i} x\hat{R}_{12}\cdot\hat{R}_{ij}\right) = D(S_2\alpha^2 + 3S_3\alpha^3 a)y^2 + S_3 D\alpha^3 y^3, \tag{13}$$

$$y = x - a, \quad a = \langle x\rangle, \tag{14}$$

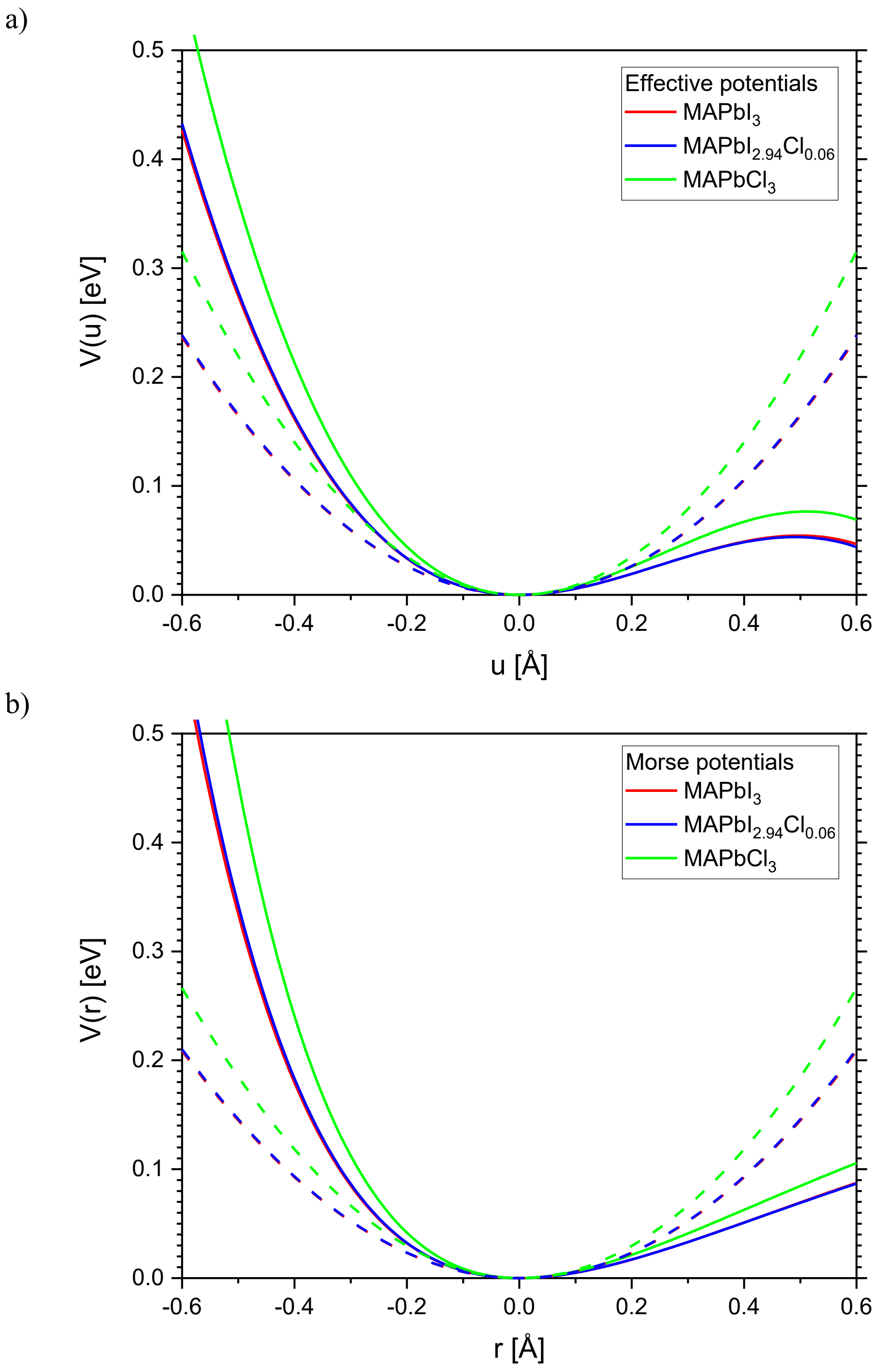

**Fig. 10** a) Effective potentials for the orthorhombic phase and b) Morse potentials of $MAPbI_3$ (red), $MAPbI_{2.94}Cl_{0.06}$ (blue) and $MAPbCl_3$ (green). Dashed line a) effective harmonic potentials and b) harmonic part of the Morse potentials.

where $M_i$ is the *i*th atomic mass, $\hat{R}$ is the bond unit vector, the sum *i* is over absorbers (i=1) and backscatters (i=2), the sum *j* is over their first nearest neighbors, excluding absorbers and backscatterers themselves, whose contributions are described by the term V(x) (see the Supplement for details on the calculation of the structure parameters $S_2$ and $S_3$). As a result, $S_2 = 1.1322$ and $S_3 = -0.9633$ were obtained for $MAPbI_3$ for the octahedral environment of the lead ($S_2$=9/8; $S_3$= -31/32 with $M_1$=$M_2$. In Table S4, the structure parameters for $MAPbI_{2.94}Cl_{0.06}$ and $MAPbCl_3$ are given). Thus, we obtained D = 0.235(6) eV and α = 1.57(4) $Å^{-1}$ for $MAPbI_3$, resulting in the Morse potential shown in Fig. 10b and 11 with a good agreement with the experimental values of Liu [28].

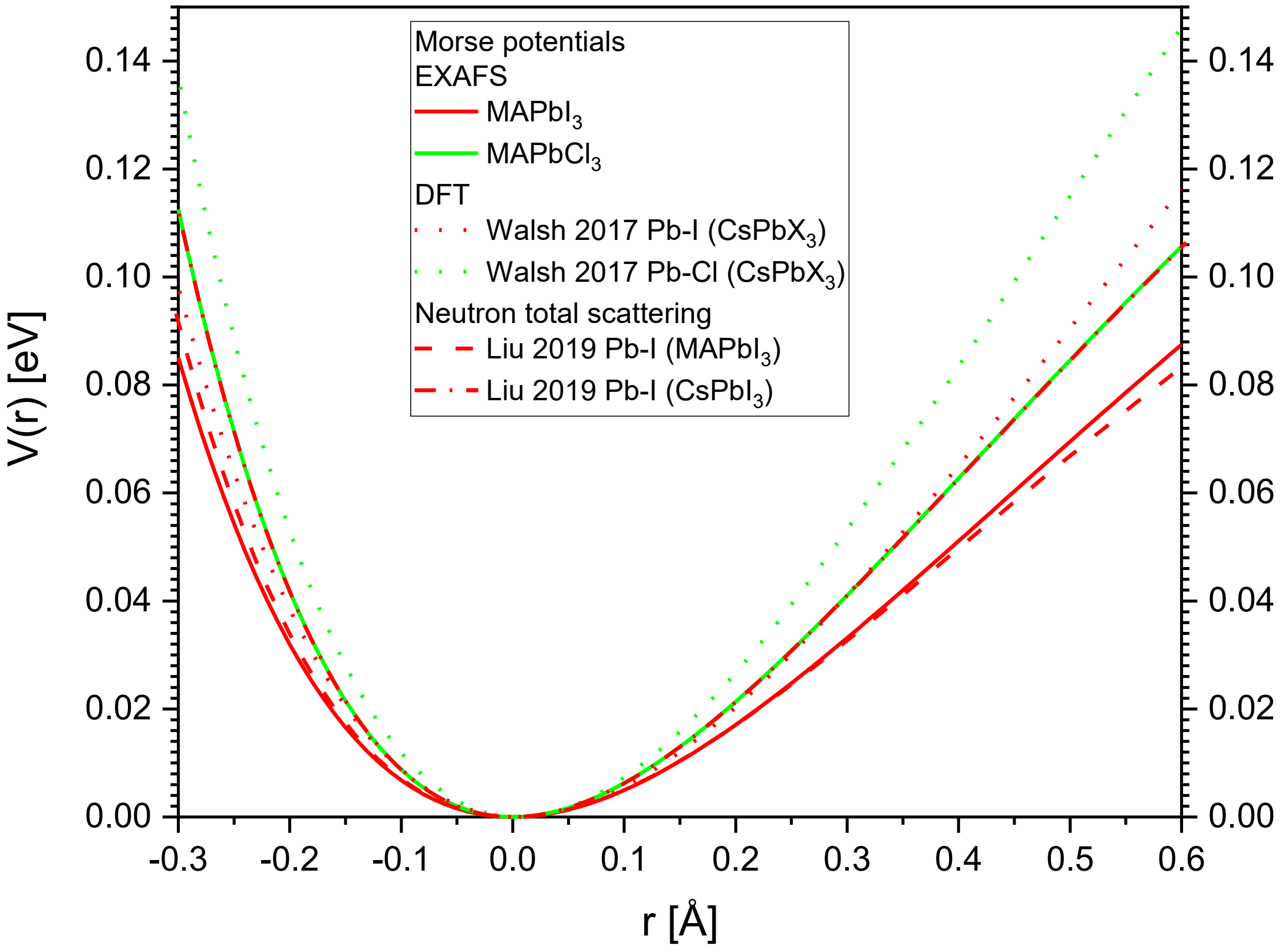

**Fig. 11** Comparison of the experimental Morse potentials for $MAPbI_3$ (red solid line) and $MAPbCl_3$ (green solid line) with experimental results from Liu 2019 for $MAPbI_3$ (red dashed line) and $CsPbI_3$ (red dash dotted line),[28,29] and with Morse potentials from DFT calculations from Walsh 2017 [82] for $CsPbI_3$ (red dotted line) and $CsPbCl_3$ (green dotted line).

For $MAPbI_{2.94}Cl_{0.06}$, we obtained D = 0.228(6) eV and α = 1.60(4) $Å^{-1}$ and for $MAPbCl_3$ D = 0.262(14) eV and α = 1.68(9) $Å^{-1}$. The Morse potential parameters allowed a comparison of the experimental results described here with other experimental methods [28,29] as well as with DFT calculations as available in the literature [82] (Fig. 11, S24). More generally, it can also be said that EXAFS results can be used as a benchmark for theoretical calculations.[70]

## 4. CONCLUSIONS

The combined temperature-dependent synchrotron XRD and EXAFS studies reported here show that the anharmonicity of the lead-halide bond changes significantly when chlorine is incorporated into $MAPbI_3$. In the tetragonal phases of $MAPbI_3$ and $MAPbI_{2.94}Cl_{0.06}$ shrinkage of the $[PbX_6]$ octahedra was observed with XRD, which, to our knowledge, has not been adequately discussed in the literature but should be considered in the anharmonic behavior of the lead-halide bond. For $MAPbCl_3$, a phase separation in the ordered orthorhombic structure was observed with XRD. This was already mentioned in the literature [20,57], but was described by us in more detail for the first time. The orthorhombic phase O2 (a ≈ 8.0 Å, b ≈ 11.3 Å, c ≈ 7.9 Å), which is obviously formed under the influence of grinding (soft mechanochemical synthesis), passes directly into the cubic phase of $MAPbCl_3$ described in the literature. An $MAPbCl_3$ intermediate tetragonal phase was not observed with XRD. From the EXAFS analysis, the perpendicular MSRD was determined, allowing a comparison of the tension and bond expansion effects in the three studied compounds. The positive bond expansion and the negative tensions effects of the lead-halide bond were roughly equal in all three samples in the orthorhombic phase. However, after transition to the disordered tetragonal/cubic phase, the balance shifted in favor of the negative tension effects in $MAPbI_3$ and $MAPbI_{2.94}Cl_{0.06}$, but in $MAPbCl_3$ the positive bond expansion seemed to predominate towards higher temperatures. The negative tension effects observed with EXAFS in the tetragonal phase of $MAPbI_3$ and $MAPbI_{2.94}Cl_{0.06}$ are correlated with the behavior of the shrinking $[PbX_6]$ octahedra observed with XRD. This correlation suggests that these two compositions have stiffer Pb-X bonds and that the [$PbX_6$] octahedra are less rigid in the tetragonal phase. The comparison of MSRD from EXAFS measurements and MSD from diffraction measurements, was only touched upon here, but is obviously very important for a better understanding of the anharmonic behavior of the lead-halide bond. It should be noted here that the behavior of the MSD as a function of temperature has hardly been considered in the literature on hybrid perovskites. The different anharmonicity contributions for the three investigated samples resulting from the EXAFS cumulant analysis showed a partially opposite behavior in the transition from the orthorhombic to the tetragonal/cubic phase. For some parameters, like bond expansion, perpendicular MSRD, $\gamma$ and - for $MAPbCl_3$ also $C_3$ - a discontinuous change after the phase transition from the orthorhombic phase to the tetragonal/cubic crystal structure was evident. For $MAPbCl_3$, a clearly different state of the lead-chlorine bond appeared in the cubic phase. From the analysis of $C_3$ it is clear that $MAPbCl_3$ has a less asymmetric pair distribution function with less weight to the longer bonds in the orthorhombic phase compared to the other two components. In line with this, our XRD results show that $MAPbCl_3$ in the orthorhombic phase exhibits a much smaller splitting of the pseudo cubic lattice parameters and smaller Pb-Cl-Pb rotation angles. All this considered together, this could also be related to the behavior of the jump rotation of the MA molecule and the associated influence on hydrogen bonding: QENS investigations [17] showed that hydrogen bonding in $MAPbI_3$ has a stronger influence in the orthorhombic phase, as much lower MA rotational jump dynamics were observed in orthorhombic $MAPbI_3$ compared to $MAPbCl_3$. But, after the phase transition to the tetragonal/cubic phase, a faster MA jump rotation was observed for $MAPbI_3$ than for $MAPbCl_3$. However, in the EXAFS investigations discussed here, all investigated parameters were lower for $MAPbCl_3$ in the orthorhombic phase than for $MAPbI_3$ and $MAPbI_{2.94}Cl_{0.06}$. $MAPbI_3$ and $MAPbI_{2.94}Cl_{0.06}$ hardly differed, whereas for the degree of anisotropy the ratio $\gamma$ slightly larger values were observed for $MAPbI_{2.94}Cl_{0.06}$. Surprisingly, the potential anharmonicity $|k_3|/k_0$ was almost the same for all three investigated samples. Our experimental results from the EXAFS analysis on chlorine-substituted $MAPbI_3$ were confirmed by the Morse potential parameters of Liu obtained from temperature-dependent neutron total scattering experiments [28] since we determined the structural parameters $S_2$ and $S_3$ required for the conversion of the effective force constants $k_0$ and $k_3$ into Morse potential parameters α and D. This in turn allowed us to compare with Morse potential parameters of $CsPbX_3$, which then allowed us to interpret that the differences in Morse potentials between $CsPbI_3$ and $MAPbI_3$ on the one hand and between $CsPbCl_3$ and $MAPbCl_3$ on the other hand are due to the presence of hydrogen bonding between the halide atoms and the hydrogen atoms of the MA molecule. In this case, the hydrogen bond has a direct influence on the lead-halide bond.

## ASSOCIATED CONTENT

**Supporting Information**. Fig. S1: cryo-EXAFS environment at KMC-2 "XANES", Fig. S2: X-ray powder diffraction pattern of $MAPbCl_3$ at temperature of 160 K, Fig. S2: XAFS energy calibration, Fig. S3: XANES spectra, Fig. S4: X-ray powder diffraction pattern of $MAPbCl_3$ at temperature of 160 K, Fig. S5: X-ray powder diffraction pattern of $MAPbCl_3$ at temperature of 100 K, Fig. S6: X-ray powder diffraction pattern of $MAPbCl_3$ at temperature of 100 K (detail), Fig. S7**:** crystal structure of $MAPbCl_3$ at 160 K in space group *Pnma* (orthorhombic phase O2), Table S1: crystal structure of $MAPbCl_3$ (orthorhombic phase O2) at 160 K in space group *Pnma* (No. 62), Fig. S8: Selected region of XRPD pattern of three different samples of $MAPbCl_3$ at 100 K, Fig. S9: X-ray powder diffraction pattern of $MAPbI_{2.94}Cl_{0.06}$ at various temperatures, Table S2: crystal structure of $MAPbI_{2.94}Cl_{0.06}$ at 150 K in space group *Pnma* (No. 62), Fig. S10: Rietveld results: Cubic and pseudo cubic unit cells, Fig. S11: Variation in the Pb–X–Pb angle as a function of temperature, Fig. S12: Fourier transforms of the EXAFS signal at 20 K and the respective best fits for $MAPbI_3$, Fig. S13: Fourier transforms of the EXAFS signal at 20 K and the respective best fits for $MAPbCl_3$, Fig. S14: temperature-dependent lead-iodine distances, Fig. S15: temperature-dependent lead-chloride distances, Fig. S16: comparison of EXAFS fit models, Fig. S17: real and imaginary parts of the Fourier transform of the EXAFS signal shown in Fig 4, Fig. S18: Isotropic temperature factors Pb $U_{iso}$ of the lead atoms as results of the Rietveld analysis of temperature-dependent synchrotron data from $MAPbI_3$, Fig. S19: Crystal structure of $MAPbI_3$ at 295 K and atomic displacement ellipsoids of the iodine, Table S3: Iodide anisotropic displacement parameters (ADP) of $MAPbI_3$, Fig. S20: Temperature-dependent MSRD and MSD of $MAPbI_3$, Fig. S21: Temperature-dependent MSRD and MSD of $MAPbI_{2.94}Cl_{0.06}$, Fig. S22: Temperature-dependent MSRD and MSD of $MAPbCl_3$, Fig. S23: Temperature-dependent behavior of $C_4$, Table S4: Morse parameter, Fig. S24: Comparison of the experimental Morse potential parameters D and α as a function of the mass ration $M_1/M_2$,

The following files are available free of charge. File type, PDF

**AUTHOR INFORMATION**

**Corresponding Author**

*Götz Schuck, e-mail: goetz.schuck@helmholtz-berlin.de

**ACKNOWLEDGEMENTS**

The authors are very grateful for the granted beam times at BESSY II, KMC-2 in Berlin. We would also like to thank Frederike Lehmann for crystal growing and quality control using lab XRD at the HZB X-Ray CoreLab. We would like to thank Marnie Bammert for her proofreading.

**TOC Graphic**

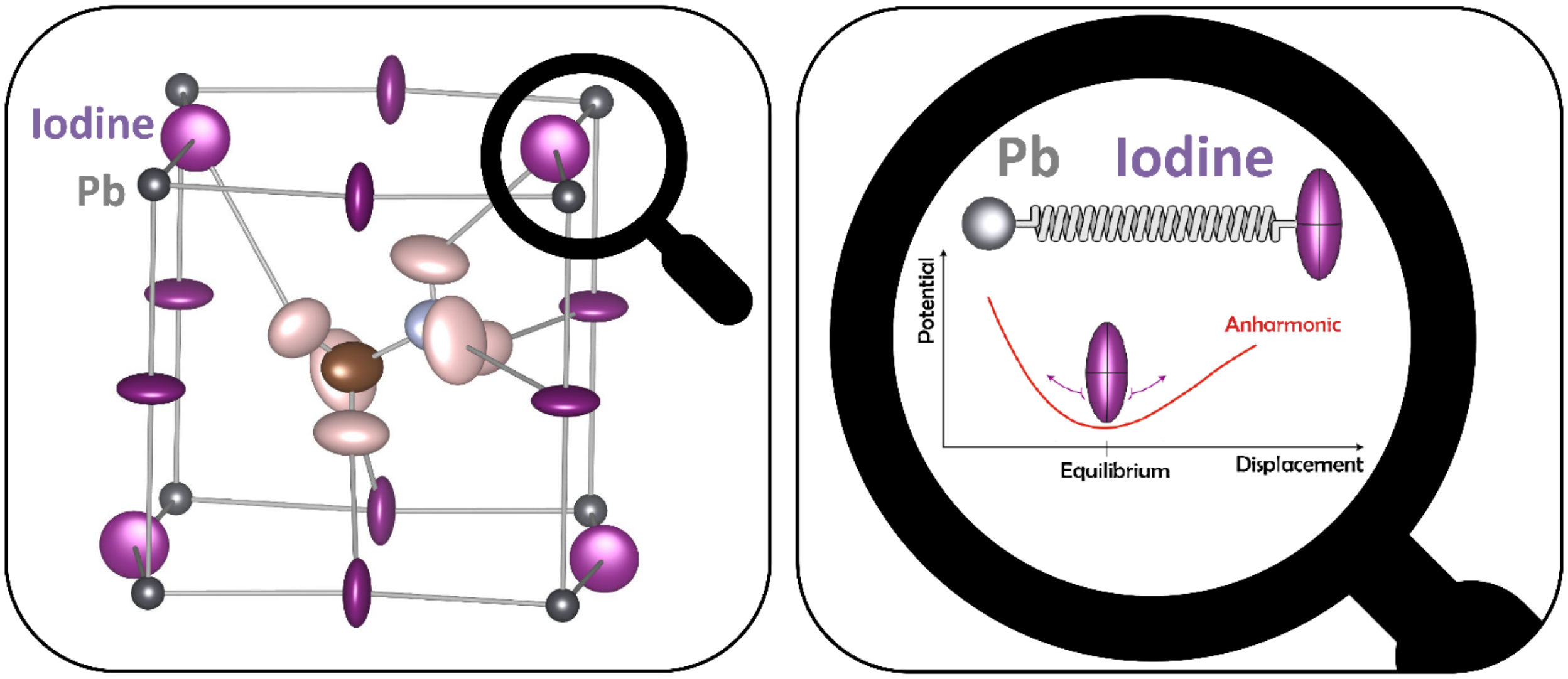

Electronic Supplementary Information for

# Temperature-dependent EXAFS measurements of the Pb L3-edge allow quantification of the anharmonicity of the lead-halide bond of chlorine-substituted methylammonium (MA) lead triiodide

Götz Schuck[1]*, Daniel M. Többens[1], Dirk Wallacher[1], Nico Grimm[1], Tong Sy Tien[2] and Susan Schorr[1,3]

[1] Helmholtz-Zentrum Berlin für Materialien und Energie, Hahn-Meitner-Platz 1, 14109 Berlin, Germany
[2] Department of Basic Sciences, University of Fire Prevention and Fighting, 243 Khuat Duy Tien, Hanoi 120602, Vietnam
[3] Institut für Geologische Wissenschaften, Freie Universität Berlin, Malteserstr. 74, 12249 Berlin, Germany

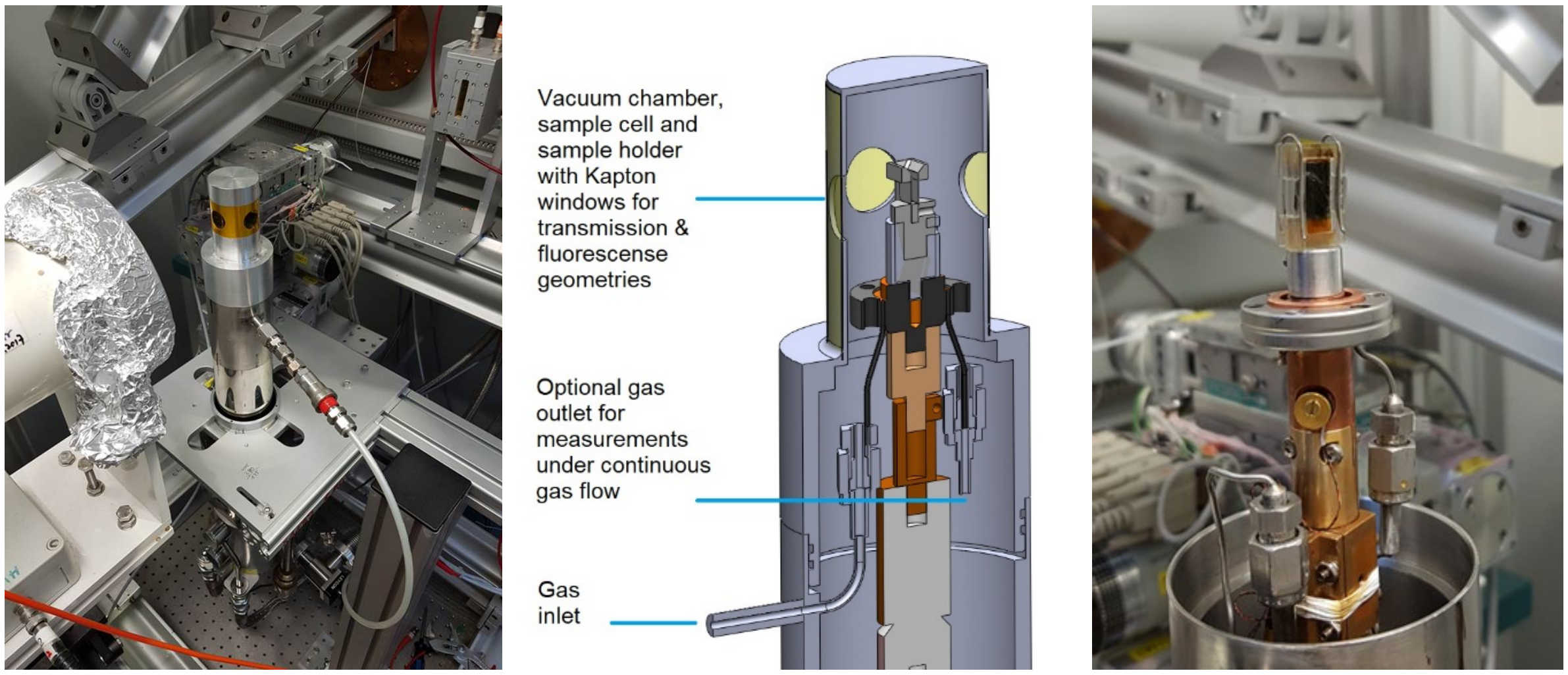

**Fig. S1** The newly build Cryo-EXAFS environment at KMC-2 "XANES" end station at BESSY II has been developed for in situ gas adsorption combined with X-ray powder diffraction experiments [1] and has been modified for use on the XANES end station. With multiple Kapton windows and a variable sample holder system, measurements in transmission and fluorescence geometries are possible. The Cryo-EXAFS consists of a Gifford-McMahon type closed cycle refrigerator (model DE202) operated by a helium compressor (model 8200) from Advanced Research Systems, Inc., USA. With an additional high temperature stage, the temperature range from 20 K to 450 K is accessible. The sample cell on the top of the temperature stage is surrounded by insulation vacuum better than 10-5 mbar. Due to a double-dome construction, the sample cell can be filled by Helium exchange gas up to an atmosphere. This guarantees a very well-defined temperature environment, especially for bad thermal conducting samples as powders. Temperature control is realized by a 4-channel temperature controller (model LS-336) from LAKE SHORE Cryotronics, USA, which is integrated into the instrument control software spec (CSS), allowing for fully automated temperature profile collection.

**XAFS energy calibration**

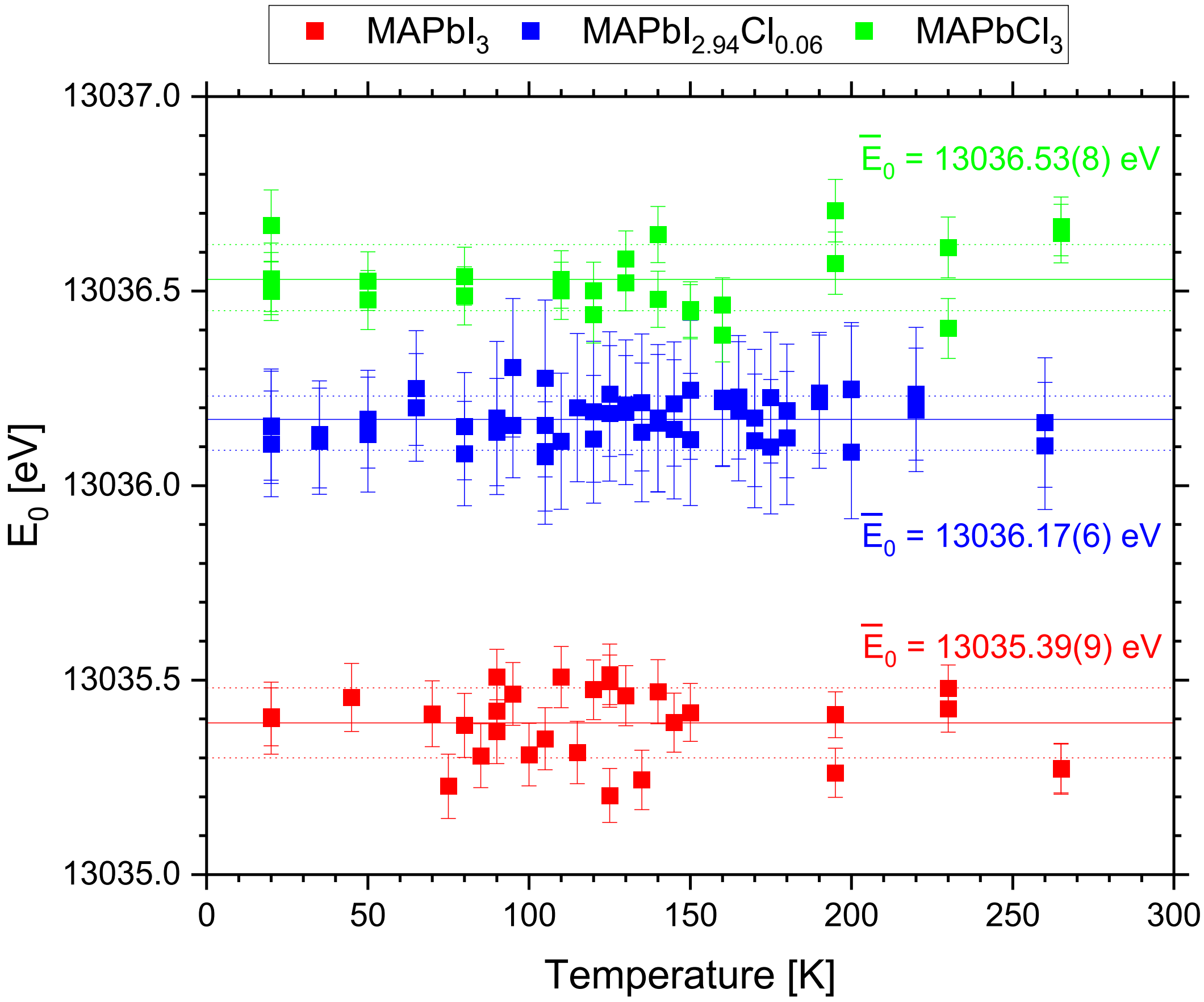

**Fig. S2**: $E_0$ values of Pb L3-edge XAS spectra (energy calibrated) of $MAPbI_3$ (red), $MAPbI_{2.94}Cl_{0.06}$ (blue) and $MAPbCl_3$ (green), averaged $\overline{E_0}$ values (solid lines) with standard deviation (dotted lines).

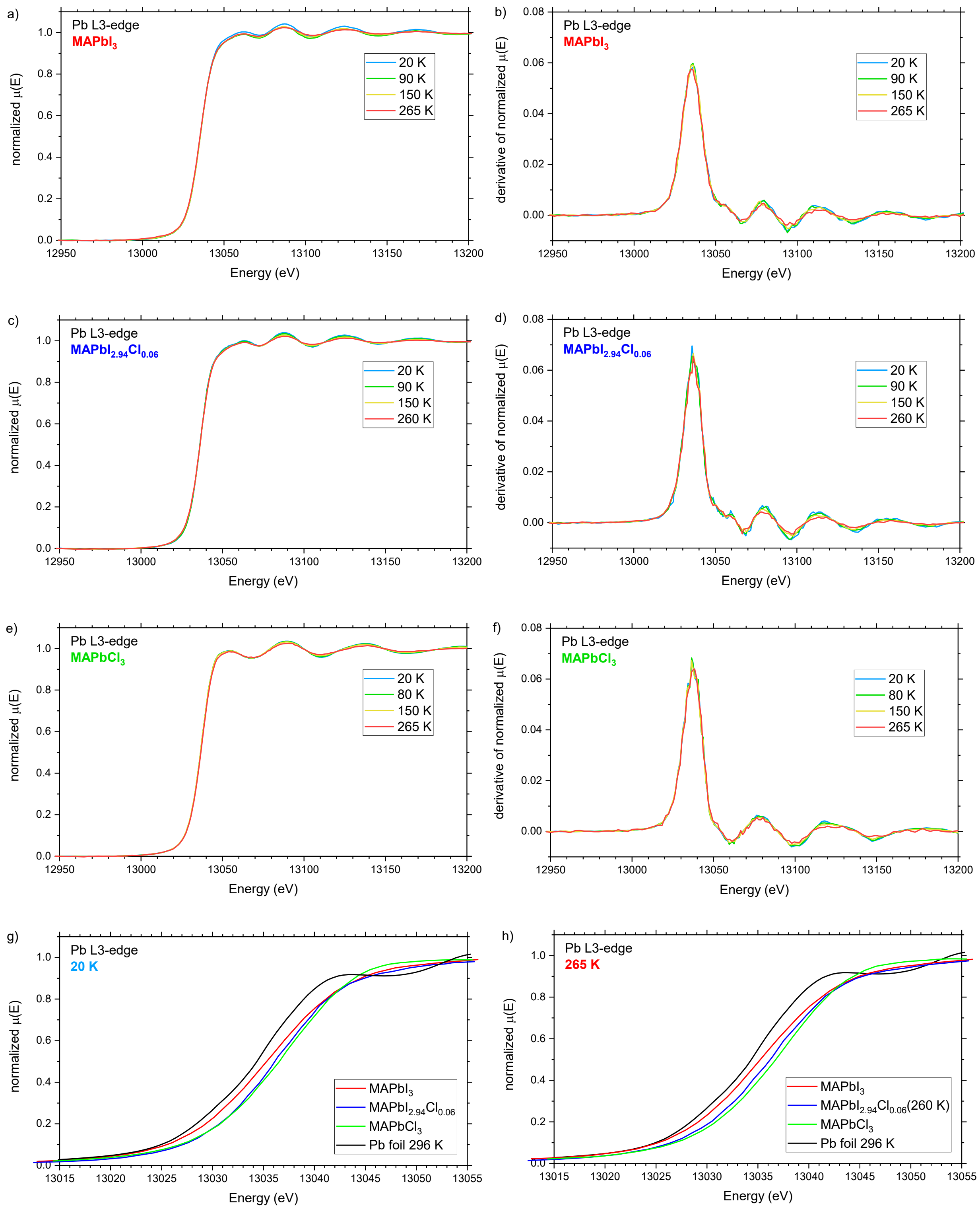

**Fig. S3**: Normalized XAS spectra (energy calibrated) of Pb L3-edge (b, d and f: derivative of normalized XAS spectra) of $MAPbI_3$ (a and b), $MAPbI_{2.94}Cl_{0.06}$ (c and d) and $MAPbCl_3$ (e and f) at 20 K (blue), 90 K (respectively for $MAPbCl_3$ 80 K, green), 150 K (dark yellow) and 265 K (respectively for $MAPbI_{2.94}Cl_{0.06}$ 260 K, red). Zoom on near edge (XANES) at g) 20 K and h) 265 K in each case compared to the reference measured at room temperature (Pb foil).

**Rietveld refinement of $MAPbCl_3$**

The two-phase nature of the low-temperature samples was established by comparing the difference between two diffraction patterns taken at the same temperature, but from samples that had formed different fractions of the two phases (samples 1 & 2 in Fig. S8). Rietveld refinement was done using FullProf.2k (version 5.30 & 7.20). The Thompson-Cox-Hastings parametrization of the pseudo-Voigt function was used for the peak shape and width. For the major phase anisotropic broadening was refined using symmetry-adapted spherical harmonics. Background was refined by linear interpolation between positions with little contribution from Bragg-peaks. Weak impurity peaks at 17.7°, 27.4°, 29.5, 29.7°, 41.7°, 42.4° were excluded from the refinement. These extra peaks appear already in the diffraction patterns of the cubic phase and are thus not related to the superstructures of the low-temperature phases. The cubic phase was modelled using the published structure of Baikie.[2] Only the heavy atoms lead and chlorine were refined, with anisotropic thermal displacement parameters. Position and thermal parameters of the light atoms forming the $NH_3CH_3$ cation were kept at fixed values. For the low temperature phase with the large 11 Å x 11 Å x 11 Å unit cell, the published structure of Chi was used.[3] As this phase only reaches a fraction of 22% the maximum, structural parameters were not refined. The structure of the intermediary phase in space group *Pnma* and medium sized 8 Å x 11 Å x 8 Å unit cell was refined using isotropic thermal displacement parameters for the chlorine atom on the 4*c* site. The chlorine on the 8*d* site was refined with anisotropic displacement, however with the off-diagonal parameters fixed to ensure a displacement ellipsoid orientation normal to the Pb-Pb line. Correlation between thermal displacement parameters and the atom positions are too high, so that independent refinement would lead to unphysical results. This is most likely caused by the high pseudo-symmetry of the structure. Position and orientation of the $NH_3CH_3$ cation were refined using a rigid-body approach based on a DFT-optimization of the molecule geometry used previously.[4] Thermal displacement parameters were kept fixed. The orientation of C-N direction and hydrogen positions were not experimentally determined, but assumed based on symmetry restrictions and geometry. Refined parameters for the single-phase structure at 160 K are given in Table S1.

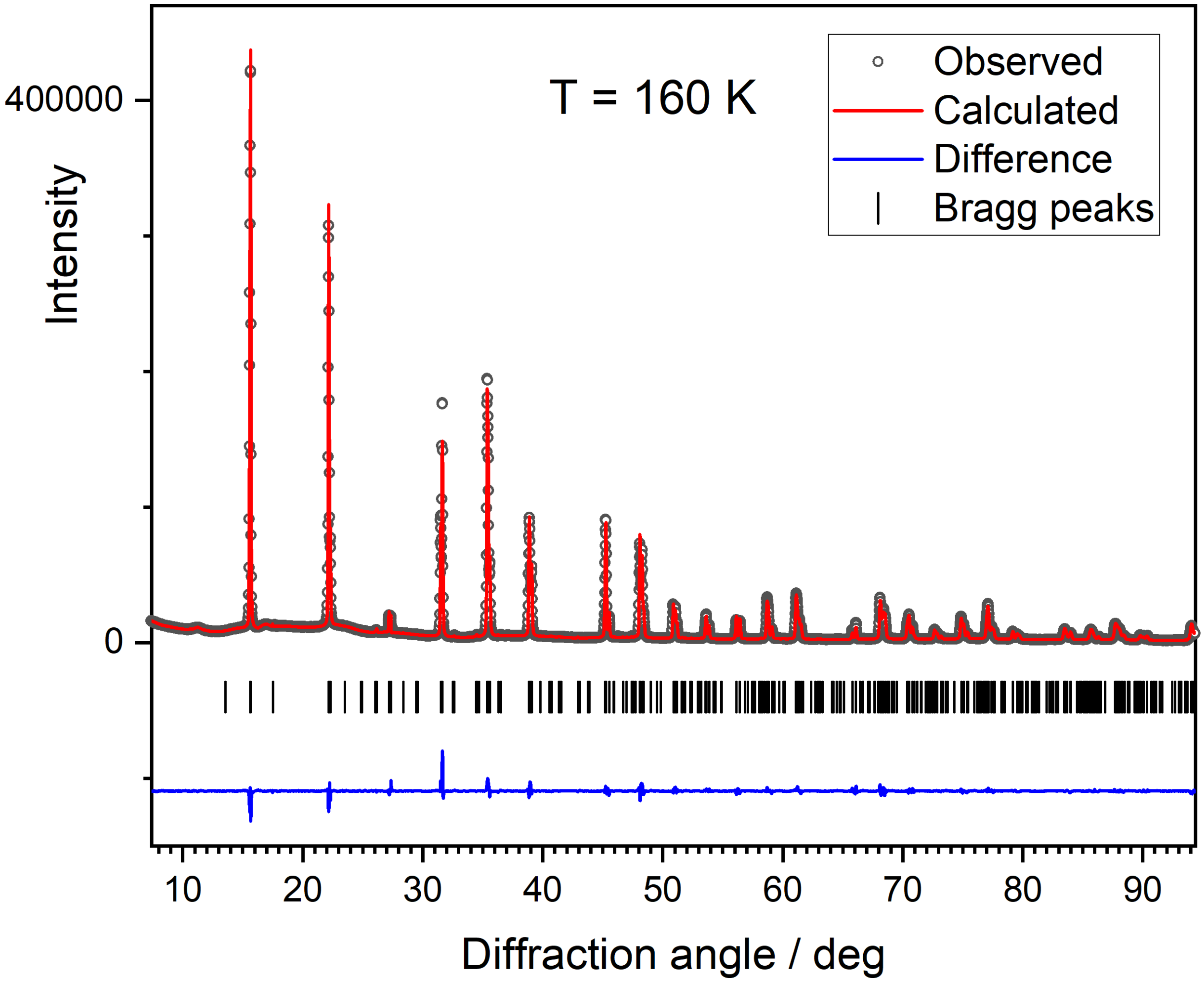

**Fig. S4**: Synchrotron X-ray powder diffraction pattern of $MAPbCl_3$ at temperature of 160 K with calculated pattern from Rietveld refinement of fully ordered structure in space group *Pnma* and lattice parameters $a = 8.018$ Å, $b = 11.344$ Å, $c = 7.961$ Å. Fit residuals are $R_{wp} = 0.055$, $\chi^2 = 15.5$, Bragg R-factor = 0.039. The comparatively high intensity deviation at 31° appears in all diffraction patterns including those of the cubic phase and is probably an artefact caused by an unknown experimental error.

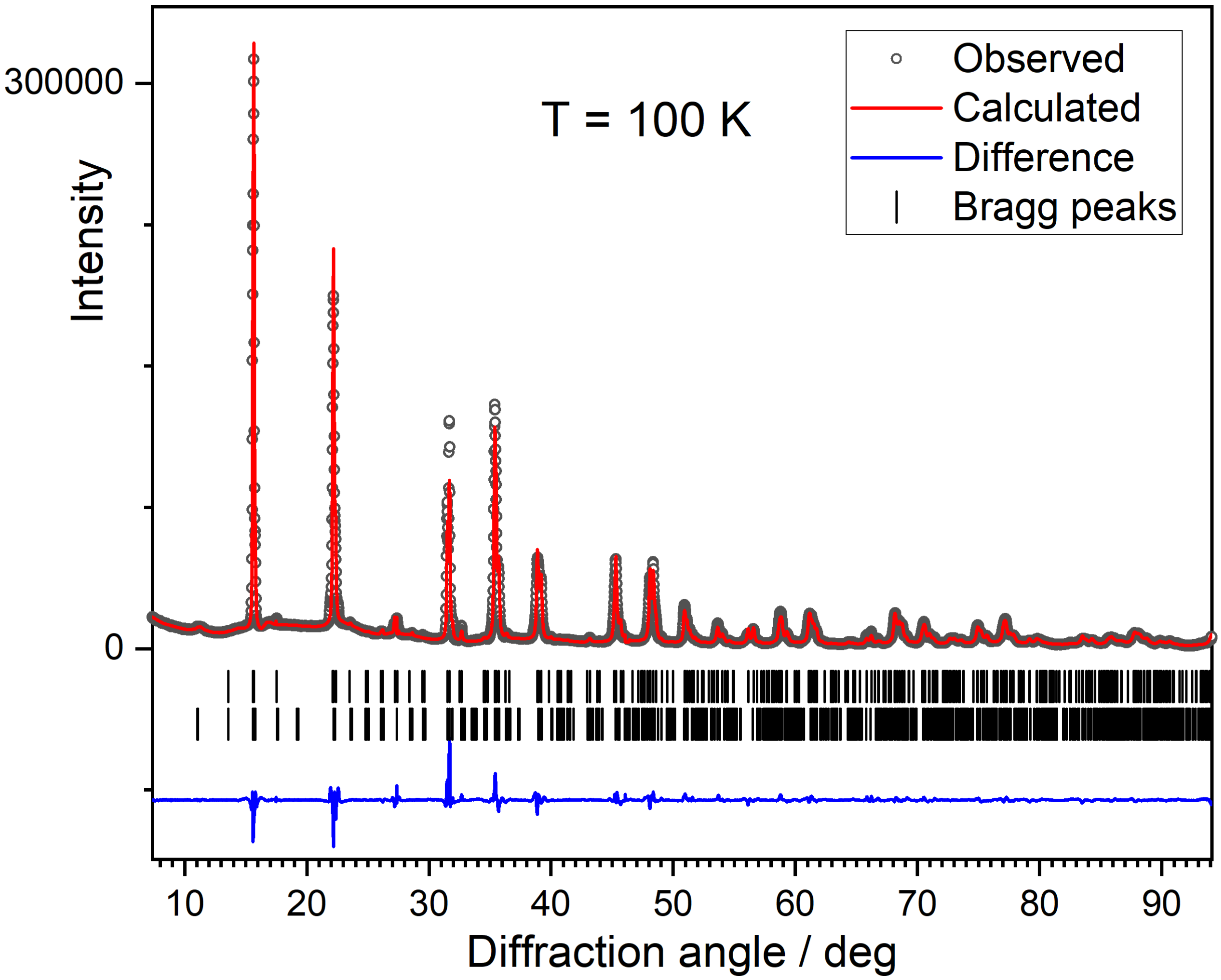

**Fig. S5**: Synchrotron X-ray powder diffraction pattern of $MAPbCl_3$ at temperature of 100 K with calculated pattern from Rietveld refinement of fully ordered structure (orthorhombic phase O2) in space group *Pnma* and lattice parameters $a$ = 8.020 Å, $b$ = 11.336 Å, $c$ = 7.933 Å (upper row of Bragg peaks) with a fraction of 17% of a second phase (orthorhombic phase O1) with the low-temperature structure given in the literature, also with *Pnma*, but $a$ = 11.195 Å, $b$ = 11.342 Å, $c$ = 11.287 Å (lower row of Bragg peaks). See also Fig. S6 for details. Fit residuals are $R_{wp}$ = 0.080, $\chi^2$ = 33.4, Bragg R-factors = 0.045 & 0.056 (first & second phase).

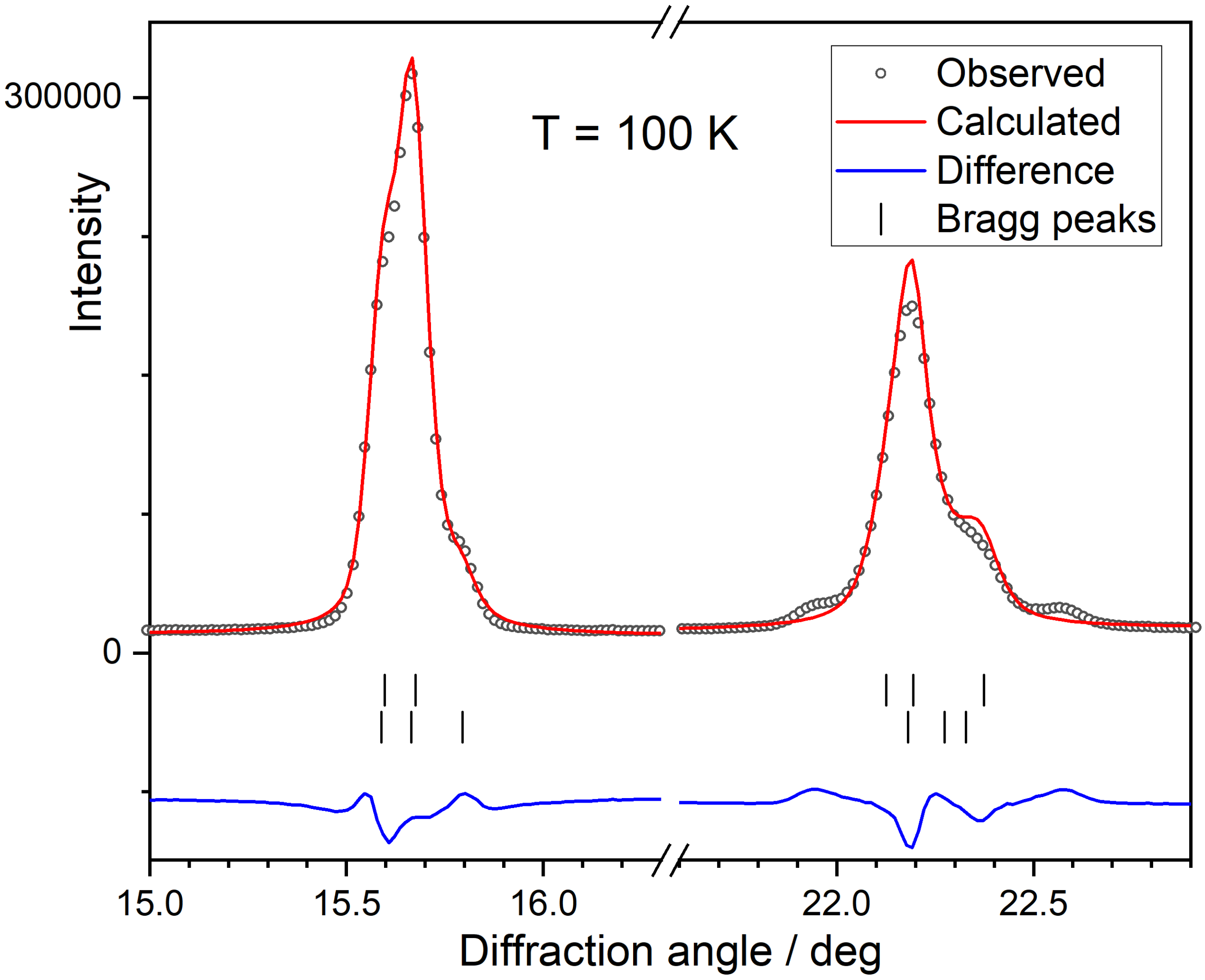

**Fig. S6**: Selected region of Fig. S5. The trifold splitting of the cubic 100 peak at 15.6° at first glance looks like it might be explained by the orthorhombic unit cell of low-temperature phase described in the literature alone (lower row of Bragg peaks). However, the relative intensities are in strong conflict with this interpretation, as are the positions of strong peaks at high diffraction angles. Relative intensities vary strongly between differently prepared samples in a manner that can only be explained by a mixture of phases with varying fractions. The group of peaks at 22.2°, resulting from the cubic 110 peak shows two additional weak peaks, which remain unexplained. Their origin is unknown, but they reliably appear in all low-temperature synchrotron XRPD measurements of $MAPbCl_3$.

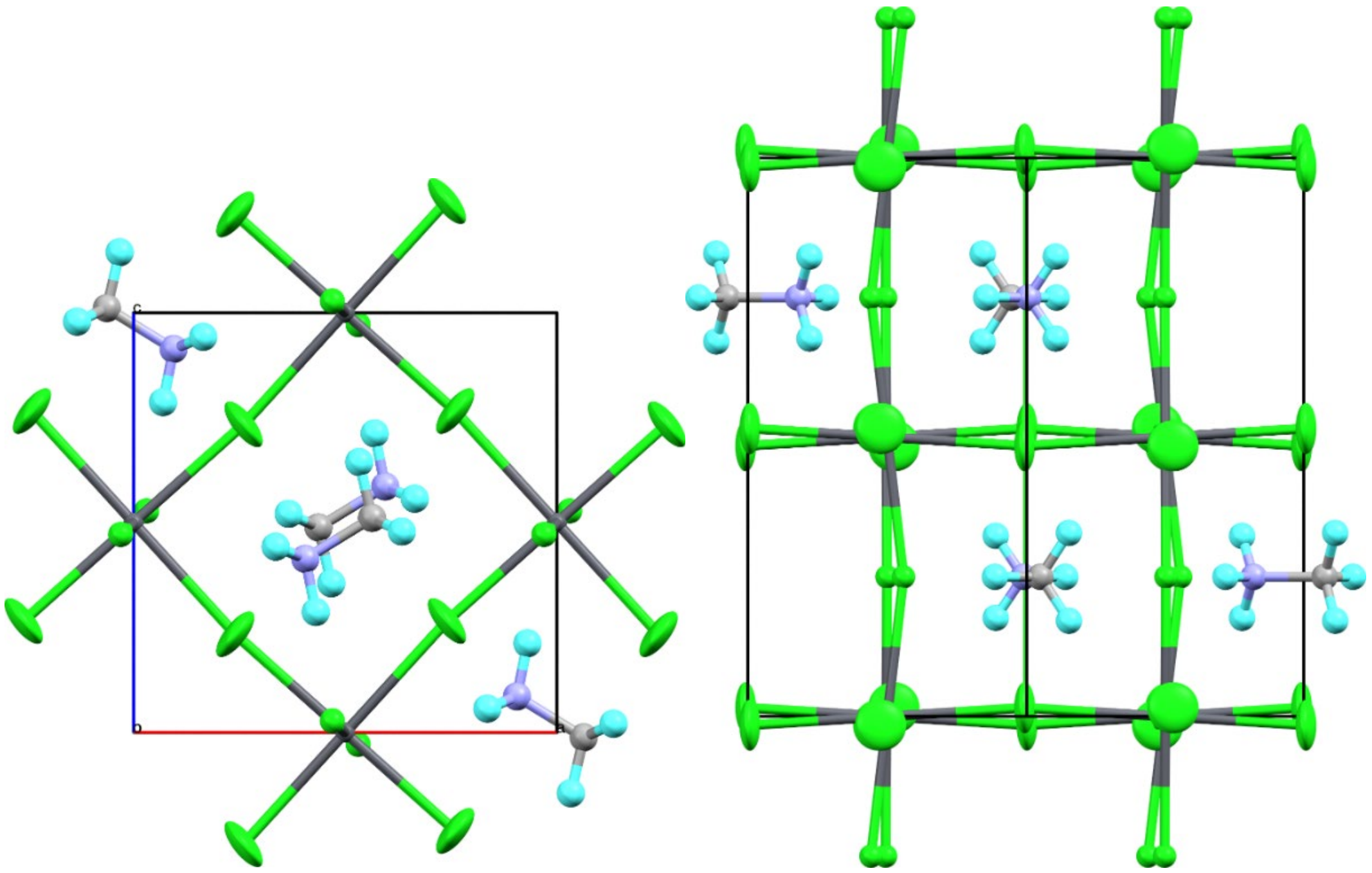

**Fig. S7:** Crystal structure of $MAPbCl_3$ at 160 K in space group *Pnma* (orthorhombic phase O2), with lattice parameters $a$ = 8.020 Å, $b$ = 11.336 Å, $c$ = 7.933 Å in projection along [010] (left) and [101] (right).

**Table S1**: Crystal structure of $MAPbCl_3$ (orthorhombic phase O2) at 160 K in space group *Pnma* (No. 62)

| Label | Type | Wyckoff | *x* | *y* | *z* | $U_{iso}$ / Å$^2$ | s.o.f. |
|---|---|---|---|---|---|---|---|
| N1 | N | 4*c* | -0.0941(73) | 0.25 | 0.0863(88) | 0.03 | 1.0 |
| H1a | H | 4*c* | -0.0689(73) | 0.25 | 0.2128(88) | 0.03 | 1.0 |
| H1b | H | 8*d* | -0.1642(73) | 0.17666 | 0.0575(88) | 0.03 | 0.5 |
| H1c | H | 8*d* | -0.1642(73) | 0.32333 | 0.0575(88) | 0.03 | 0.5 |
| C2 | C | 4*c* | 0.0632(73) | 0.25 | -0.0116(88) | 0.03 | 1.0 |
| H2a | H | 4*c* | 0.0325(73) | 0.25 | -0.1447(88) | 0.03 | 1.0 |
| H2b | H | 8*d* | 0.1337(73) | 0.17139 | 0.0205(88) | 0.03 | 0.5 |
| H2c | H | 8*d* | 0.1337(73) | 0.32862 | 0.0205(88) | 0.03 | 0.5 |
| Pb3 | Pb | 4*b* | 0.0 | 0.0 | 0.5 | 0.01523(19) | 1.0 |
| Cl4 | Cl | 4*c* | 0.0307(10) | 0.25 | 0.5281(9) | 0.024(2) | 1.0 |
| Cl5 | Cl | 8*d* | 0.2600(17) | 0.0154(8) | 0.7341(19) | 0.083(3) | 1.0 |
| Anisotropic displacement parameters U / Å$^2$ | | | | | | | |
| Cl5 | $U_{11}$ = 0.0548(20), $U_{22}$ = 0.114(4), $U_{33}$ = 0.079(4), $U_{12}$ = 0.0, $U_{13}$ = -0.0482, $U_{23}$ = 0.0 | | | | | | |
| Lattice parameters / Å | | | | | | | |
| $a$ = 8.01772(3) Å, $b$ = 11.34375(4) Å, $c$ = 7.96138(3) Å | | | | | | | |

The isotropic U given for Cl5 is the equivalent value calculated from the anisotropic displacement. Hydrogen atoms H1b, H1c H2b, H2c, respectively, are symmetry equivalent in the crystal structure.

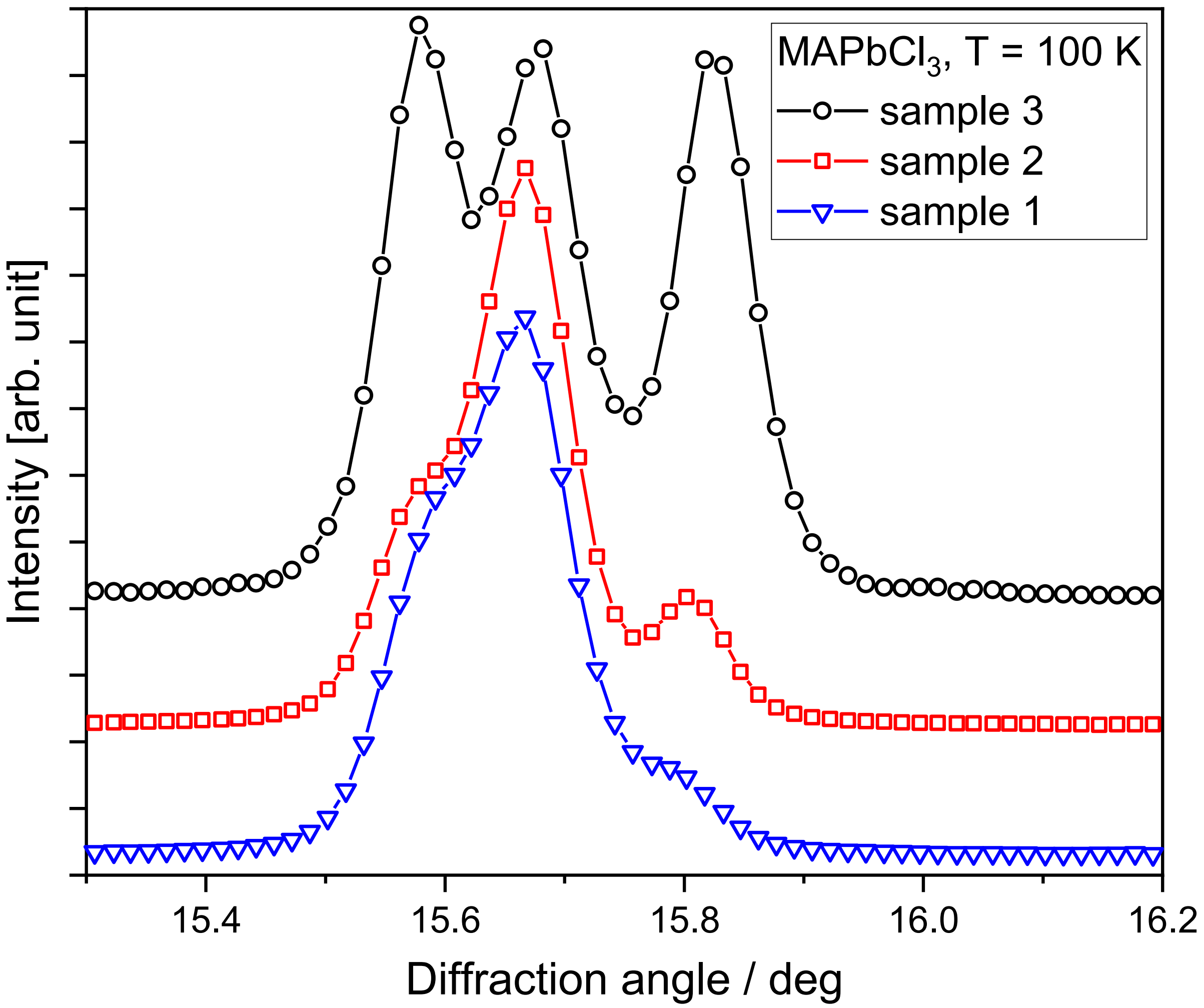

**Fig. S8**: Selected region of synchrotron XRPD pattern of three different samples of $MAPbCl_3$ at 100 K. Sample 1 (bottom) is the sample from Fig. S5, which was heavily ground and consists mostly of orthorhombic phase O2, containing only 17 % of the phase described in the literature as low-temperature phase (phase O1). Sample 2 is the same batch, but only lightly ground, and contains about 30 % of O1 phase. Sample 3 is a different batch and very lightly ground only, to the degree typically used in laboratory XRD. This sample consists entirely of the literature low-temperature phase O1. However, low divergence of synchrotron radiation leads to bad particle statistics, preventing Rietveld refinement. EXAFS experiments discussed in this paper were done on samples ground even harder than sample 1; pure phase O2 was present in the powders used for EXAFS measurements. For $MAPbI_3$ no such effect was observed.

# Rietveld refinement of $MAPbI_{2.94}Cl_{0.06}$

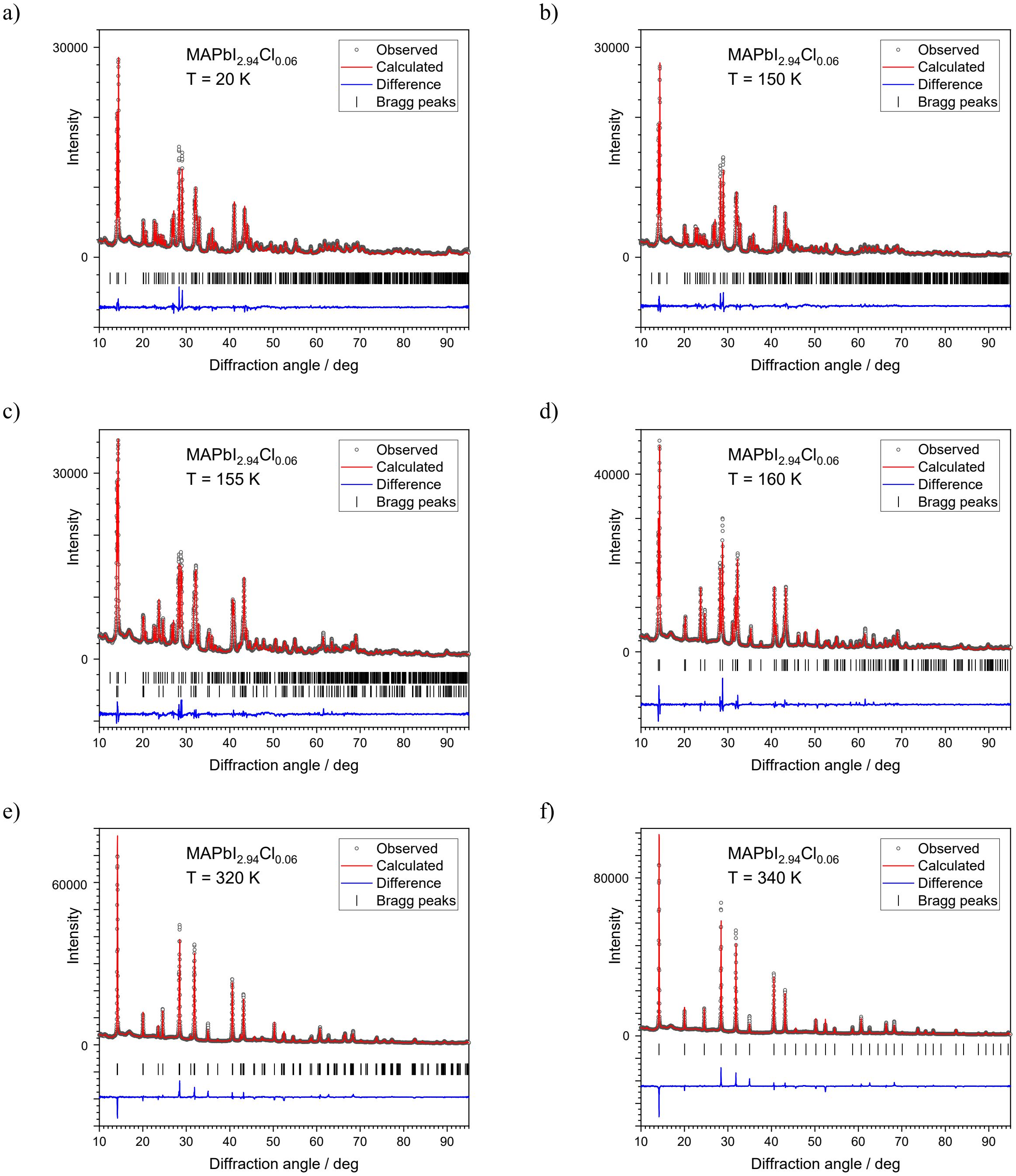

**Fig. S9**: Synchrotron X-ray powder diffraction pattern of $MAPbI_{2.94}Cl_{0.06}$ at various temperatures with calculated pattern from Rietveld refinement of fully ordered structure in space group *Pnma* (a and b), orthorhombic/tetragonal structure (c), in the tetragonal structure (d and e), and in the cubic structure (f).

**Table S2**: Crystal structure of $MAPbI_{2.94}Cl_{0.06}$ at 150 K in space group *Pnma* (No. 62)

| Label | Type | Wyckoff | *x* | *y* | *z* | $U_{iso}$ / Å$^2$ | s.o.f. |
|---|---|---|---|---|---|---|---|
| Pb1 | Pb | 4b | 0.5 | 0.0 | 0.0 | 0.0211(6) | 1.0 |
| I2 | I | 4c | 0.4886(6) | 0.25 | 0.9484(3) | 0.0416(8) | 0.98 |
| Cl2 | Cl | 4c | 0.4886(6) | 0.25 | 0.9484(3) | 0.0416(8) | 0.02 |
| I3 | I | 8d | 0.1953(2) | 0.0171(2) | 0.1901(2) | 0.0416(8) | 0.98 |
| Cl3 | Cl | 8d | 0.1953(2) | 0.0171(2) | 0.1901(2) | 0.0416(8) | 0.02 |
| N4 | N | 4c | 0.57301(11) | 0.25 | 0.52340(10) | 0.0416(8) | 1.0 |
| C5 | C | 4c | 0.42049(11) | 0.25 | 0.45237(10) | 0.0416(8) | 1.0 |
| H6 | H | 4c | 0.4328(5) | 0.25 | 0.32652(16) | 0.0416(8) | 1.0 |
| H7 | H | 8d | 0.3618(4) | 0.32107(6) | 0.4906(3) | 0.0416(8) | 1.0 |
| H8 | H | 8d | 0.6321(3) | 0.31606(6) | 0.4875(2) | 0.0416(8) | 1.0 |
| H9 | H | 4c | 0.5668(5) | 0.25 | 0.64324(13) | 0.0416(8) | 1.0 |
| Lattice parameters / Å | | | | | | | |
| *a* = 8.85749(12) Å, *b* = 12.60386(14) Å, *c* = 8.59400(12) Å | | | | | | | |

Fit residuals are $R_{wp}$ = 0.125, $\chi^2$ = 4.674, Bragg R-factor = 0.0451.

## Cubic and pseudo cubic unit cells

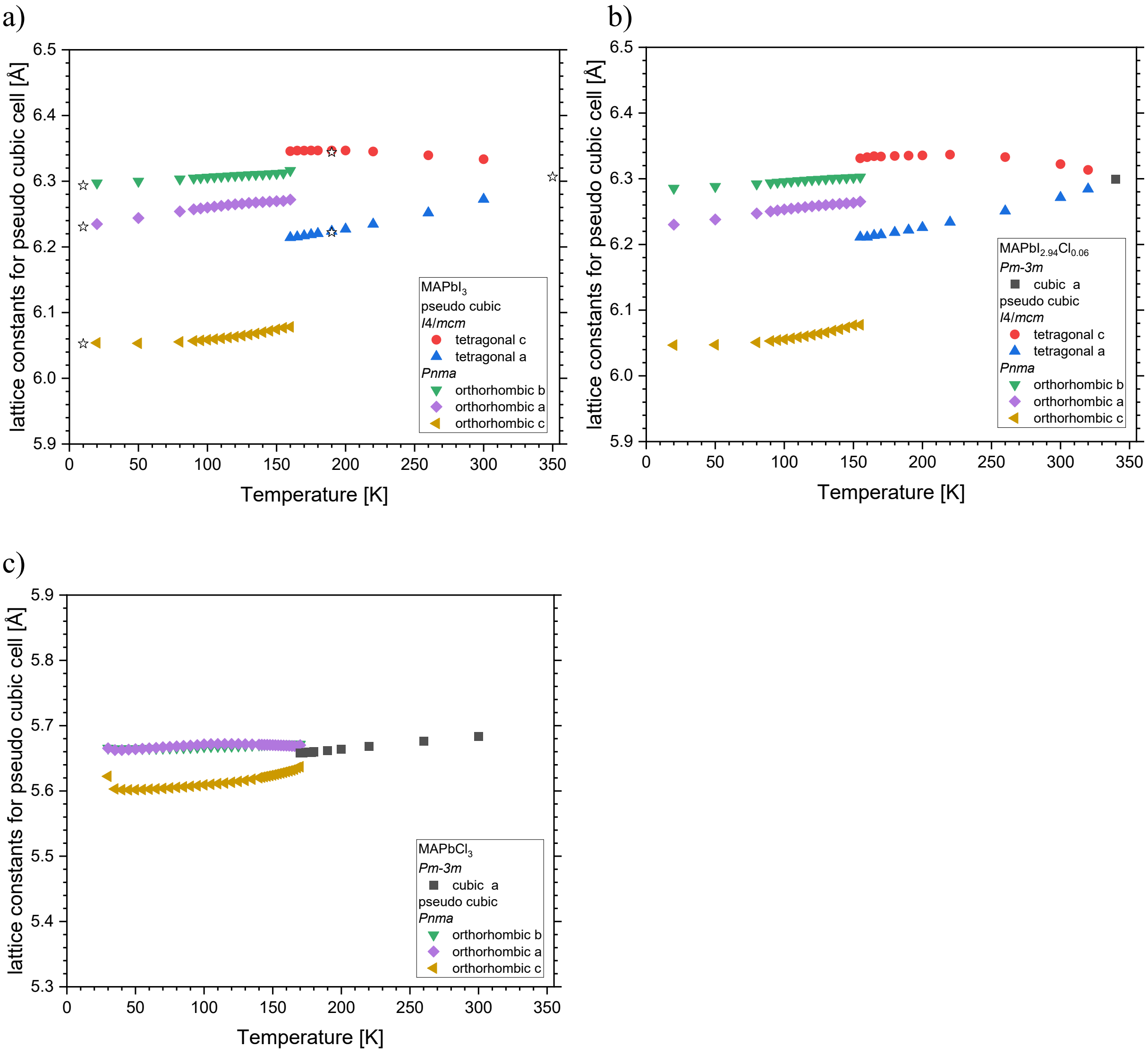

**Fig. S10**: Rietveld results from synchrotron X-ray powder diffraction of a) $MAPbI_3$ (open stars: Whitfield et al. [8]), b) $MAPbI_{2.94}Cl_{0.06}$, and c) $MAPbCl_3$ (O2 phase) at various temperatures. Cubic and pseudo cubic unit cells.

**Pb-X-Pb angle**

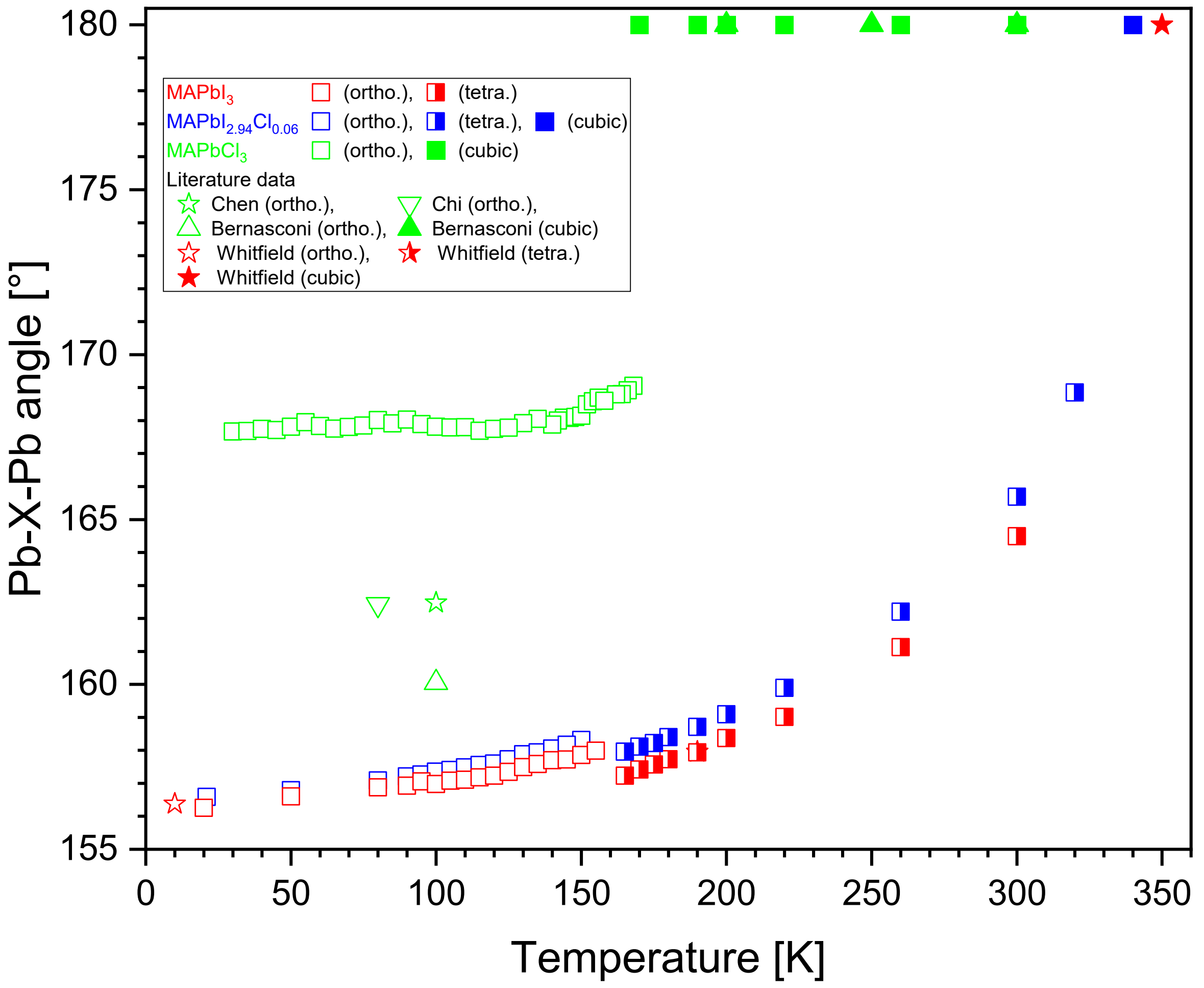

**Fig. S11**: Results of the Rietveld analysis of synchrotron XRD data of $MAPbI_3$ (red color), $MAPbI_{2.94}Cl_{0.06}$ (blue color), and $MAPbCl_3$ (green color). Variation in the Pb–X–Pb angle as a function of temperature in the orthorhombic (open symbols, averaged values), tetragonal (half filed symbols), and cubic (solid symbols) phase. Literature values are also given for comparison.[3,8,20,21]

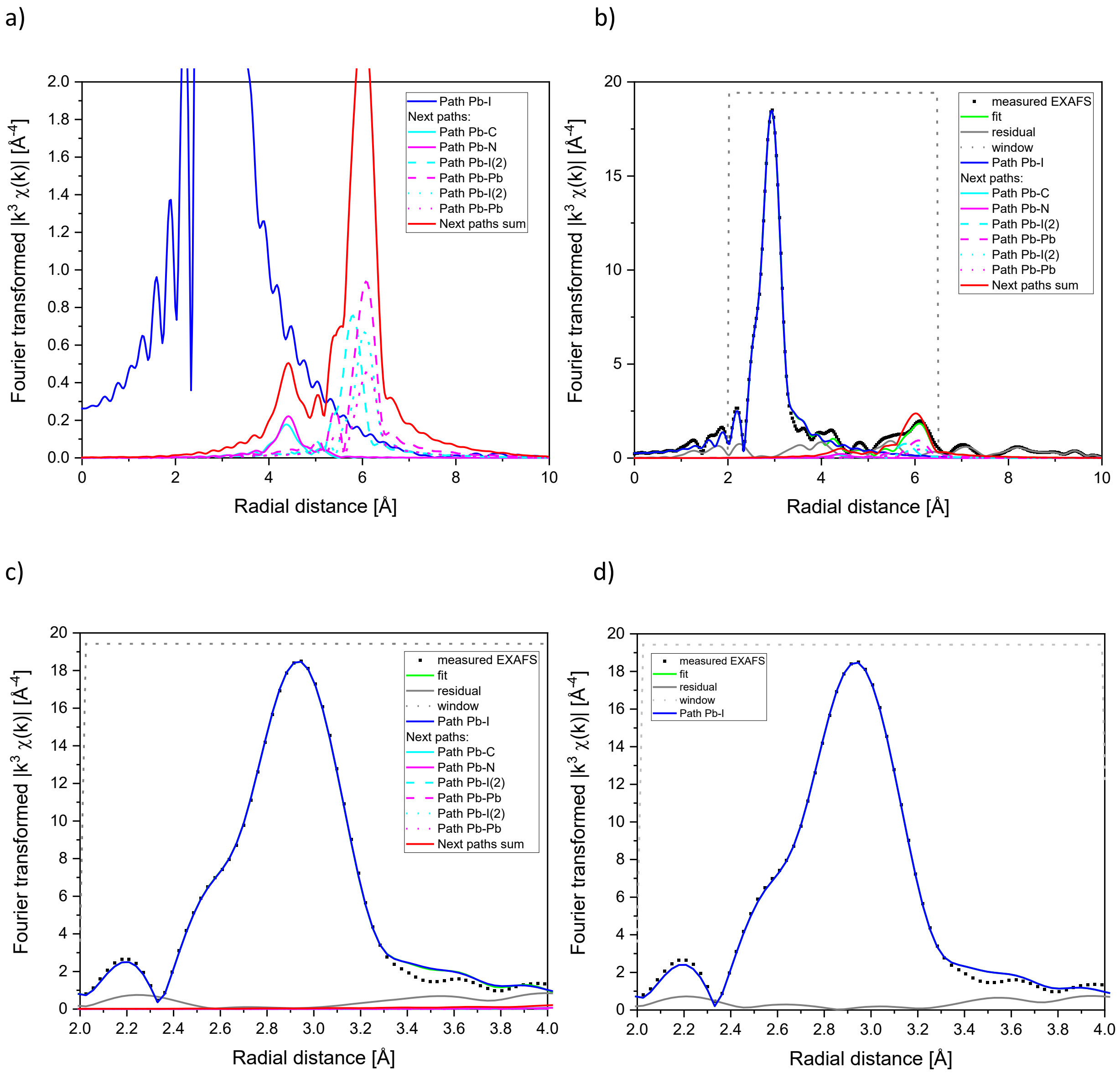

**Fig. S12**Fourier transforms of the EXAFS signal at 20 K (black squares are the moduli) and the respective best fits for $MAPbI_3$. The Fourier transforms were made in the interval k = 3 - 14 $Å^{-1}$, with a $k^3$ weighting and a Hanning window for all spectra. a), b), and c) fit was performed with a R range from 2 - 6.5 Å with Pb-I(1) (first shell) (blue) and following single paths (red colors). The sum of the next paths (red dashed line) corresponds to a fraction of 1.0 % of the total |(χ(R)| in the range of 2 - 4 Å. d) comparison with fit with Pb-I (first shell) path only (R range from 2 - 4 Å). Fit results for path Pb-I(1) with next paths (a, b, and c): ΔR = -0.017(4) Å, $\sigma^2$ = 0.0031(4) $Å^2$, $C_3$ = -0.00009(4) $Å^3$, and $C_4$ = -0.000000(9) $Å^4$ for comparison the results with one path only (d): ΔR = -0.019(7) Å, $\sigma^2$ = 0.0028(9) $Å^2$, $C_3$ = -0.00010(7) $Å^3$, and $C_4$ = -0.000008(15) $Å^4$.

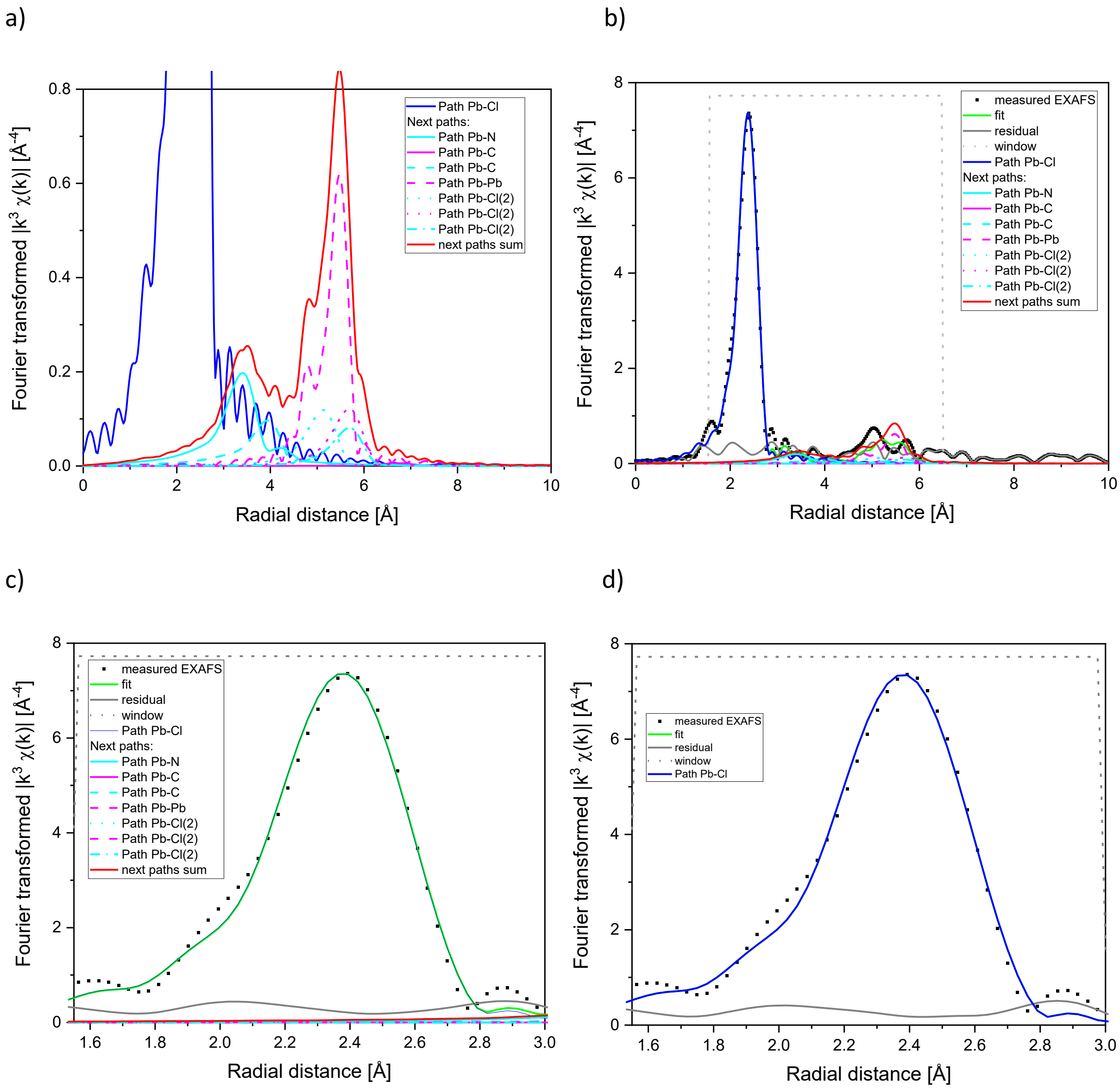

**Fig. S13**: Fourier transforms of the EXAFS signal at 20 K (black squares are the moduli) and the respective best fits for $MAPbCl_3$. The Fourier transforms were made in the interval k = 3 - 12 $Å^{-1}$, with a $k^3$ weighting and a Hanning window for all spectra. a), b), and c) fit was performed with a R range from 1.55 - 6.5 Å with Pb-Cl(1) (first shell) (blue) and following single paths (red colors). The sum of the next paths (red dashed line) corresponds to a fraction of 1.8 % of the total $|(\chi(R)|$ in the range of 1.55 - 3 Å. d) comparison with fit with Pb-Cl(1) (first shell) path only (R range from 1.55 - 3 Å). Fit results for path Pb-Cl(1) with next paths (a, b, and c): $\Delta R = 0.028(3)$ Å, $\sigma^2 = 0.0054(5)$ $Å^2$, $C_3 = 0.00038(8)$ $Å^3$, and $C_4 = -0.00004(2)$ $Å^4$ for comparison the results with one path only (d): $\Delta R = 0.027(3)$ Å, $\sigma^2 = 0.0054(5)$ $Å^2$, $C_3 = 0.00035(9)$ $Å^3$, and $C_4 = -0.00004(2)$ $Å^4$.

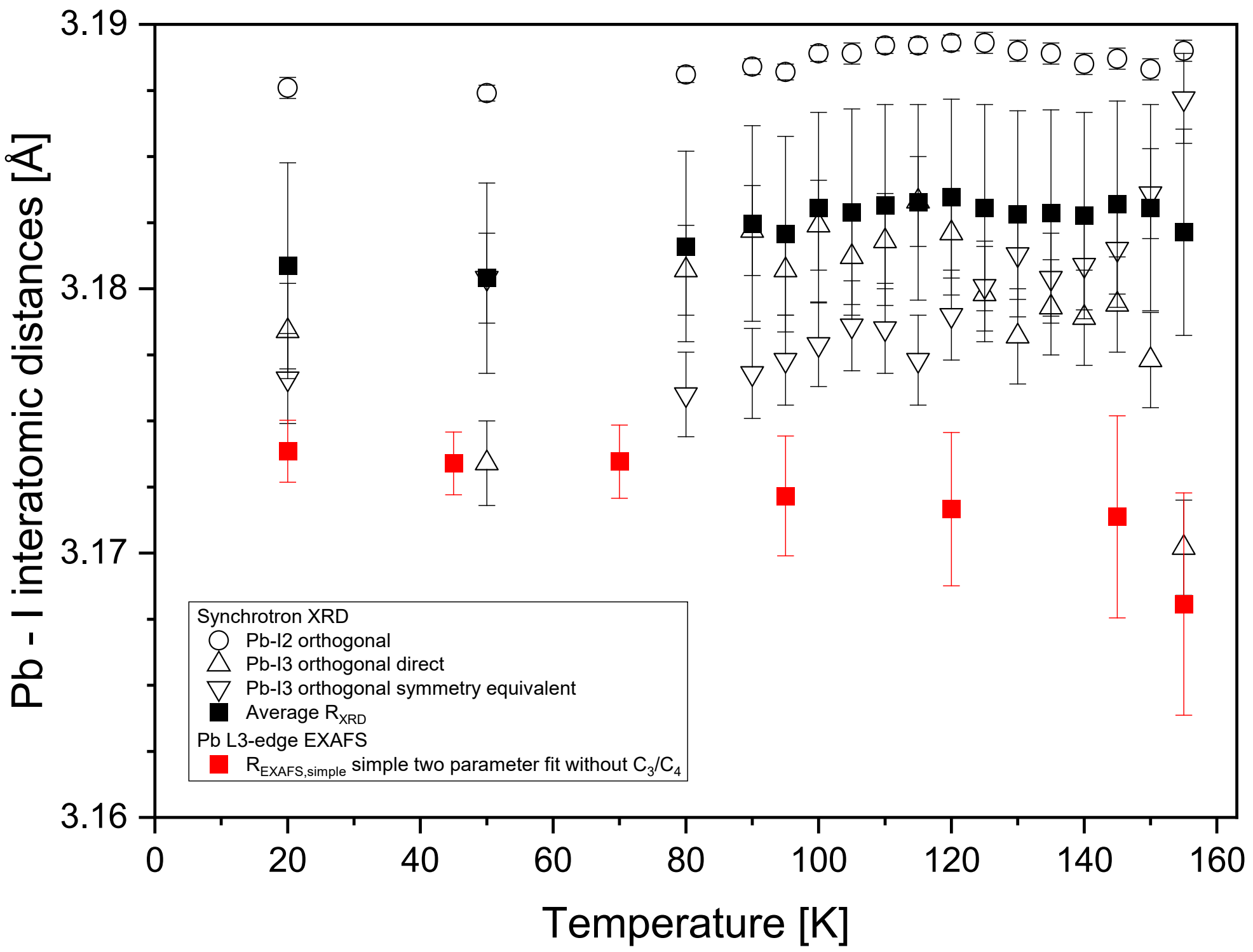

**Fig. S14** Temperature-dependent lead-iodine distances $R_{XRD}$ in the orthorhombic phase of $MAPbI_3$ (black symbols). The three different lead-iodine distances (black open symbols) and the resulting averaged lead-iodine distance (black closed symbols) are shown. Red squares: for comparison, the $R_{EXAFS}$ values resulting from a simple EXAFS fit without the cumulants $C_3$ and $C_4$ are shown (fit parameter: $R_{EXAFS}$ and the EXAFS Debye Waller value ($\sigma^2$); see also Fig. S16).

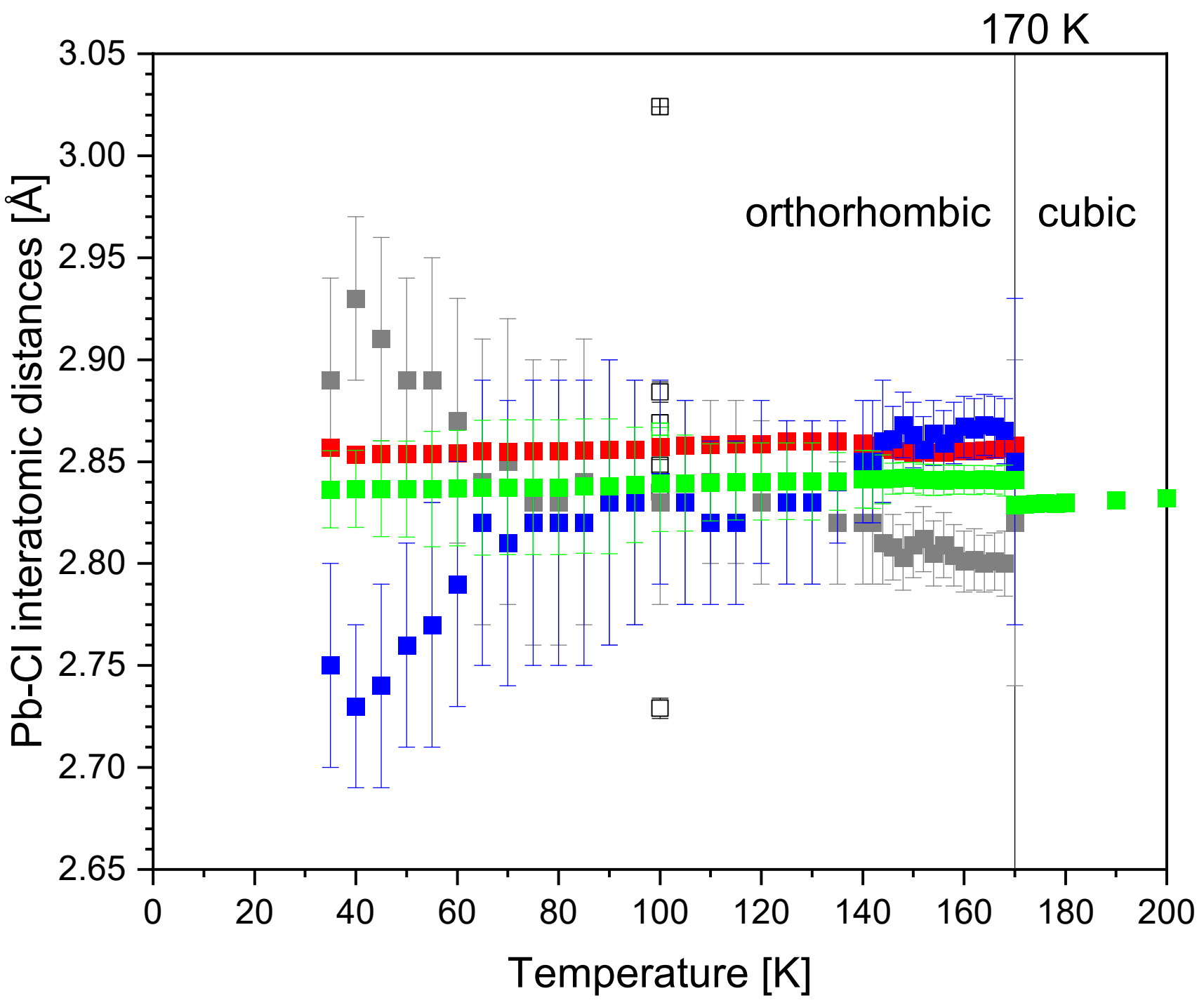

**Fig. S15** Temperature-dependent lead-chloride distances in the orthorhombic and cubic phase of $MAPbCl_3$. The three different lead-iodine distances in the orthorhombic phase O2 from synchrotron XRD Rietveld refinement (closed symbols) are compared with the averaged lead-chloride distance (green symbols), and with the six different lead-chloride distances from Chi et al. [3] (open symbols, averaged lead-chloride distance: green open symbol).

a)

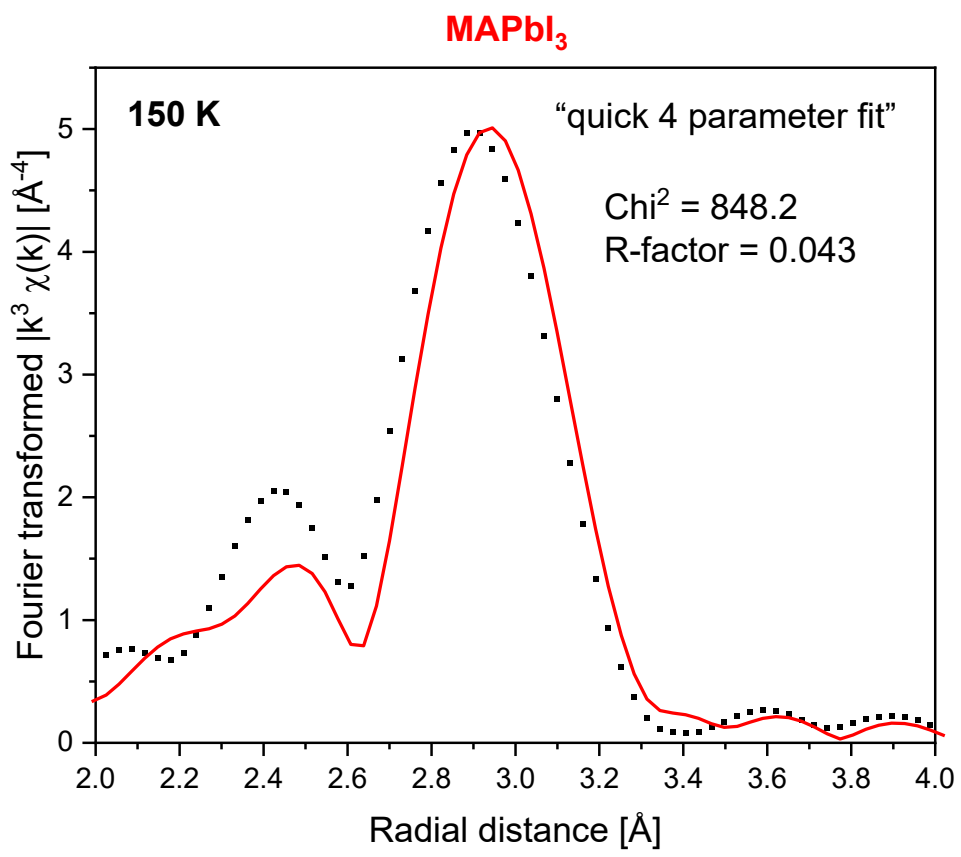

b)

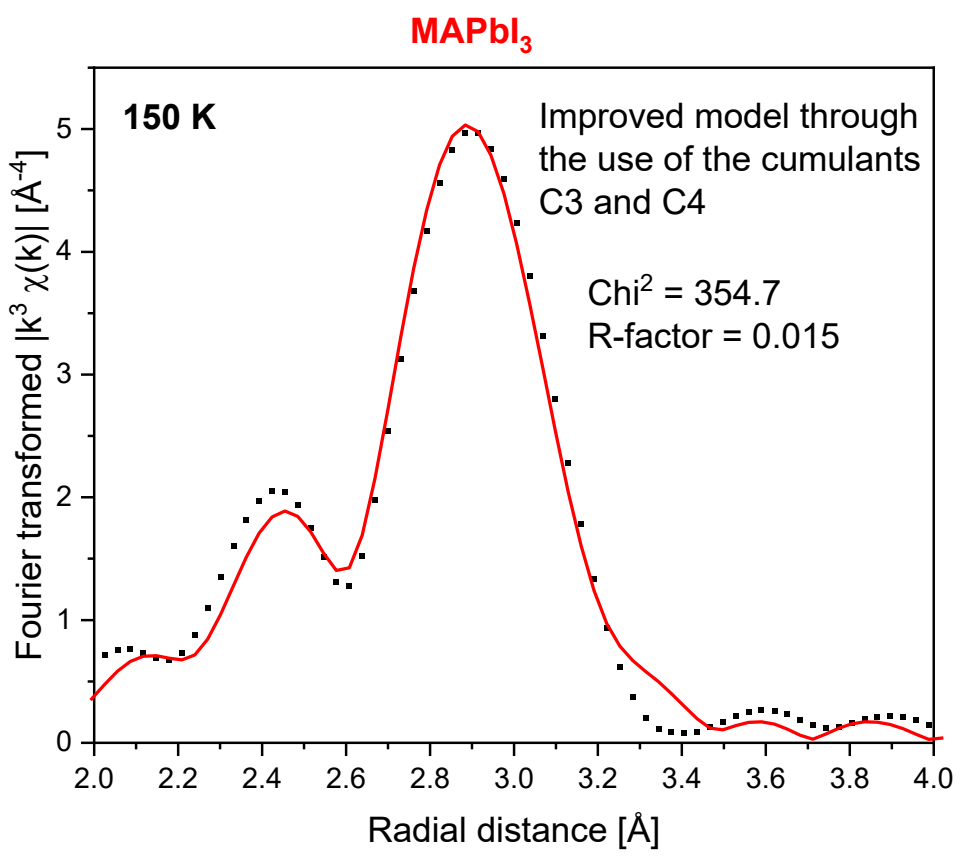

**Fig. S16** Comparison of (a) a simple fit model ("quick 4 parameter fit" with ΔE0 and $S0^2$ determined at 20 K) with (b) a model which considers the cumulants $C_3$ and $C_4$. The Fourier transforms of the EXAFS signal from $MAPbI_3$ at 150 K (black squares are the moduli) and the respective best fits (red lines). The Fourier transforms have been made in the interval k = 3–14 $Å^{-1}$, with a $k^3$ weighting and a Hanning window.

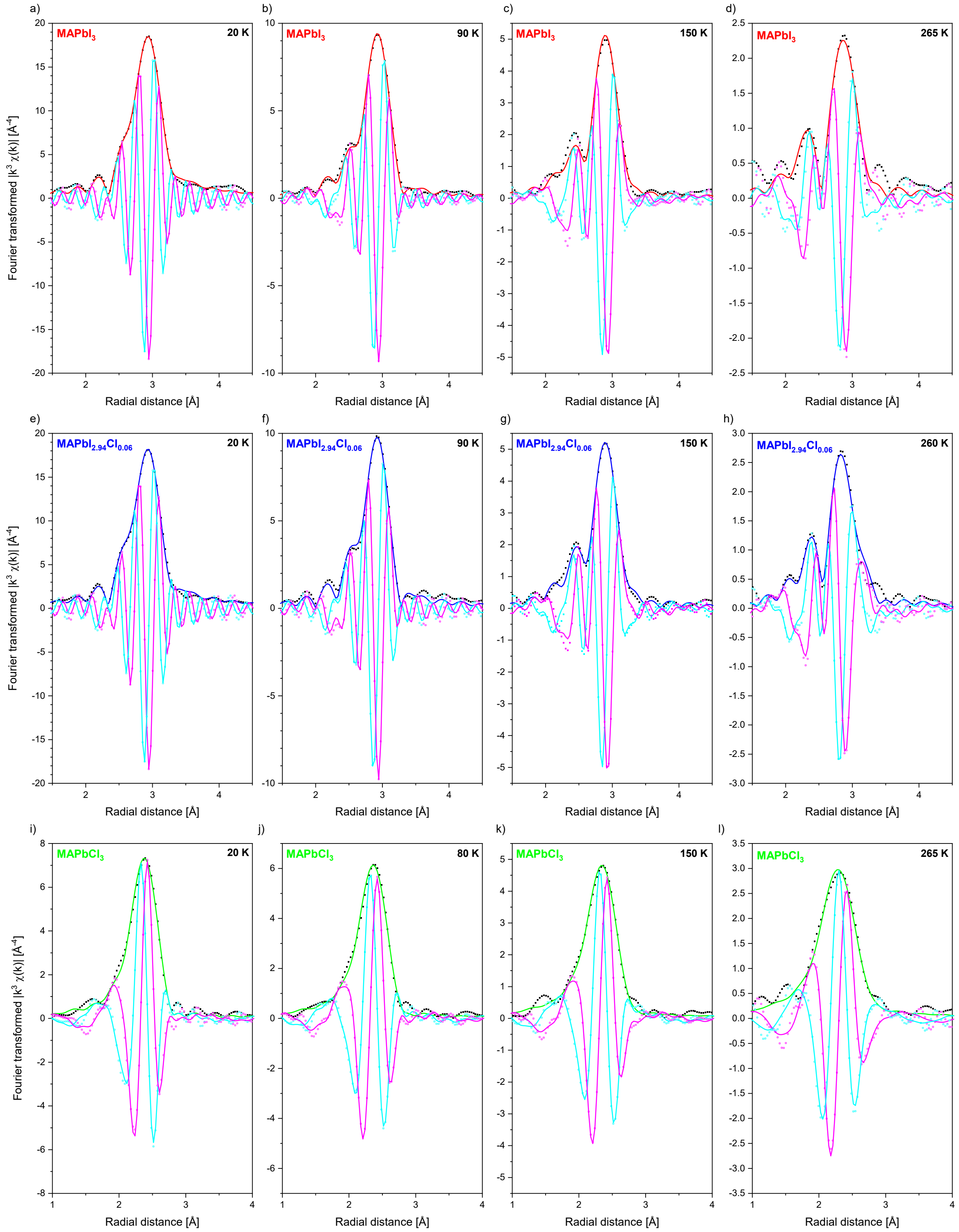

**Fig. S17** Real (magenta) and imaginary (cyan) parts of the Fourier transform of the EXAFS signal (squares are the moduli) shown in Fig. 4 and the respective best fits (solid lines) for $MAPbI_3$ (a, b, c, d), for $MAPbI_{2.94}Cl_{0.06}$ (e, f, g, h) and for $MAPbCl_3$ (i, j, k, l).

**MSRD and MSD**

In general, the comparison of mean square relative displacement (MSRD) from cumulant analysis of temperature-dependent EXAFS data and mean square displacement (MSD) from diffraction allows deeper insights into the vibrational anisotropy of the bonds under consideration.[11] For our case, this means that we have to consider the sum of the temperature factors of the lead and the halide atom for the MSD from the diffraction, and the parallel MSRD and the perpendicular MSRD for the MSRD from the EXAFS data. For our investigated hybrid perovskites, however, an isotropic temperature factor is present for the lead atom in the diffraction and a discus-shaped atomic displacement ellipsoid is present for the halide atoms (Fig. S19a). It should be noted here that the comparison of partially anisotropic temperature factors for the MSDs and MSRD from EXAFS were already carried out for a very similar system, namely $ReO_3$.[12]

However, the atomic displacement ellipsoid was parameterized in the Rietveld refinement by the six anisotropic displacement parameters (ADP). The determination of the ADP in the orthorhombic phase was not readily possible from Rietveld refinement of our powder diffraction data. Single crystal diffraction data in the orthorhombic phase was only available for $MAPbI_3$ at 20K. For the chlorine component, only single crystal diffraction data was available in the cubic structure, and for the orthorhombic phase O2 there was no ADP data at all. Moreover, there is no literature data for $MAPbI_{2.94}Cl_{0.06}$. The ordering processes in the low-temperature structure of $MAPbX_3$ lead to effects which complicate the analysis of the temperature factors with diffraction (twinning, segregation, and or mis-ordered phases). We tried to circumvent the difficulties in the ADP determination with the Rietveld method by reducing the free parameters of the Rietveld refinement. For example, we parameterized the isotropic temperature parameters of the lead atom with an Einstein behavior. Analogous to the cumulant analysis of the EXAFS data, it was assumed that the relative temperature change of the isotropic temperature factors is unaffected by other Rietveld refinement parameters (for example: the course of the background, or profile parameters). For the isotropic temperature parameters of the lead atoms, an Einstein temperature of 87.4 K could be determined for MAPI and an Einstein temperature of 86.3 K for $MAPbI_{2.94}Cl_{0.06}$ (Fig. S18). In a subsequent Rietveld refinement on the synchrotron data of $MAPbI_{2.94}Cl_{0.06}$, ADPs for the halide atoms could be refined. Here, the isotropic temperature factors of the lead atoms were no longer free parameters, but the parameters given by Einstein's behavior were used (Fig. S21). For the synchrotron data of $MAPbI_3$, this approach did not allow refinement of the ADPs for the iodine atoms, so that we could only draw a comparison with the literature data available to us (Table S3 and Fig. S20). Furthermore, the analysis of ADPs is only possible to a limited extent for $MAPbCl_3$. For example, although a refinement of the anisotropic temperature factors for the chlorine atom on the Wyckoff position 8d is possible for the synchrotron data at 160 K (Table S1 and Fig. S7), only isotropic temperature factors can be refined for the chlorine atom on 4c (Fig. S22). The analysis of the isotropic temperature factors for the lead atom is also difficult. In some cases, lower temperatures result in negative temperature factors, so that we refrain from a detailed discussion for the time being. From our preliminary compilation (Fig. S20-S22), however, we can already note for $MAPbI_3$ and $MAPbI_{2.94}Cl_{0.06}$ (probably also for $MAPbCl_3$) that parallel MSD = Pb Uiso + halide $\overline{U}_3$ (Fig. S19) is larger than parallel MSRD (the difference then corresponds to the displacement correlation function (DCF)[13,14]). The DCF is similar to that for $ReO_3$.[15] The differences between perpendicular MSD = Pb $U_{iso}$ + halide $\overline{U}_1$ (Fig. S18) and the perpendicular MSRD seem to be smaller for $MAPbI_3$ and $MAPbI_{2.94}Cl_{0.06}$. Overall, however, it can be stated that there is not yet sufficient information regarding the ADPs from the diffraction, especially in the orthorhombic phase. One possibility would be to carry out a parameterized Rietveld refinement.[16] This is beyond the scope of this work and could be a future approach to this issue.

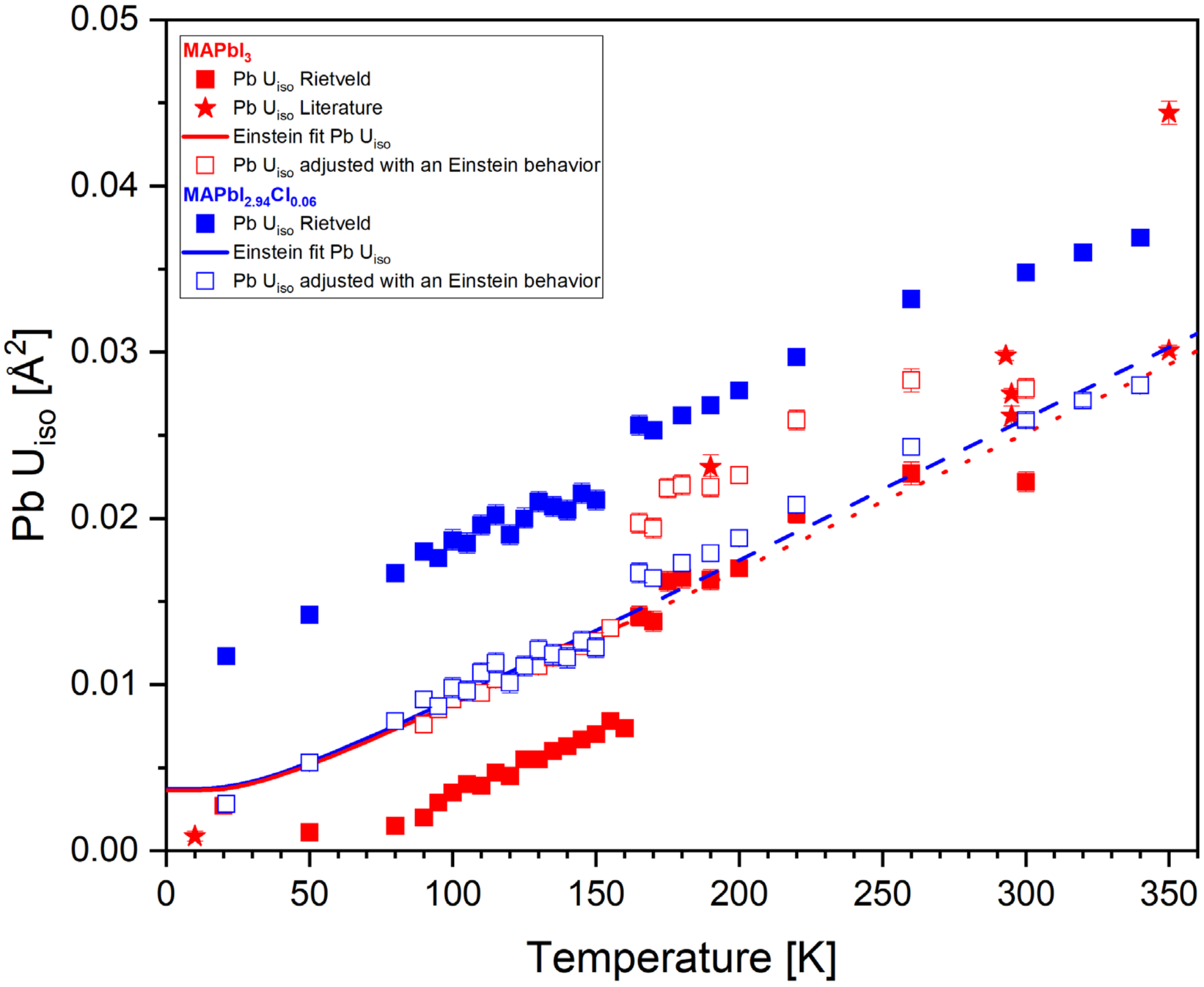

**Fig. S18** Isotropic temperature factors Pb $U_{iso}$ of the lead atoms as results of the Rietveld analysis of temperature-dependent synchrotron data from $MAPbI_3$ (red filled squares) and $MAPbI_{2.94}Cl_{0.06}$ (blue filled squares). For comparison, literature values are also given for $MAPbI_3$ (filled stars). The experimental isotropic temperature factors were fitted with an Einstein behavior (open symbols and drawn lines).

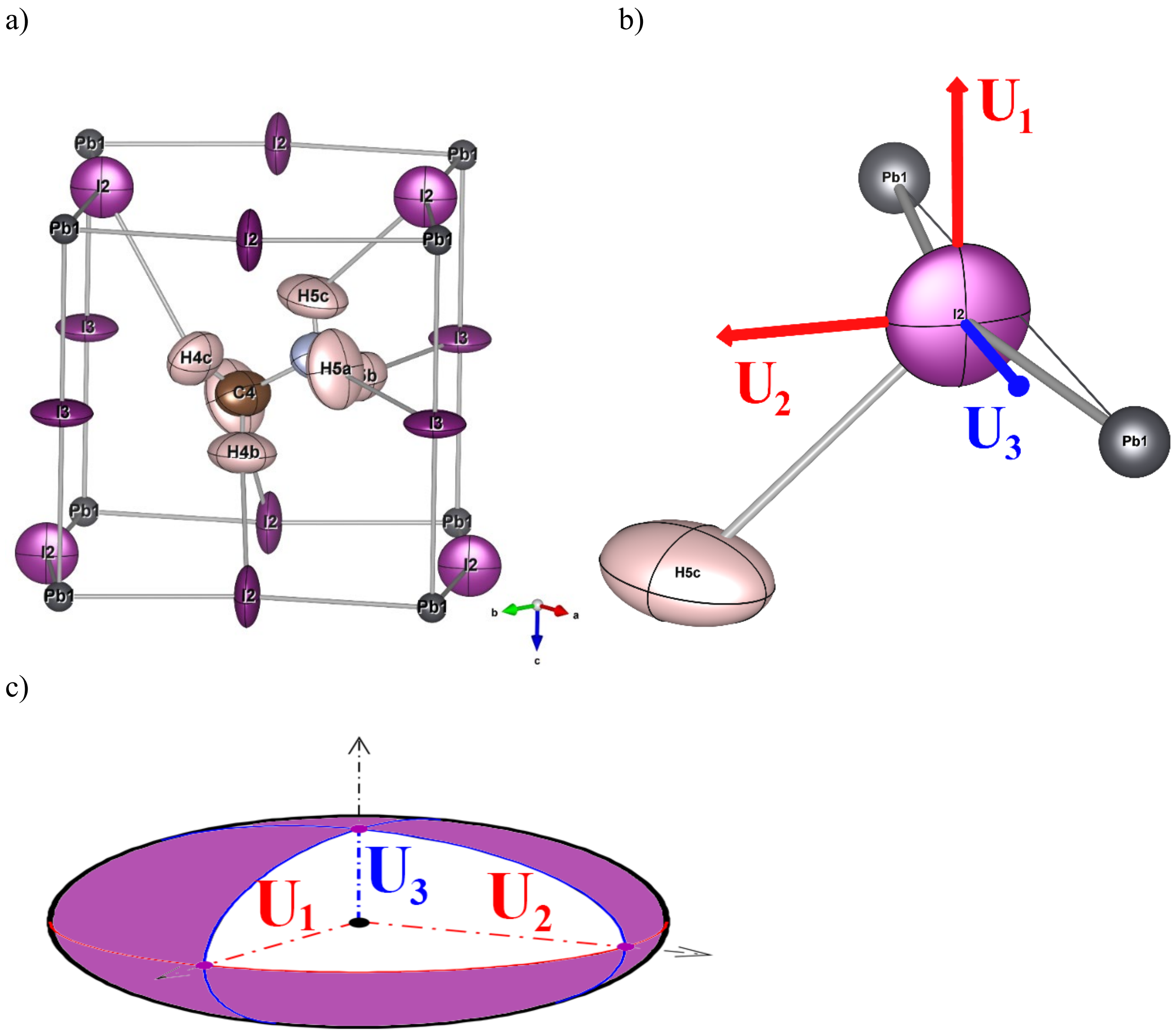

**Fig. S19** a) Crystal structure of $MAPbI_3$ at 295 K in space group *I*4/*mcm* (Franz et al. 2016. Note that the orientation of I2 is not drawn correctly in Fig. 3 of the original work).[5] Orientation of the $CH_3NH_3^+$ cation in the pseudo-cubic $[PbI_3]$ cage. The discus-shaped atomic displacement ellipsoids of the iodine atoms (in violet) are represented at 50 % probability. b) Enlargement of the atomic displacement ellipsoid of I2 from a) with principal axes of the anisotropic atomic displacement parameters. c) Principle representation of the discus-shaped atomic displacement ellipsoids of the iodine atom I2. The atomic displacement ellipsoid can be represented by the mean square displacement (MSD) parameters $U_1$, $U_2$ and $U_3$ (principal axes of the anisotropic atomic displacement parameters).[10] The averaged mean square displacement (MSD) $\bar{U}_1$ (perpendicular MSD) was determined from averaging the $U_1$ and $U_2$ values of the corresponding atom positions. $\bar{U}_3$ (parallel MSD) was determined from the averaged value from $U_3$.

**Table S3**: Iodide anisotropic displacement parameters (ADP) of $MAPbI_3$ at various temperatures from different diffraction experiments.[5-8] From the ADP's, the principal axes of the ADP (U1, U2, and U3) were calculated with Vesta.[9] The eigenvectors of $U_{cart}$ (ADP's with Cartesian basis) define the principal directions of the anisotropic displacement ellipsoid.[10] The averaged mean square displacement (MSD) $\overline{U}_1$ (perpendicular MSD) was then determined from the $U_1$ and $U_2$ values of the corresponding atom positions. For $\overline{U}_3$ (parallel MSD), the averaged value from $U_3$ was determined. The last column shows the isotropic displacement parameters for lead.

| **Iodide Anisotropic displacement parameters** [Å²] | | | | | | **Principal axes of the anisotropic atomic displacement parameters** (MSD) [Å²] for Iodide | | | **Averaged MSD** [Å²] ($\overline{U}_1$ averaged MSD $U_1$ and $U_2$ for each site; $\overline{U}_2$ averaged $U_3$ for each site) for Iodide | | **Isotropic** displacement parameter for **Pb** [Å²] |
|---|---|---|---|---|---|---|---|---|---|---|---|
| $U_{11}$ | $U_{22}$ | $U_{33}$ | $U_{12}$ | $U_{13}$ | $U_{23}$ | $U_1$ | $U_2$ | $U_3$ | $\overline{U}_{1Iodide}$ | $\overline{U}_{3Iodide}$ | $U_{isoPb}$ |
| Franz et al. (2016),[5] 295 K | | | | | | | | | | | |
| Site I2 | | | | | | | | | | | |
| 0.0686(8) | 0.0686(8) | 0.117(1) | −0.0490(9) | 0 | 0 | 0.118 | 0.117 | 0.020 | | | |
| Site I3 | | | | | | | | | | | |
| 0.116(1) | 0.116(1) | 0.020(1) | 0 | 0 | 0 | 0.116 | 0.116 | 0.020 | 0.117(1) | 0.020(1) | 0.0275(3) |
| Breternitz et al. (2020),[6] 293 K | | | | | | | | | | | |
| Site I1 | | | | | | | | | | | |
| 0.112(2) | 0.112(2) | 0.0190(9) | 0 | 0 | 0 | 0.112 | 0.112 | 0.019 | | | |
| Site I2 | | | | | | | | | | | |
| 0.0705(9) | 0.0705(9) | 0.112(2) | 0.048(1) | 0.010(3) | 0.010(3) | 0.130 | 0.101 | 0.023 | 0.114(2) | 0.021(1) | 0.0298(3) |
| Ren et al. (2016),[7] 350 K | | | | | | | | | | | |
| Site I1 | | | | | | | | | | | |
| 0.172(2) | 0.172(2) | 0.0225(7) | 0 | 0 | 0 | 0.172 | 0.172 | 0.023 | 0.172(2) | 0.023(1) | 0.0301(3) |
| Ren et al. (2016),[7] 295 K | | | | | | | | | | | |
| Site I1 | | | | | | | | | | | |
| 0.093(3) | 0.093(3) | 0.021(2) | 0 | 0 | 0 | 0.093 | 0.093 | 0.021 | | | |
| Site I2 | | | | | | | | | | | |
| 0.058(2) | 0.058(2) | 0.090(3) | 0 | 0 | 0.039(2) | 0.097 | 0.090 | 0.019 | 0.093(3) | 0.020(1) | 0.0262(4) |
| Whitfield et al. (2016),[8] 350 K | | | | | | | | | | | |
| Site I | | | | | | | | | | | |
| 0.025(2) | 0.176(2) | 0.176(2) | 0 | 0 | 0 | 0.176 | 0.176 | 0.025 | 0.176(2) | 0.025(2) | 0.0444(7) |
| Whitfield et al. (2016),[8] 190 K | | | | | | | | | | | |
| Site I1 | | | | | | | | | | | |
| 0.053(1) | 0.053(1) | 0.013(1) | 0 | 0 | 0 | 0.053 | 0.053 | 0.013 | | | |
| Site I2 | | | | | | | | | | | |
| 0.037(1) | 0.0367(8) | 0.057(1) | -0.0209(8) | 0 | 0 | 0.058 | 0.057 | 0.016 | 0.055(1) | 0.014(1) | 0.0231(7) |
| Whitfield et al. (2016),[8] 10 K | | | | | | | | | | | |
| Site I1 | | | | | | | | | | | |
| 0.0077(7) | 0.0004(5) | 0.0029(6) | 0 | 0 | 0 | 0.008 | 0.003 | 0.000 | | | |
| Site I2 | | | | | | | | | | | |
| 0.0025(4) | 0.0024(4) | 0.0025(4) | -0.0007(3) | 0.001(1) | 0.001(1) | 0.004 | 0.002 | 0.001 | 0.004(1) | 0.001(1) | 0.0008(3) |

a)

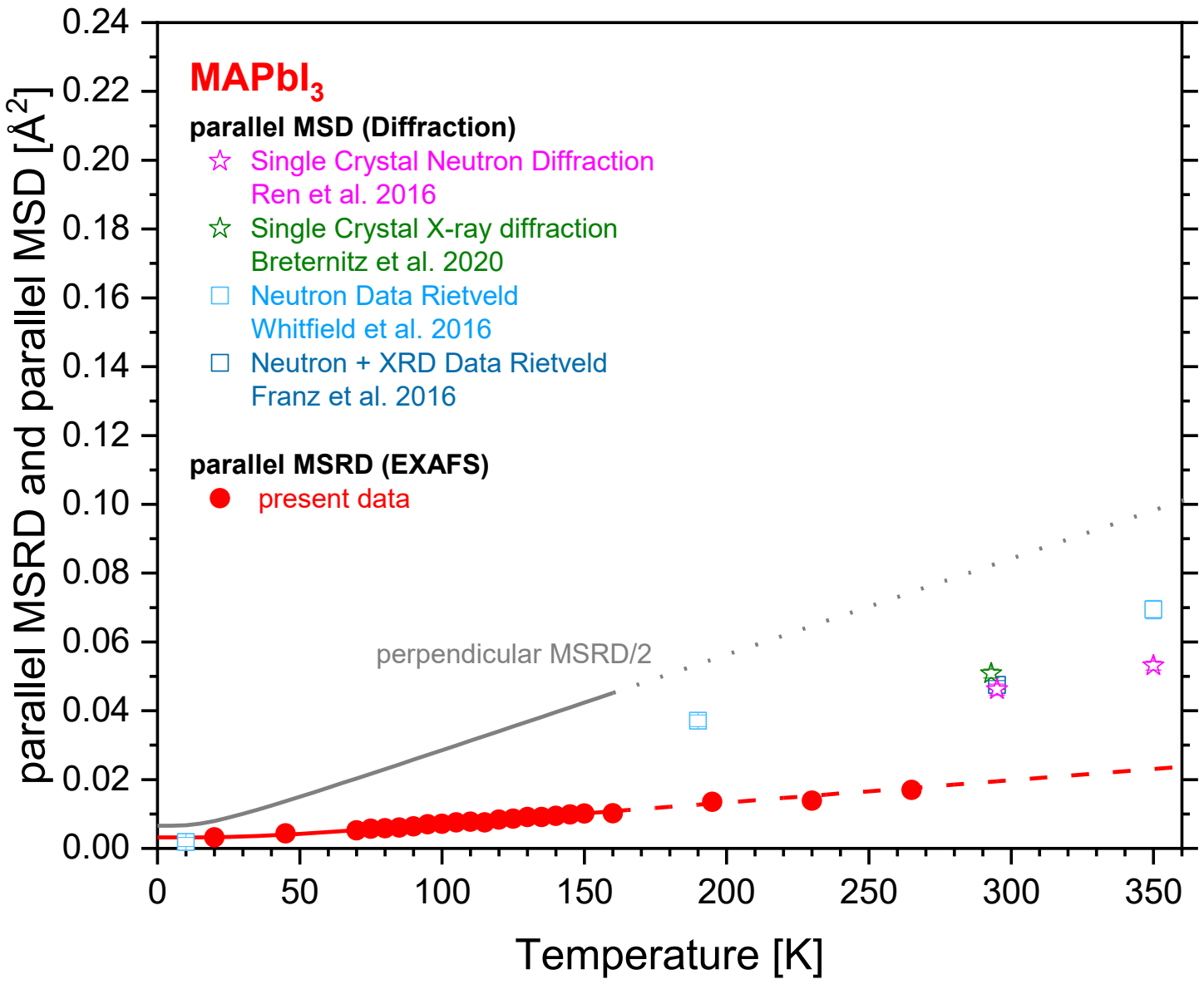

b)

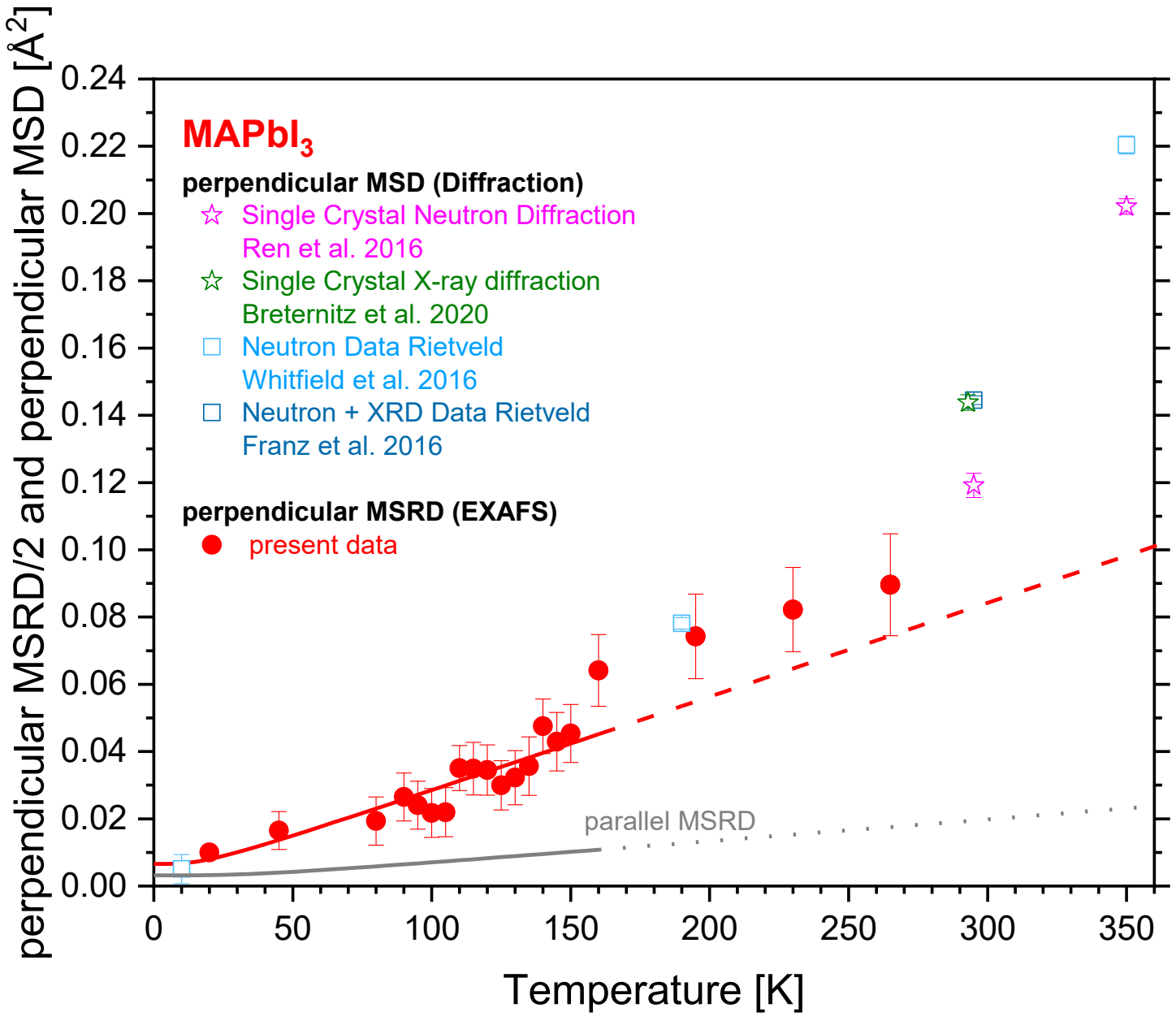

**Fig. S20** Temperature-dependent MSRD and MSD of $MAPbI_3$. a) solid red circles: $C_2$ (parallel MSRD); open symbols: the sum of the averaged mean square displacement $\overline{U}_3$ (parallel MSD) of the iodine atoms, and Pb $U_{iso}$ of the lead atom derived from diffraction literature values (Table S3). b) solid red circles: perpendicular MSRD/2; open symbols: the sum of the averaged mean square displacement $\overline{U}_1$ (perpendicular MSD), and the isotropic temperature factor $U_{iso}$ of the lead atom derived from diffraction literature values (Table S3). Red lines: a) parallel MSRD, b) perpendicular MSRD/2 Einstein behavior; grey lines: a) perpendicular MSRD/2, and b) parallel MSRD Einstein behavior; dashed and dotted lines: extrapolations of the low-temperature, Einstein model fits to higher temperatures.

a)

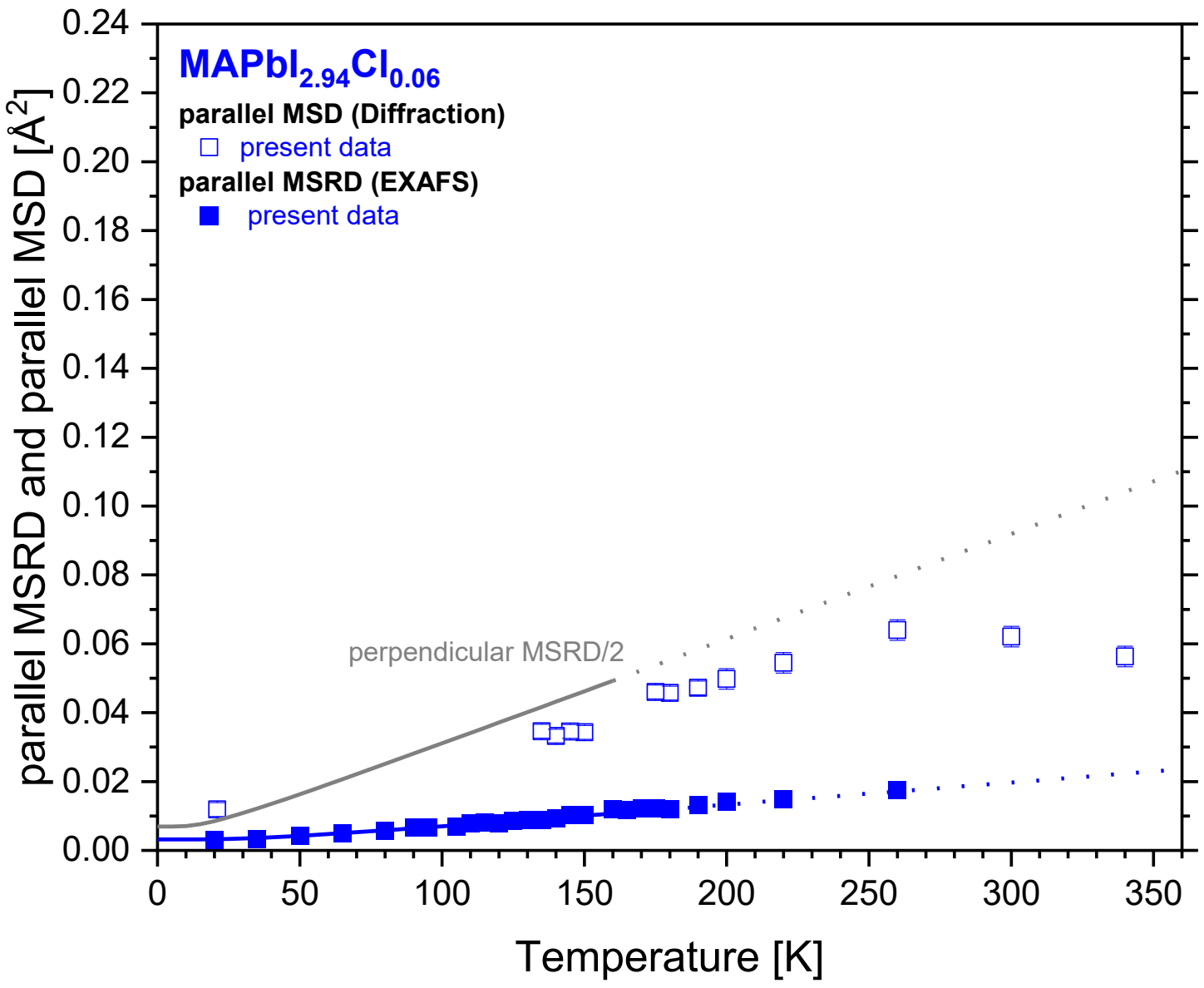

b)

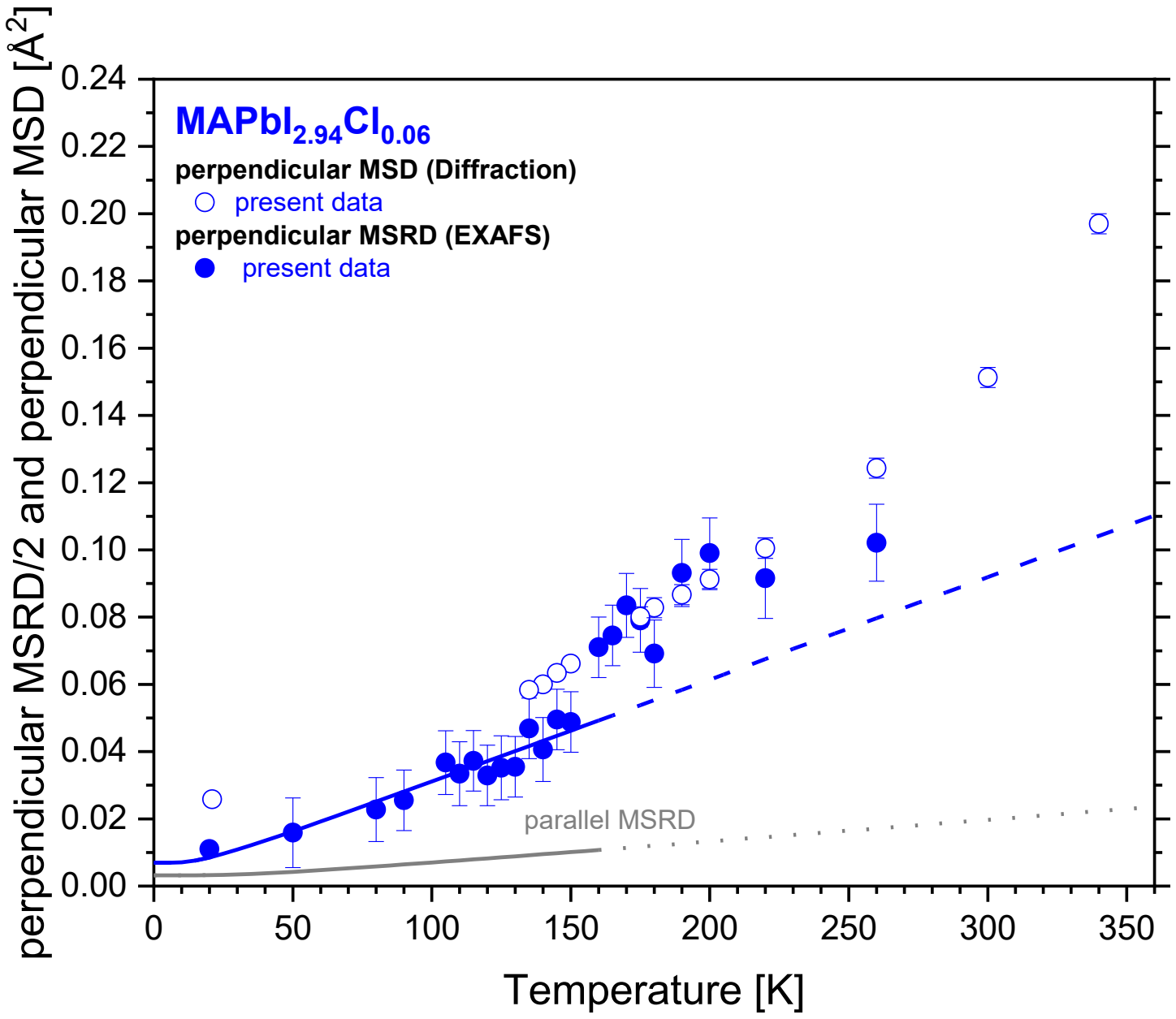

**Fig. S21** Temperature-dependent MSRD of $MAPbI_{2.94}Cl_{0.06}$ (blue), a) in addition to $C_2$ (parallel MSRD) the sum of the averaged mean square displacement $\overline{U}_3$ (parallel MSD) of the iodine atoms and the isotropic temperature factor $U_{iso}$ of the lead atom derived from diffraction is shown. In b), the perpendicular MSRD is shown, which was calculated based on $\Delta R_{XRD}$ and $\Delta R_{EXAFS}$. In addition, the sum of the averaged mean square displacement $\overline{U}_1$ (parallel MSD) and Pb $U_{iso}$ derived from diffraction is shown. Blue lines: a) parallel MSRD, b) perpendicular MSRD/2 Einstein behavior; grey lines: a) perpendicular MSRD/2, and b) parallel MSRD Einstein behavior; dashed and dotted lines: extrapolations of the low-temperature, Einstein model fits to higher temperatures.

a)

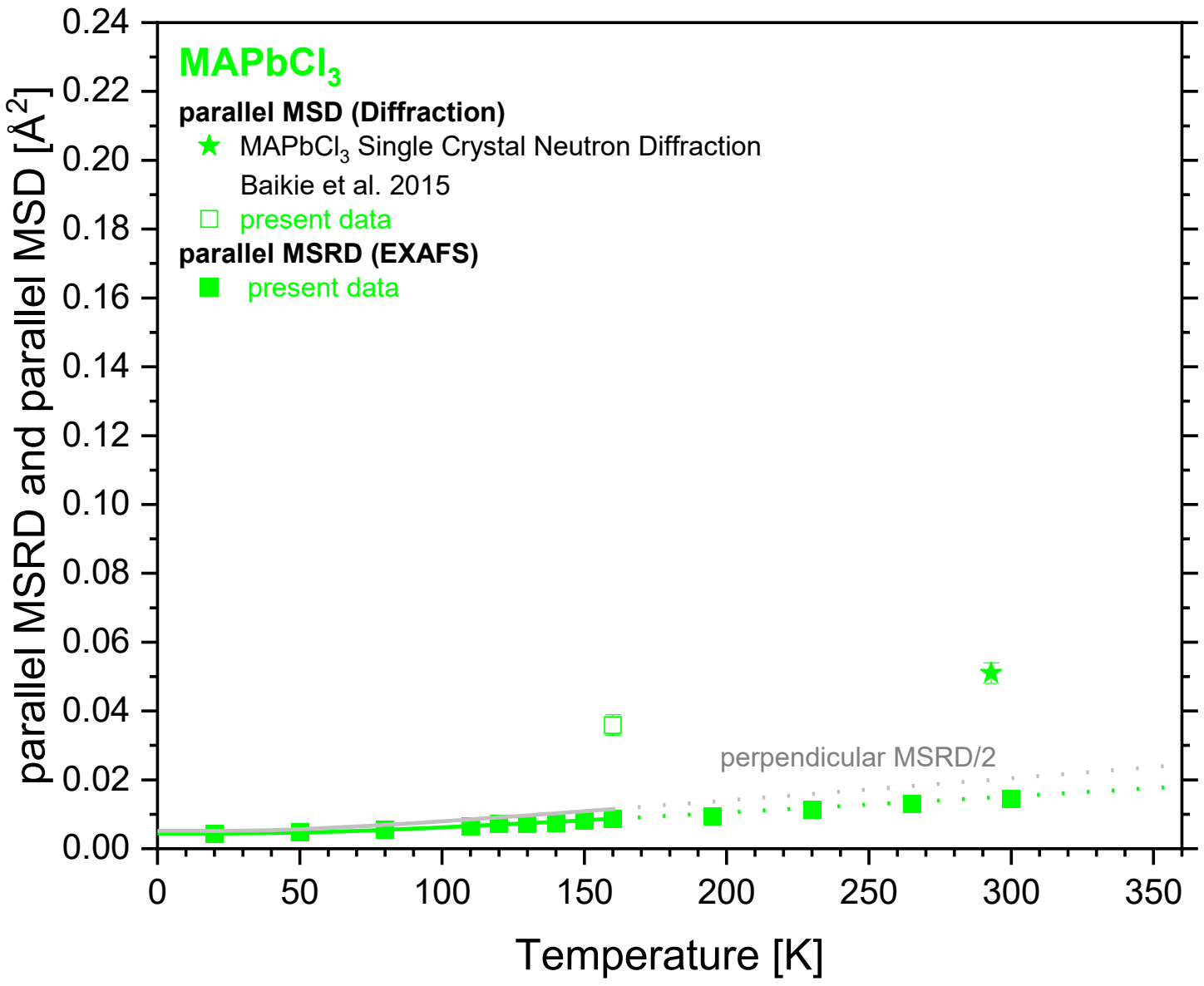

b)

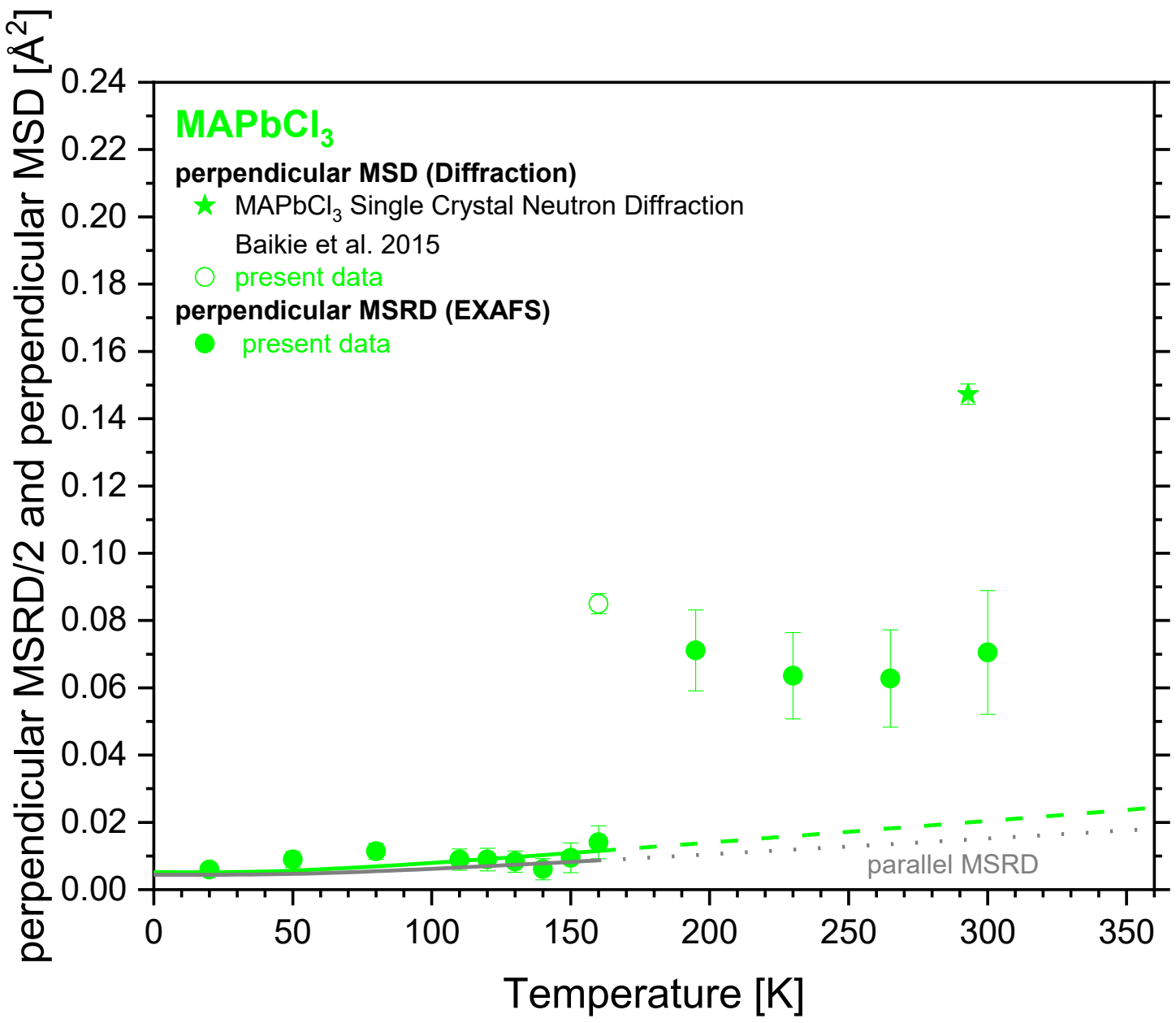

**Fig. S22** Temperature-dependent MSRD of $MAPbCl_3$ (green), a) in addition to $C_2$ (parallel MSRD) the sum of the averaged mean square displacement $\overline{U}_3$ (parallel MSD) of the iodine atoms and the isotropic temperature factor $U_{iso}$ of the lead atom derived from diffraction (Table S1) and Baikie et al. [2] is shown. In b), the perpendicular MSRD is shown, which was calculated based on $\Delta R_{XRD}$ and $\Delta R_{EXAFS}$. In addition, the sum of the averaged mean square displacement $\overline{U}_1$ (parallel MSD) and Pb $U_{iso}$ derived from diffraction (Table S1) and Baikie et al. [2] is shown. Green lines: a) parallel MSRD, b) perpendicular MSRD/2 Einstein behavior; grey lines: a) perpendicular MSRD/2, and b) parallel MSRD Einstein behavior; dashed and dotted lines: extrapolations of the low-temperature, Einstein model fits to higher temperatures.

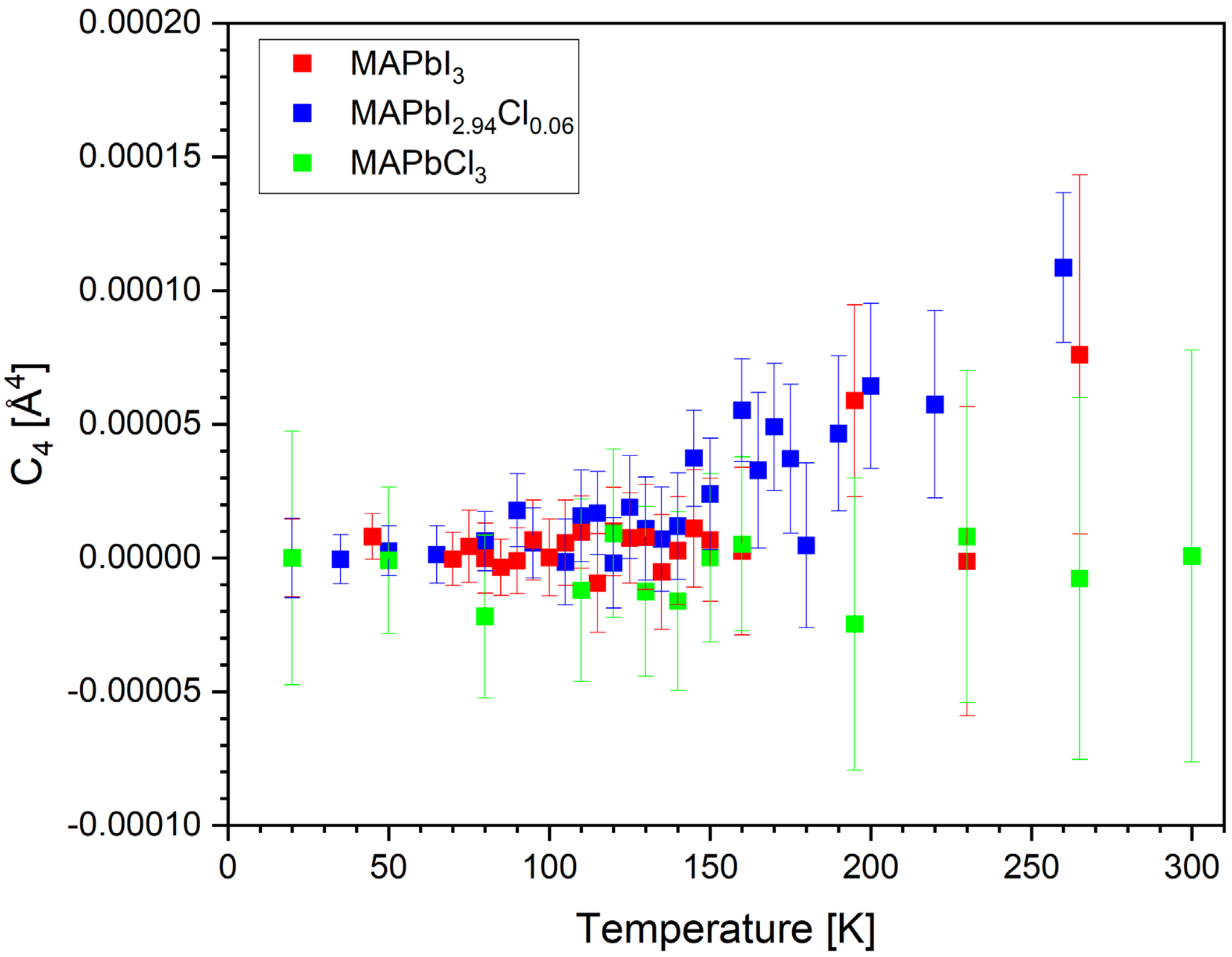

**Fig. S23** Temperature-dependent behavior of $C_4$ of $MAPbI_3$ (red), $MAPbI_{2.94}Cl_{0.06}$ (blue) and $MAPbCl_3$ (green).

**Calculation of the structural parameters $S_2$ and $S_3$, and local force constants for the orthorhombic crystal structures at 150 K of $MAPbCl_3$, $MAPbI_3$, and $MAPbI_{2.94}Cl_{0.06}$**

Using anharmonic effective potential

$$V_{eff}(x) = V(x) + \sum_{i=1,2}\sum_{j\neq i} V\left(\frac{\mu}{M_i}x\hat{R}_{12}\hat{R}_{ij}\right), \mu = \frac{M_1M_2}{M_1+M_2}, M_1 \equiv M_{Pb}, M_2 \equiv M_{I,Cl}, \hat{R}_{ij}$$

$$\equiv \hat{R}_{Pb-(I,Cl)}(i=1) or \hat{R}_{(I,Cl)-Pb}(i=2)$$

$$= V(x) + \sum_{i=1,2}\left[\varepsilon\sum_{j\neq i} V\left(\frac{\mu}{M_i}x\hat{R}_{12}\hat{R}_{ij}\right) + (1-\varepsilon)\sum_{j\neq i} V\left(\frac{\mu}{M_i}x\hat{R}_{12}\hat{R}_{ij}\right)\right], \varepsilon = \frac{N_{Pb-I}}{N_{Pb-I}+N_{Pb-Cl}}$$

$$\simeq V(x) + \varepsilon V\left(-\frac{M_{Pb}}{2(M_{Pb}+M_I)}x\right) + \varepsilon V\left(-\frac{M_I}{2(M_{Pb}+M_I)}x\right) + (1-\varepsilon)V\left(-\frac{M_{Pb}}{2(M_{Pb}+M_{Cl})}x\right)$$

$$+ (1-\varepsilon)V\left(-\frac{M_{Cl}}{2(M_{Pb}+M_{Cl})}x\right) + \theta(x)$$

where $\theta(x)$ is the negligible potential energy caused by the interaction of other surrounding atoms on the pair bond Pb-I, Pb-Cl.

Ignoring $\theta(x) << V_{eff}(x)$ and using the Morse potential $V(x) = D(e^{-2\alpha x} - 2e^{-\alpha x}) \simeq -D + D\alpha^2x^2 - D\alpha^3x^3$.

Then ignoring overall constant in $V_{eff}(x)$ and transforming $y = x - a = x - \langle x\rangle$:

$V_{eff}(y) \simeq \frac{1}{2}k_{eff}y^2 + k_{eff}y^3$,

$$k_{eff} = 2D(S_2\alpha^2 + 3S_3\alpha^3a) \approx 2D\alpha^2S_2 = 2D\alpha^2\left(1 + \frac{\varepsilon(M_{Pb}^2+M_I^2)}{4(M_{Pb}+M_I)^2} + \frac{(1-\varepsilon)(M_{Pb}^2+M_{Cl}^2)}{4(M_{Pb}+M_{Cl})^2}\right)$$

$$k_{3eff} = -D\alpha^3S_3 = -D\alpha^3\left(1 - \frac{\varepsilon(M_{Pb}^3+M_I^3)}{8(M_{Pb}+M_I)^3} - \frac{(1-\varepsilon)(M_{Pb}^3+M_{Cl}^3)}{8(M_{Pb}+M_{Cl})^3}\right)$$

Calculating for $MAPbI_3$ ($M_{Pb}$=207.20; $M_I$=126.90): $S_2 \simeq 1.1322, S_3 \simeq -0.9633$

$k_{eff} \simeq 2 \times 1.1322 \times D\alpha^2 = 1.3116$ eV/Å$^2$, $k_{3eff} \simeq -0.9633 \times D\ \ \alpha^3 = -0.8760$ eV/Å$^3$

Calculating for $MAPbI_{2.94}Cl_{0.06}$ ($M_{Pb}$=207.20; $M_{Cl}$=35.45; $M_I$=126.90): $S_2 \simeq 1.1333, S_3 \simeq -0.9625$

$k_{eff} \simeq 2 \times 1.1333 \times D\alpha^2 = 1.3229$ eVx/Å$^2$, $k_{3eff} \simeq -0.9625 \times D \quad \alpha^3 = -0.8989$ eV/Å$^3$

Calculating for $MAPbCl_3$ ($M_{Pb}$=207.20; $M_{Cl}$=35.45): $S_2 \simeq 1.1876, S_3 \simeq -0.9218$

$k_{eff} \simeq 2 \times 1.1876 \times D\alpha^2 = 1.7564$ eV/Å$^2$, $k_{3eff} \simeq -0.9218 \times D\alpha^3 = -1.1452$ eV/Å$^3$

**Table S4**: Morse parameter.

| | Mass ratio $M_1/M_2$ (Pb/Halid) | $S_2$ | $S_3$ | D [eV] | α [Å$^{-1}$] |
|---|---|---|---|---|---|
| $M_1=M_2$ | 1 | 1.125 | -0.9688 | - | - |
| $MAPbI_3$ | 1.63 | 1.1322 | -0.9633 | 0.235(6) | 1.57(4) |
| $MAPbI_{2.94}Cl_{0.06}$ | 1.66 | 1.1333 | -0.9625 | 0.228(6) | 1.60(4) |
| $MAPbCl_3$ | 5.84 | 1.1876 | -0.9218 | 0.262(14) | 1.68(9) |

Compare the calculated results with experimental data

| | $MAPbI_3$ | Calculation | $MAPbI_{2.94}Cl_{0.06}$ | Calculation | $MAPbCl_3$ | Calculation |
|---|---|---|---|---|---|---|
| μ | 78.7017 | | 77.9944 | | 30.2709 | |
| $\Theta_{E\parallel}$ [K] | 97.0(3) | 96.8745 | 97.8(4) | 97.7594 | 180.5(8) | 180.7781 |
| $\nu_{E\parallel}$ [THz] | 2.021(5) | 2.0184 | 2.037(9) | 2.0362 | 3.76(2) | 3.7658 |
| $k_\parallel$ [eV/Å$^2$] | 1.315(3) | 1.3116 | 1.324(6) | 1.3229 | 1.751(9) | 1.7564 |
| $k_3$ [eV/Å$^3$] | -0.88(2) | -0.8760 | -0.90(2) | -0.8989 | -1.14(5) | -1.1452 |
| $\lvert k_3 \rvert / k_0$ | 0.67(2) | 0.6679 | 0.68(2) | 0.6795 | 0.65(3) | 0.6520 |

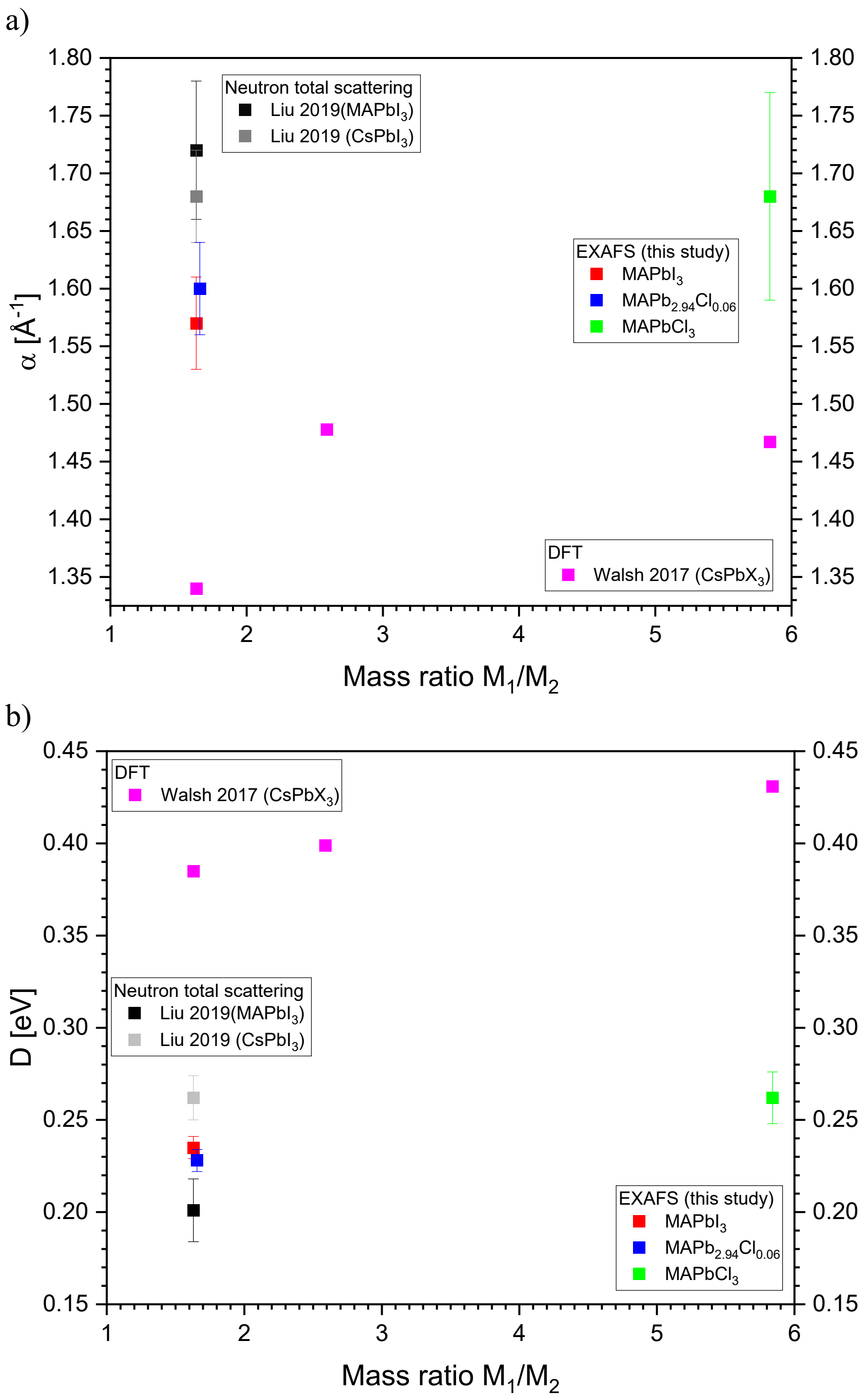

**Fig. S24** Comparison of the experimental Morse potential parameters D and α as a function of the mass ration $M_1/M_2$ of the Lead-Halide bond for $MAPbI_3$ (red), $MAPbI_{2.94}Cl_{0.06}$ (blue), and $MAPbCl_3$ (green) with experimental results from Liu 2019 [17,18] for $MAPbI_3$ (black) and $CsPbI_3$ (gray) with Morse potentials from DFT calculations from Walsh 2017 [19] for $CsPbX_3$ (magenta).

## AUTHOR INFORMATION

**Corresponding Author**

*Götz Schuck, e-mail: goetz.schuck@helmholtz-berlin.de